\documentclass[11pt]{article}

\pdfoutput=1

\usepackage{epsfig}
\usepackage{cite}
\usepackage{amsmath}
\usepackage[footnotesize]{caption}
\usepackage{slashed}
\usepackage[utf8]{inputenc}

\numberwithin{equation}{section}

\newcommand{\be}{\begin{equation}}
\newcommand{\ee}{\end{equation}}
\newcommand{\beq}{\begin{eqnarray}}
\newcommand{\eeq}{\end{eqnarray}}

\newcommand{\lsim}{\raisebox{-0.13cm}{~\shortstack{$<$ \\[-0.07cm]
      $\sim$}}~}

\begin{document}

\title{
\textbf{Scalar mass dependence of angular variables in $t\bar t\phi$ production \\[4mm]}}

\date{}
\author{
Duarte Azevedo$^{1\,}$\footnote{E-mail:
  \texttt{drpazevedo@fc.ul.pt}} ,
Rodrigo Capucha$^{1\,}$\footnote{E-mail:
\texttt{rodrigocapucha@hotmail.com}} ,
Ant\'{o}nio Onofre$^{2\,}$\footnote{E-mail:
\texttt{antonio.onofre@cern.ch}} ,
Rui Santos$^{1,3\,}$\footnote{E-mail:
  \texttt{rasantos@fc.ul.pt}} 
\\[5mm]
{\small\it $^1$Centro de F\'{\i}sica Te\'{o}rica e Computacional,
    Faculdade de Ci\^{e}ncias,} \\
{\small \it    Universidade de Lisboa, Campo Grande, Edif\'{\i}cio C8
  1749-016 Lisboa, Portugal} \\[3mm]
{\small\it
$^2$ Departamento de F\'{\i}sica , Universidade do Minho, 4710-057 Braga, Portugal} \\[3mm]
{\small\it
$^3$ISEL -
 Instituto Superior de Engenharia de Lisboa,} \\
{\small \it   Instituto Polit\'ecnico de Lisboa
 1959-007 Lisboa, Portugal} \\[3mm]
}

\maketitle

\begin{abstract}
\noindent
In this paper we explore CP discrimination in the associated production of top-quark pairs ($t\bar{t}$) with a generic scalar boson ($\phi$) at the LHC. We probe the CP-sensitivity of several observables for a varying scalar boson mass and CP-number, either CP-even ($\phi=H$) or CP-odd ($\phi=A$), using dileptonic final states of the $t\bar{t}\phi$ system, with $\phi\rightarrow b\bar{b}$. We show that CP-searches are virtually impossible for $\phi$ boson masses above a few hundred GeV in this channel. A full phenomenological analysis was performed, using Standard Model background and signal events generated with \texttt{MadGraph5\_aMC@NLO} and reconstructed using a kinematic fit. The most sensitive CP-observables are used to compute Confidence Levels (CLs), as a function of luminosity, for the exclusion of different signal hypotheses with scalar and pseudoscalar boson masses that range from $m_\phi$ = 40~GeV up to 200~GeV. We finalize by analysing the impact of a measurement (or limit) of the CP-violating angle in the parameter space of a complex two-Higgs doublet model known as the C2HDM.
\end{abstract}
\thispagestyle{empty}
\vfill
\newpage
\setcounter{page}{1}

\section{Introduction}
\hspace{\parindent} 
 
The CP-nature of the discovered Higgs boson is still an open issue. Although the ATLAS and CMS collaborations established that 
the discovered 125 GeV Higgs~\cite{Aad:2012tfa, Chatrchyan:2012ufa} cannot be a pure pseudoscalar with more than 99\% confidence level (CL), a mixed state
with a significant CP-odd component is still possible. The need for further sources of CP-violation was first discussed by Sakharov as one 
of the three conditions for baryogenesis to occur~\cite{Sakharov:1967dj}. This is an important motivation to look for physics
Beyond the Standard Model (BSM) and, in particular, to models with extra sources of CP-violation. One of the simplest extensions of 
the Standard Model (SM) we can build  with a CP-violating scalar sector is to add an extra scalar doublet to the SM field content
while keeping the same gauge symmetries. The CP-conserving version of the model is commonly referred to as
two-Higgs doublet model (2HDM), while the simplest CP-violating version of the model is known as complex 2HDM (C2HDM)
and has been the subject of many studies~\cite{Lee:1973iz, Ginzburg:2002wt, Khater:2003wq, ElKaffas:2007rq, Grzadkowski:2009iz, Arhrib:2010ju, Barroso:2012wz, 
Inoue:2014nva,Cheung:2014oaa, Fontes:2014xva, Fontes:2015mea, Chen:2015gaa, Muhlleitner:2017dkd, Fontes:2017zfn}.
Due to its simplicity, the C2HDM is seen as an ideal benchmark model to test the CP quantum numbers of scalars at the LHC.
In the C2HDM, all three neutral scalars may have a mixture of CP-even and CP-odd components and there is no restriction to the mass of these
scalars. Although one of the scalars has to be the discovered 125 GeV Higgs boson, it can be any of the three neutral scalars predicted by the theory, from the lightest to the heaviest.

The search for BSM physics is a major goal of the LHC experiments. The measurement of the Yukawa couplings has become a primary target, since it is decisive to establish the CP-nature of the scalars in case
a new scalar is discovered. 
The relation between the scalar and the pseudoscalar components in the Yukawa couplings can be directly probed either in the production or in the decays of the scalars, 
depending on the final state fermions. Examples of the use of asymmetries to probe the CP-nature of the
Higgs boson in the top quark Yukawa coupling were discussed in~\cite{Gunion:1996xu, Boudjema:2015nda, Santos:2015dja, AmorDosSantos:2017ayi}, 
while the decays of the tau leptons were used to probe the $\tau$-lepton Yukawa coupling~\cite{Berge:2008wi, Berge:2008dr, Berge:2011ij, Berge:2014sra, Berge:2015nua}.
These studies for the top-quark and for the $\tau$-lepton are now being discussed in detail by the ATLAS and CMS collaborations. In the case of the top quark we will be probing the Yukawa coupling directly in the production
process. For the $\tau$-leptons, the Higgs decay is used. It is important to point out that there is still room for very large pseudoscalar components in the couplings 
to $b$-quarks and $\tau$-leptons, for any of the scalars in some Yukawa versions of the C2HDM, considering the recently announced ACME collaboration's constraint on the electron EDM,
\begin{equation}
	|d_e|< 1.1 \times 10^{-29}  \text{ e cm},
\end{equation}
from measurements of the $\mathrm{ThO}$ molecule\cite{Andreev:2018ayy},
as recently reported in \cite{talk_DF, talk_RS}.

In this work we will examine in detail how the asymmetries and the angular variables distributions previously proposed change when the mass of the scalar differs from the measured Higgs mass. Although many studies have examined several angular variables in $t\bar t\phi$ production (with $m_\phi=125$ GeV), a detailed study for a scalar with a mass either below or above this value is still not available in the literature.
We will build on a series of papers where the issue of the CP-nature of the discovered Higgs boson was thoroughly studied in associated production with top quark pairs~\cite{AmorDosSantos:2017ayi, Santos:2015dja, Azevedo:2017qiz, Ferroglia:2019qjy}.  We will discuss the same set of angular variables with several goals in mind. The first one is to answer the question: if a new scalar or pseudoscalar boson exists, what is the confidence level to exclude a signal hypothesis (either CP-even or CP-odd) assuming the SM holds, as a function of the LHC luminosity and $\phi$ boson mass? The second one focuses on determining the confidence level for exclusion of a pure CP-odd signal in case a new massive scalar boson is found, as a function of the $\phi$ boson mass and the LHC luminosity. The third one relates to setting the confidence level for the exclusion of the SM (once again as a function of mass and luminosity) assuming a new CP-even scalar particle signal is found.

The outline of the paper is as follows.
In section~\ref{sec:TH}, we describe the $\phi$ boson mass dependence of the several angular distributions to be studied.
In section \ref{sec:disc}, we present and discuss our main results. In section~\ref{sec:C2HDM}, we consider the impact of the discovery of a new Higgs boson on the parameter space of the main benchmark model for CP-violation studies, the complex version of the two-Higgs doublet model (C2HDM). Our conclusions are presented in section~\ref{sec:concl}.

\section{Theoretical limitations on asymmetries measurements \label{sec:TH}}
\hspace{\parindent} 

The most general Yukawa interaction of a boson ($\phi$), with no definite value of CP, to a top quark pair can be written as 
\begin{equation}
{\cal L} = \kappa_t y_t  \bar t (\cos \alpha + i \gamma_5 \sin \alpha) t \phi \, ,
\label{eq:higgscharacter}
\end{equation}
where $y_t$ is the SM Yukawa coupling, $\kappa_t$ parametrises the total coupling strength relative to the SM and the angle $\alpha$ parametrises the CP-phase, which is related to the parameters in the Higgs potential. We will refer to $\phi = H$ for the pure CP-even scenario and $\phi=A$ for the pure CP-odd case. The pure CP-even case is recovered by setting $\cos \alpha = \pm 1$ while the pure CP-odd case is obtained by fixing $\cos \alpha = 0$.

In previous works~\cite{Gunion:1996xu, Boudjema:2015nda, Santos:2015dja, AmorDosSantos:2017ayi} several angular variables were proposed, not only to increase the sensitivity in discriminating signals from irreducible backgrounds at the LHC in $t\bar t\phi$ final states, but also as a means to probe the CP nature of the Yukawa coupling in $t\bar t\phi$ production at the LHC. The results in~\cite{Santos:2015dja, AmorDosSantos:2017ayi} showed that we can define a minimal set of variables
to obtain the best possible sensitivity, to achieve both goals in a very effective way. While these studies assumed a mass of 125~GeV for the $\phi$ boson, in this paper we extend their use to a wider mass range, from 40~GeV to 500~GeV. This is discussed in the following sections. 

\subsection{$t\bar{t}H$ and $t\bar{t}A$ angular distributions}
\hspace{\parindent} 

A first set of variables is introduced~\cite{Santos:2015dja} using $\theta^X_Y$,  defined as the angle between the direction of the $Y$ system 3-momentum (in the rest frame of $X$) with respect to the momentum direction of the $X$ system (in the rest frame of its parent system). When reconstructing the signal angular distributions, we consider successive two body decays of the $t\bar{t}\phi$ system down to the final state particles i.e., the quarks (or jets), the charged and the neutral leptons, which originated from the decays of the $t$, $\bar{t}$ quarks and $\phi$ boson. If the decay chain of the $t\bar{t}\phi$ system is labelled $(123)$, the successive decays considered include all possible combinations of the type $(123)\rightarrow1+(23)$, $(23)\rightarrow2+(3)$ and $(3)\rightarrow4+5$. We then build three families of observables: $f(\theta^{123}_1)g(\theta^{3}_4)$, $f(\theta^{123}_1)g(\theta^{23}_3)$ and $f(\theta^{23}_3)g(\theta^{3}_4)$, with $f,g=\{\sin,\cos\}$. 
The momentum direction of the $(123)$ system is measured with respect to the laboratory (LAB) frame, where the net 3-momentum of the protons colliding is zero. Particles 1 to 3 are either the $t$ or the $\bar{t}$ quark, or the Higgs boson, while
particle 4 can be any of the products of the decay of the top quarks and the Higgs boson, including the intermediate $W$ bosons. We use two ways of computing particle 4 Lorentz vector in the centre-of-mass of particle 3. One is by using the laboratory four-momentum of both particles 3 and 4, and boost particle 4 directly to the centre-of-mass frame of particle 3 ({\it direct} boost). The other, is to boost particles 3 and 4 sequentially through all intermediate centre-of-mass systems until particle 4 is evaluated in the centre-of-mass frame of particle 3 ({\it sequential} boost or {\it seq.} boost). 

We will also use the variables $b_2$ and $b_4$ as defined in~\cite{Gunion:1996xu, Ferroglia:2019qjy} in the LAB and $t\bar{t}\phi$ centre-of-mass systems ($b_2^{t\bar{t}\phi}$ and $b_4^{t\bar{t}\phi}$, respectively),
\begin{equation}
b_2  = ( \vec{p}_{t} \times \hat{k}_z ).( \vec{p}_{\bar{t}} \times \hat{k}_z )/ (|\vec{p}_{t}| .  |\vec{p}_{\bar{t}}|), \ \ \ \  b_4 = (p^z_t . p^z_{\bar{t}}) / (|\vec{p}_{t}| . |\vec{p}_{\bar{t}}| ),
\end{equation}
where the $z$-direction corresponds to the beam line. 
It is worth noting that $b_2$ and $b_4$ have a natural physics interpretation. They depend on the $t$ and $\bar{t}$ polar angles, $\theta_t$ and $\theta_{\bar{t}}$ respectively, with respect to the $z$-direction, and can be expressed as  $b_2=\sin{\theta_t} \times \sin{\theta_{\bar{t}}}$ and $b_4=\cos{\theta_t} \times \cos{\theta_{\bar{t}}}$.

Forward-backward asymmetries associated to each of the observables under study were defined according to~\cite{Santos:2015dja}
\begin{eqnarray}
\ensuremath{A^Y_{FB}= \frac  {    \sigma(x_Y>x_Y')-\sigma(x_Y<x_Y')   }{ \sigma( x_Y>x_Y')+\sigma(x_Y<x_Y') } },
\label{equ:AFB}
\end{eqnarray}
where $\sigma(x_Y>x_Y')$ and $\sigma(x_Y<x_Y')$ correspond to the total cross section for $x_Y$ above and below $x_Y'$. The latter is the central value of the $x_Y$ domain.

The reason why these distributions allow us to probe the CP-nature of a scalar in the $t \bar t \phi$ coupling lies
ultimately in the behaviour of the cross section as a function of the particle's CP value. In fact, as discussed 
in~\cite{Gunion:1996xu}, the amplitude for the process $pp \to t \bar t \phi$ has two terms: one that
does not depend on the mixing angle, $\alpha$, and another that is proportional to $\cos 2 \alpha$. Hence, only
the latter is sensitive to a CP-odd component of the Yukawa coupling. This term is proportional to the top quark mass 
and therefore its contribution is important, as long as the Higgs boson mass is of the same order of magnitude. One could ask if the 
process  $pp \to b \bar b \phi$ could be used to probe the Yukawa structure of the $b \bar b \phi$ vertex.
The answer is clearly negative because the interference term is now proportional to $m_b^2$, that is,
at least three orders of magnitude smaller.
\begin{figure}[h!]
\begin{tabular}{ccc}
\hspace*{-5mm}\includegraphics[height = 5.1cm]{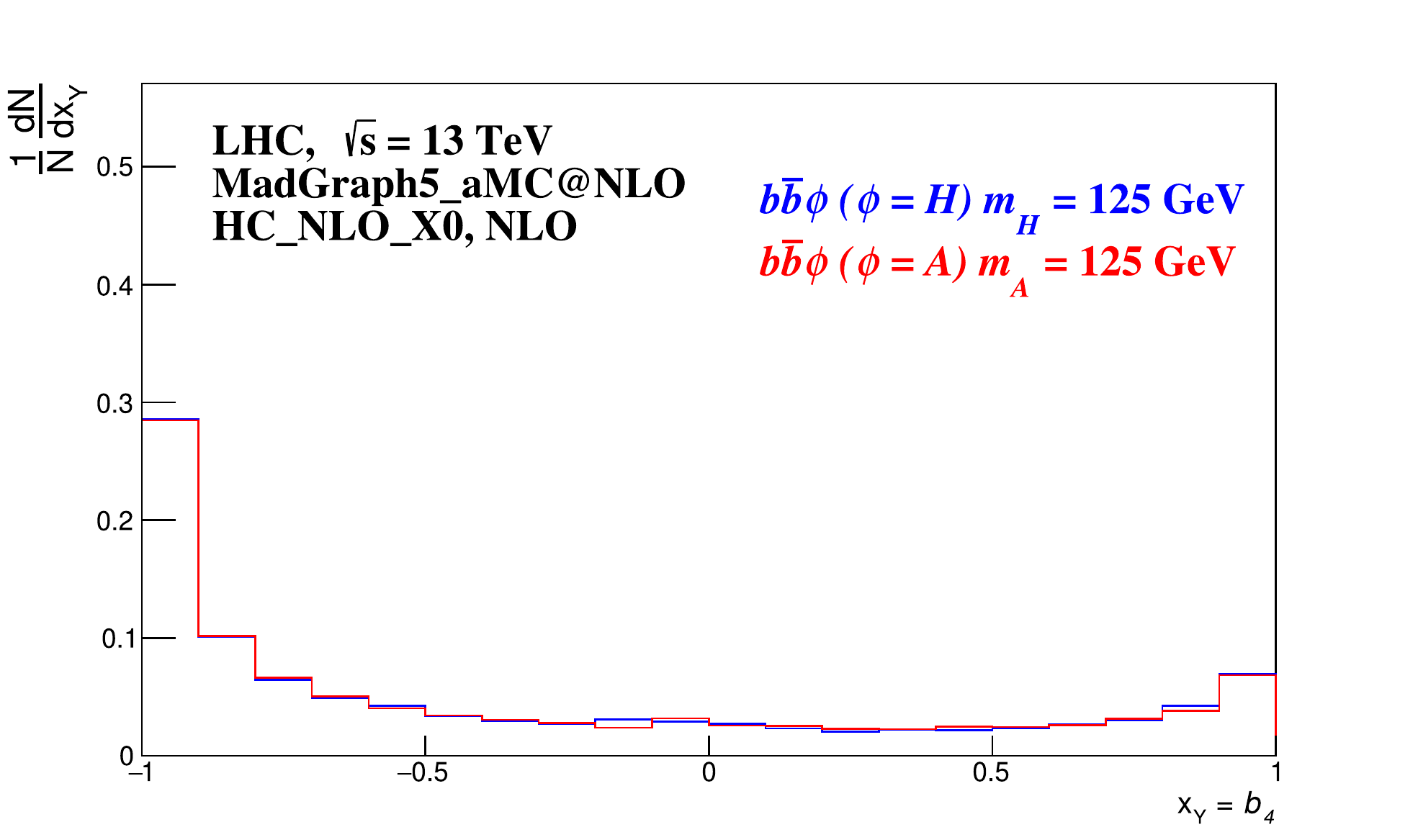} \hspace*{-5mm} \includegraphics[height = 5.1cm]{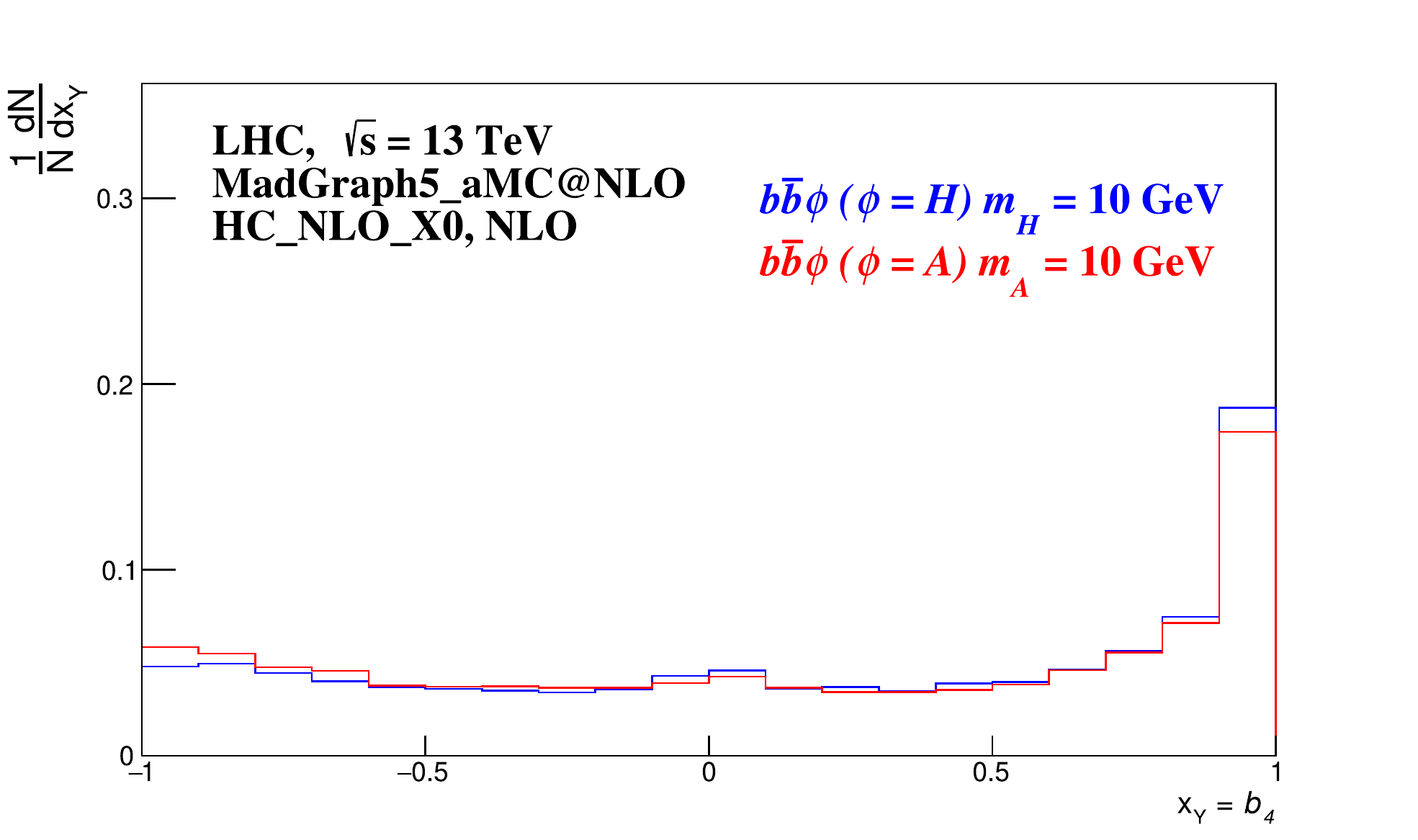}
\end{tabular}
\caption{Parton level $b_4$ distributions at Next-to-leading order (NLO), normalized to unity, for $m_\phi= 125$ GeV (left) and $m_\phi= 10$ GeV (right).}
\label{fig:bbphi}
\end{figure}
In the left panel of Figure~\ref{fig:bbphi}, we present the $b_4$ distribution, at parton level, for the process  $pp \to b \bar b  \phi$ for $m_\phi= 125$ GeV.
In blue, we present the pure scalar case while in red we show the pure pseudoscalar one. As expected no difference is found in the distributions.
We have checked that the distributions of all other angular variables follow the same trend and again no difference was seen. 
Finally we repeated the procedure for a very light scalar, with a mass of $m_\phi = 10 $ GeV,  with similar null results as we show on the right side of the same figure.
The case of the $b\bar b\phi$ final state has been previously discussed in~\cite{Gritsan:2016hjl, Ghosh:2019dmv}.

\begin{figure}[h!]
	\centering
\includegraphics[height=5.5cm,angle=0]{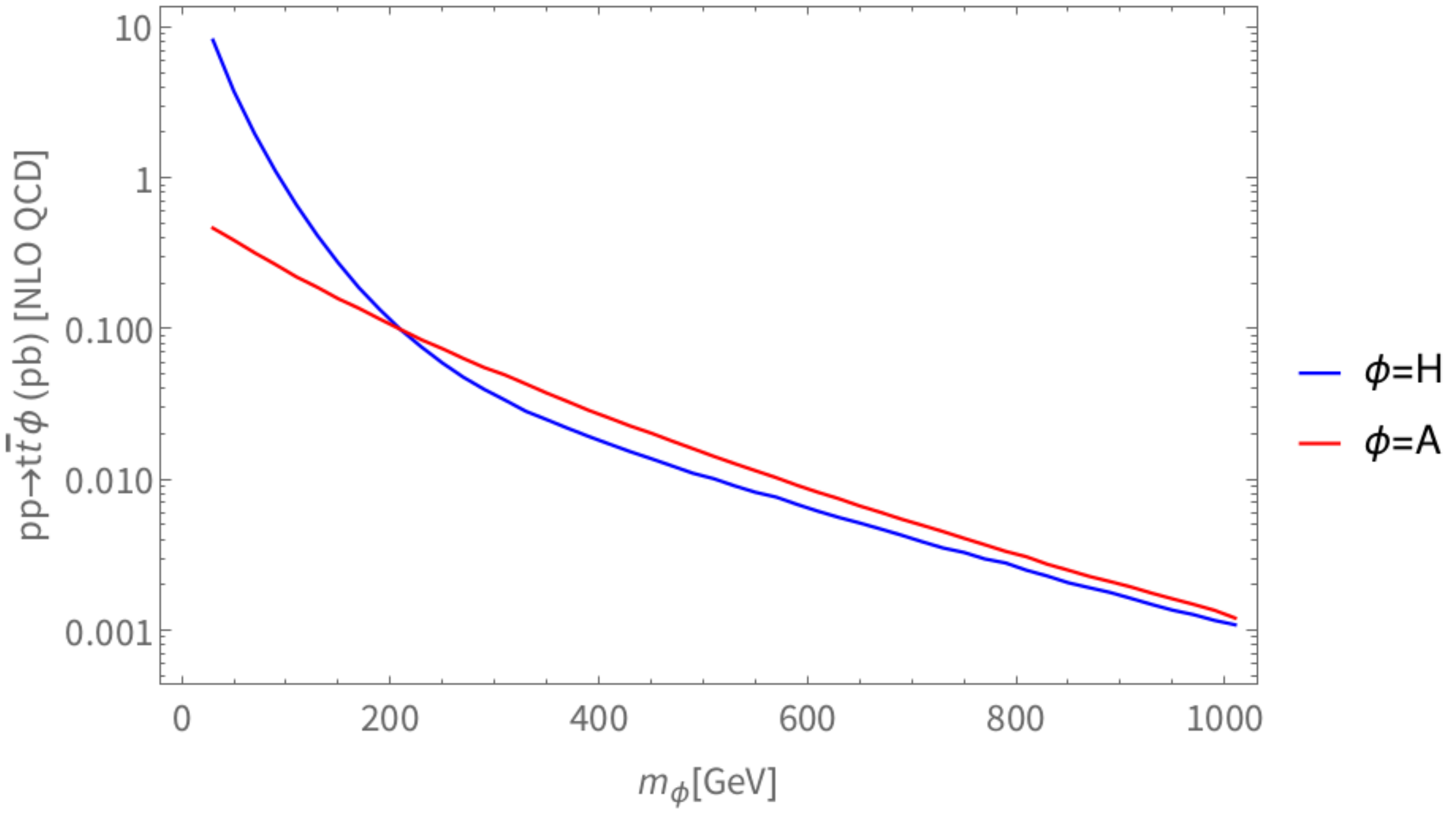}
\caption{Total cross section for the process $pp \to t\bar tH$ (blue) and $pp \to t\bar tA$ (red) as a function of the $\phi$ boson mass, at NLO, for a centre-of-mass energy of 13 TeV, at the LHC. Details about the generation of the events represented through Figures \ref{fig:mass}, \ref{fig:b2b4}, \ref{fig:b2b4cm} and \ref{fig:afb}, can be found in section 3.1.}
\label{fig:mass}
\end{figure}
\begin{figure}[h!]
	\begin{center}
		\begin{tabular}{ccc}
			\hspace*{-5mm}\includegraphics[height=5.1cm]{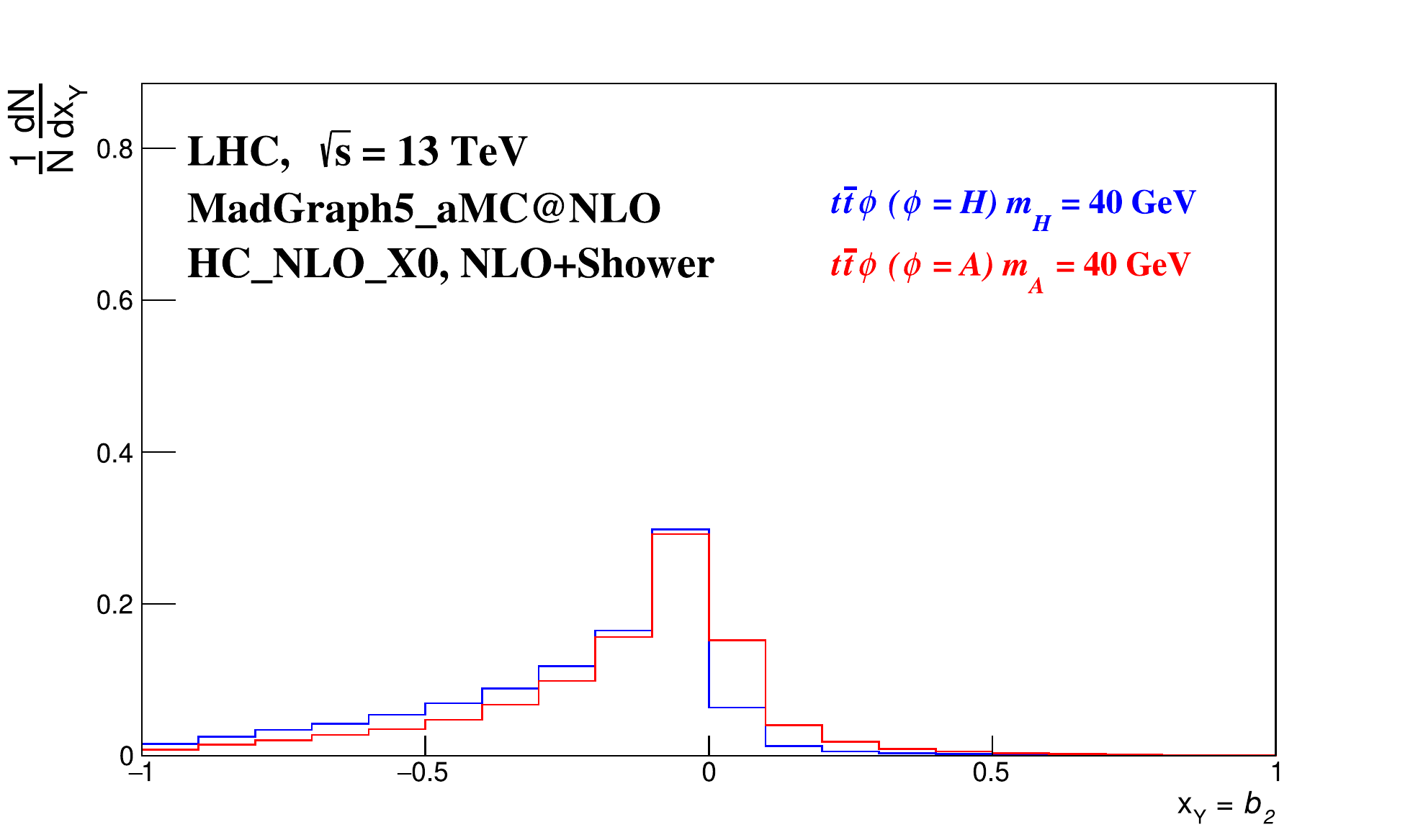}
			\hspace*{-5mm}\includegraphics[height=5.1cm]{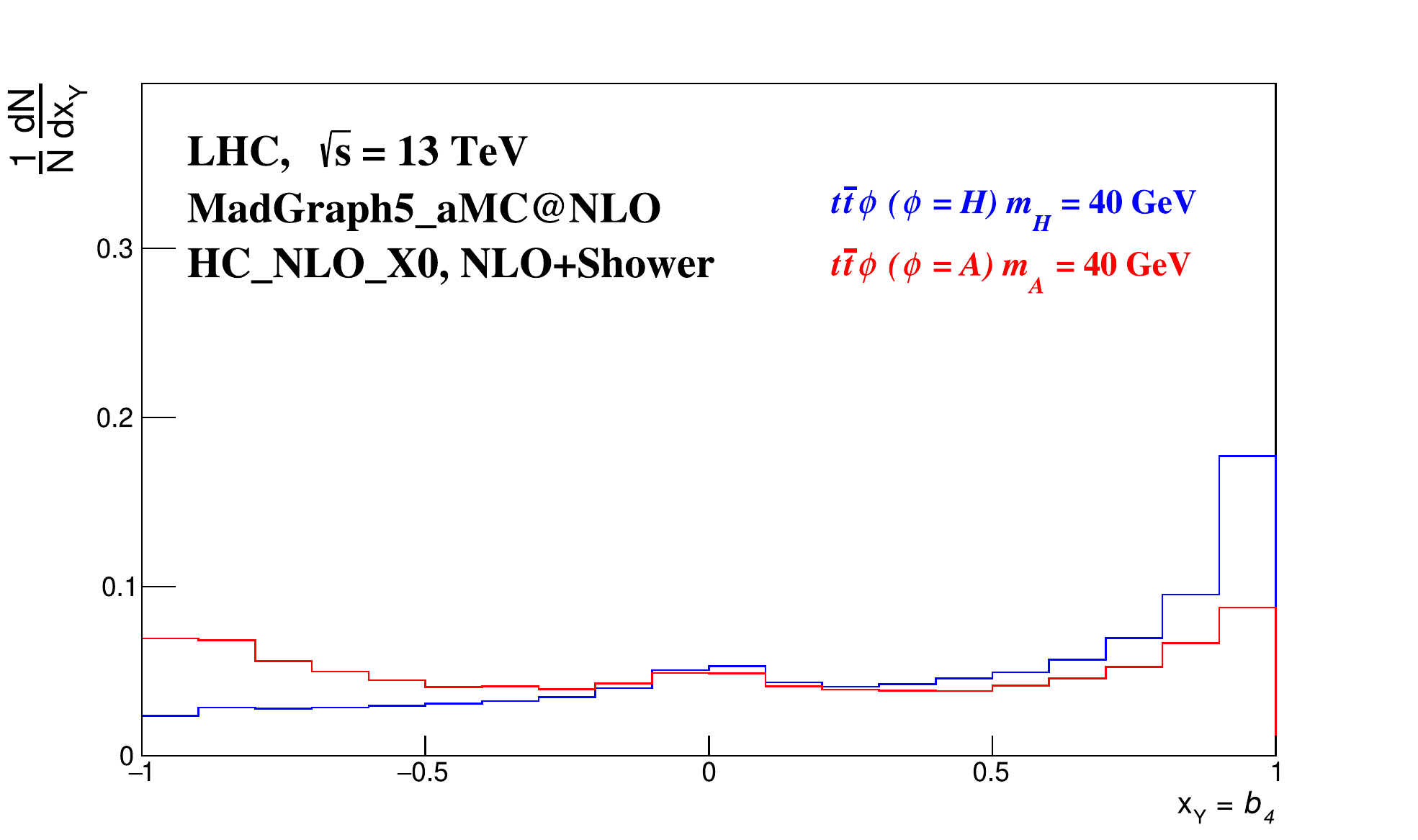}
			\\
			\hspace*{-5mm}\includegraphics[height=5.1cm]{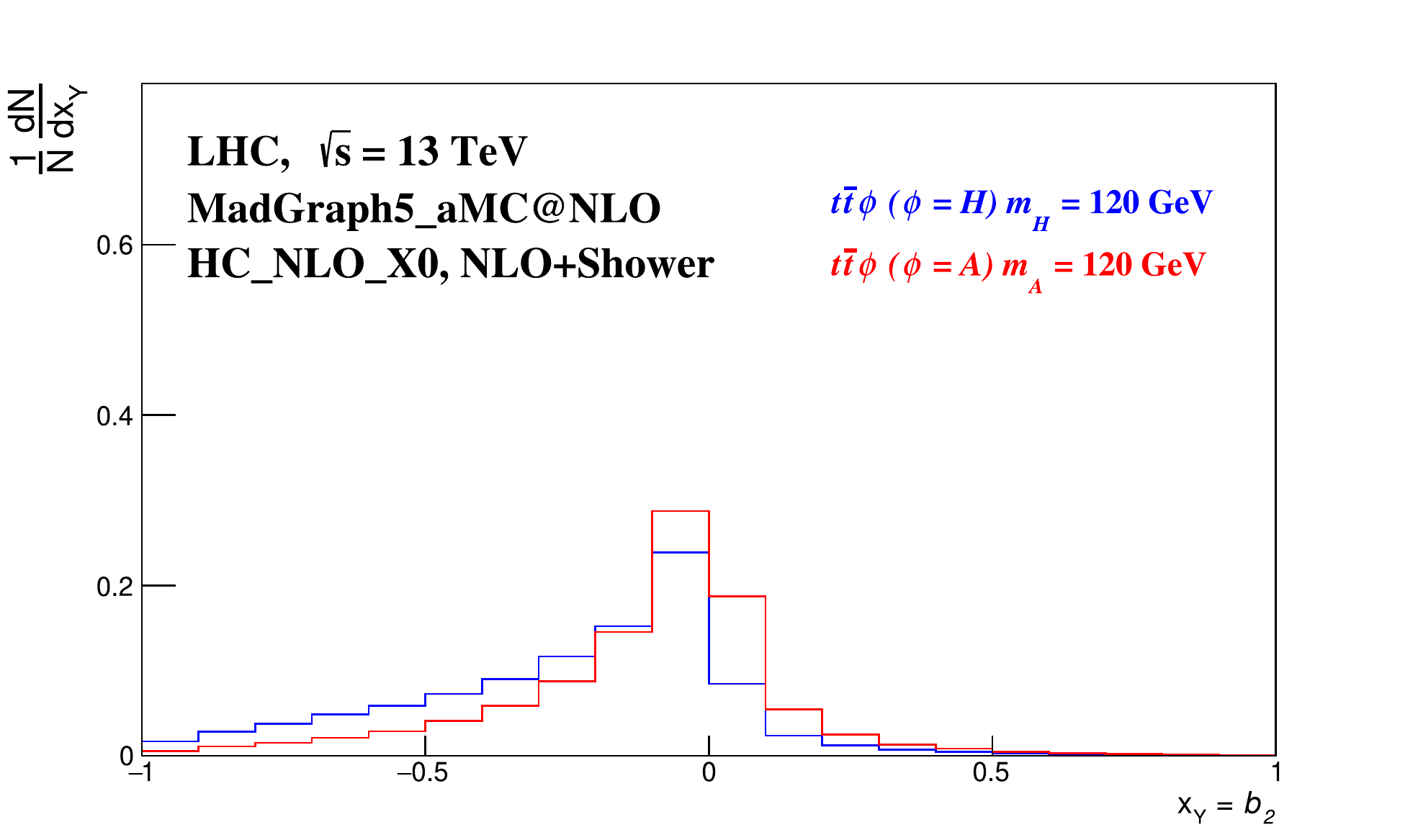}
			\hspace*{-5mm}\includegraphics[height=5.1cm]{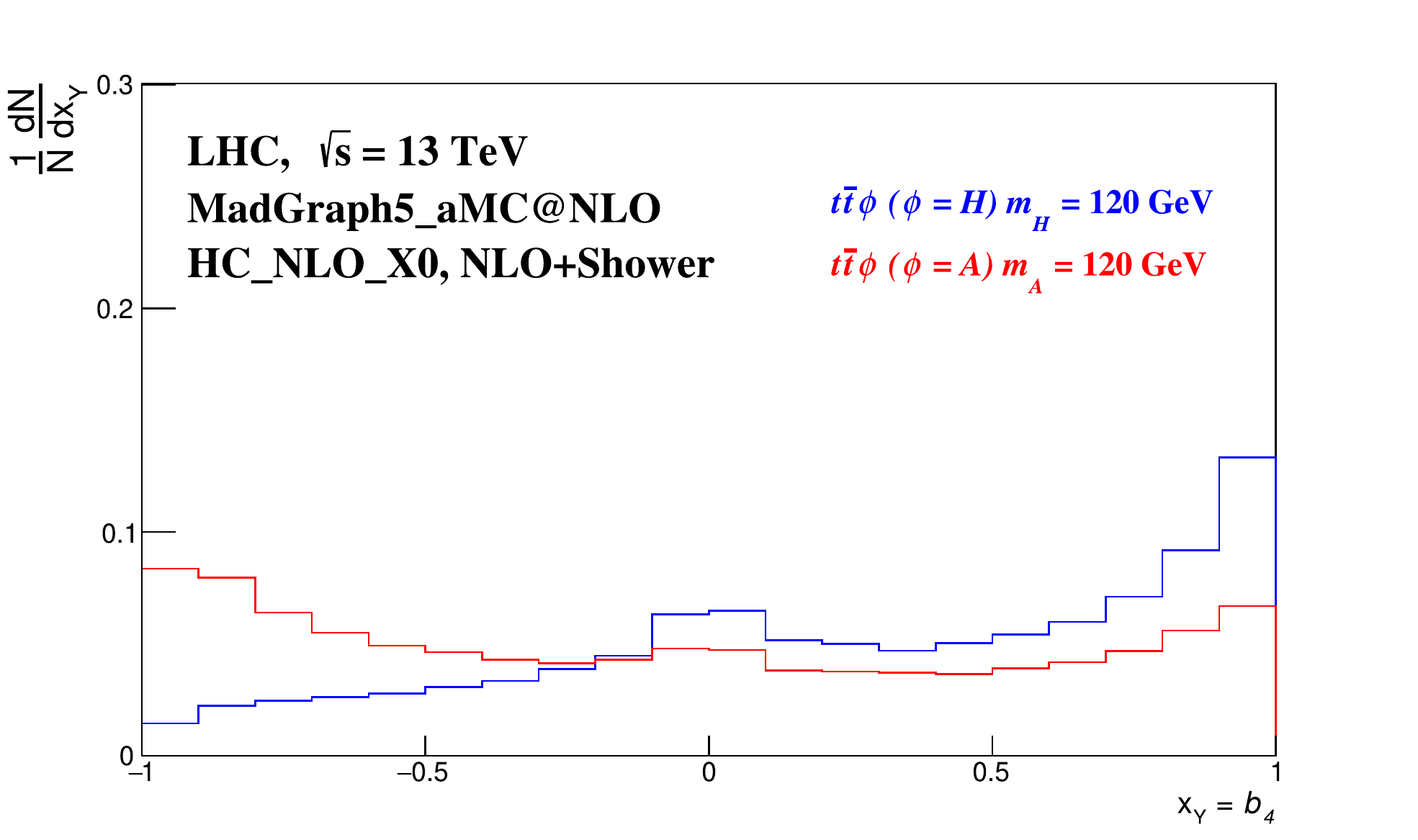}
			\\
			\hspace*{-5mm}\includegraphics[height=5.1cm]{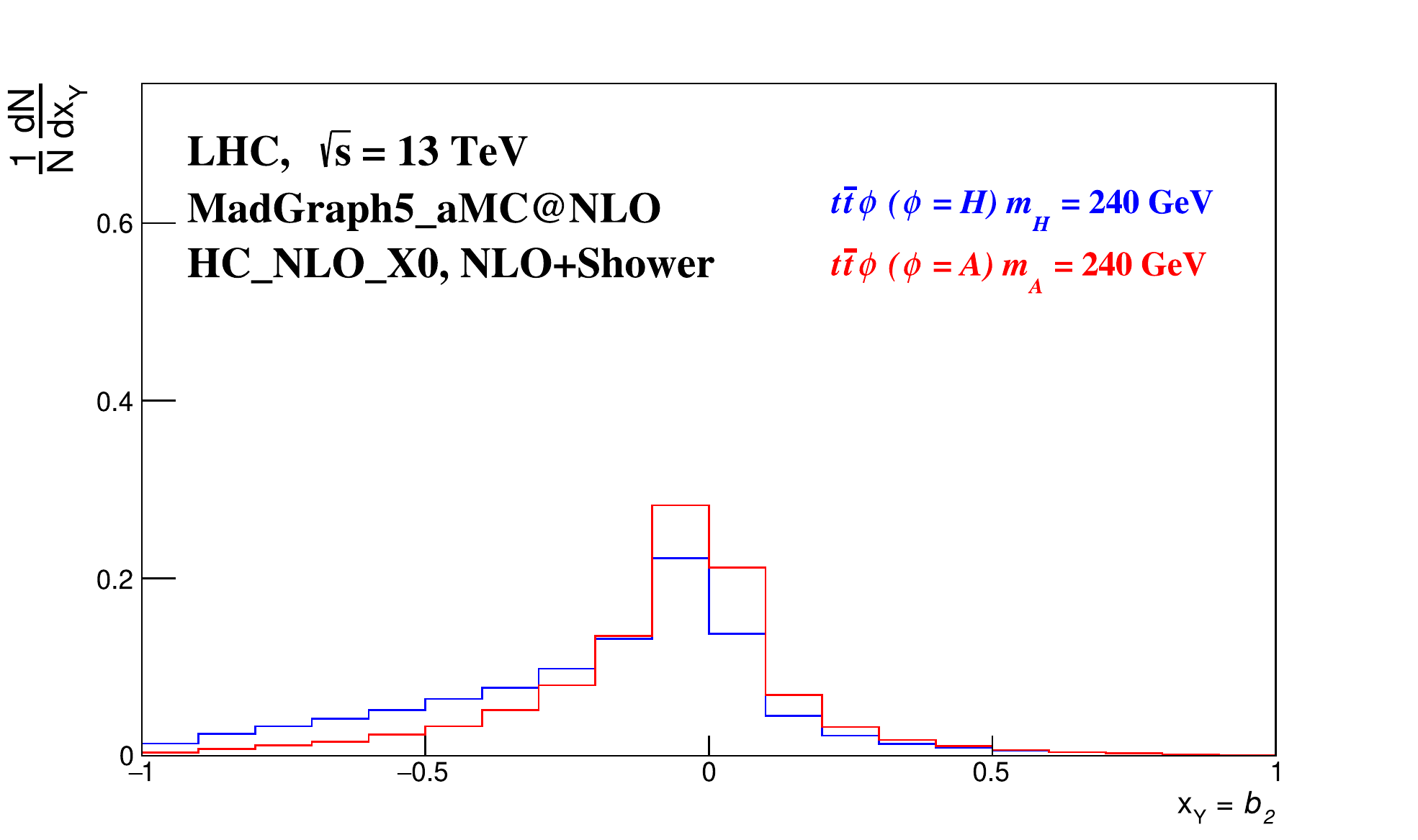}
			\hspace*{-5mm}\includegraphics[height=5.1cm]{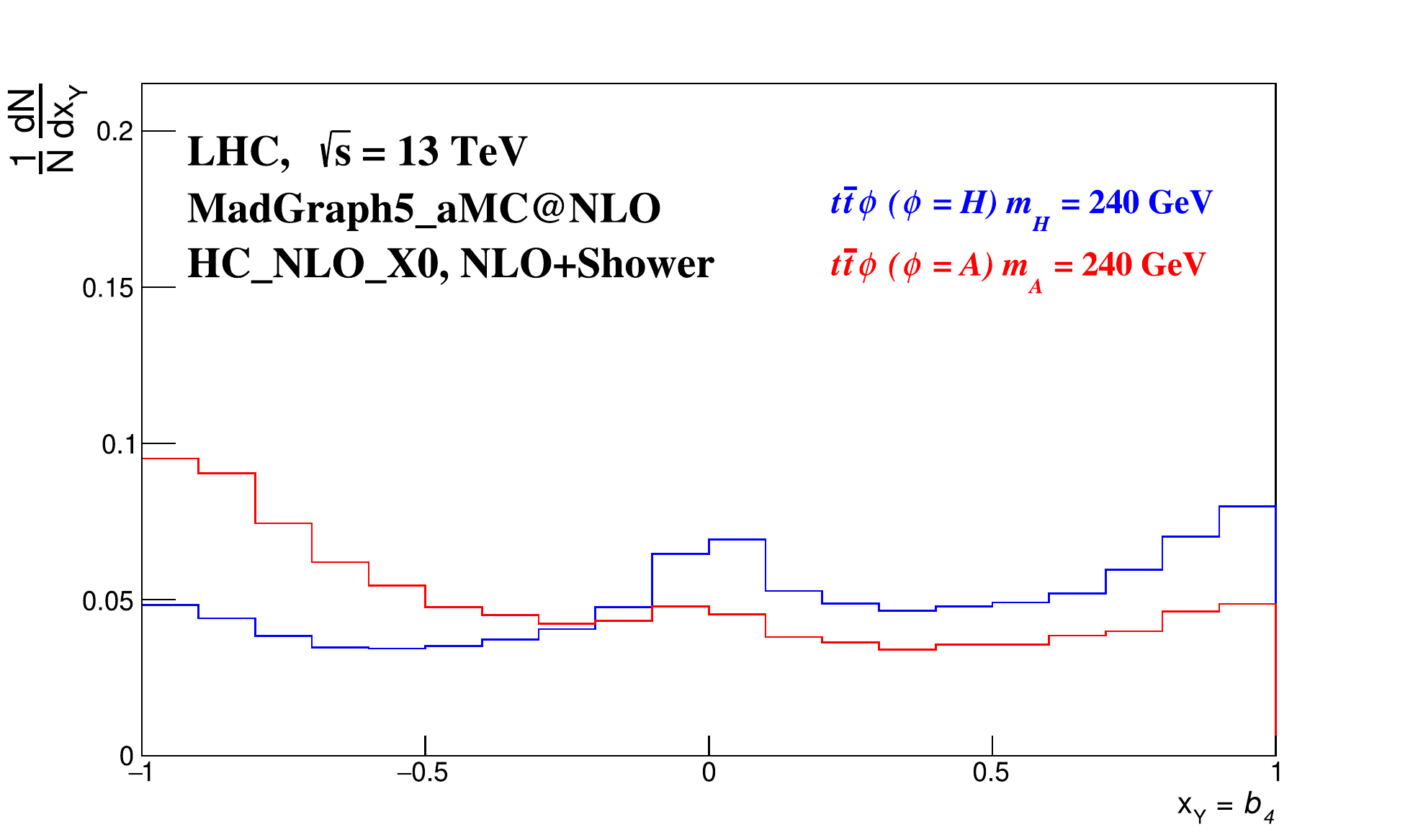}
			
		\end{tabular}
		\caption{Full normalized distributions at NLO+Shower for the variables $b_2$ (left) and $b_4$ (right) in the LAB frame, without any selection cuts nor reconstruction, for both the pure scalar (blue line) and pure pseudoscalar (red line) signals with $m_\phi = $ 40, 120 and 240 GeV.}
		\label{fig:b2b4}
	\end{center}
\end{figure}
\begin{figure}[h!]
	\begin{tabular}{ccc}
		\hspace*{-5mm}\includegraphics[height=5.1cm]{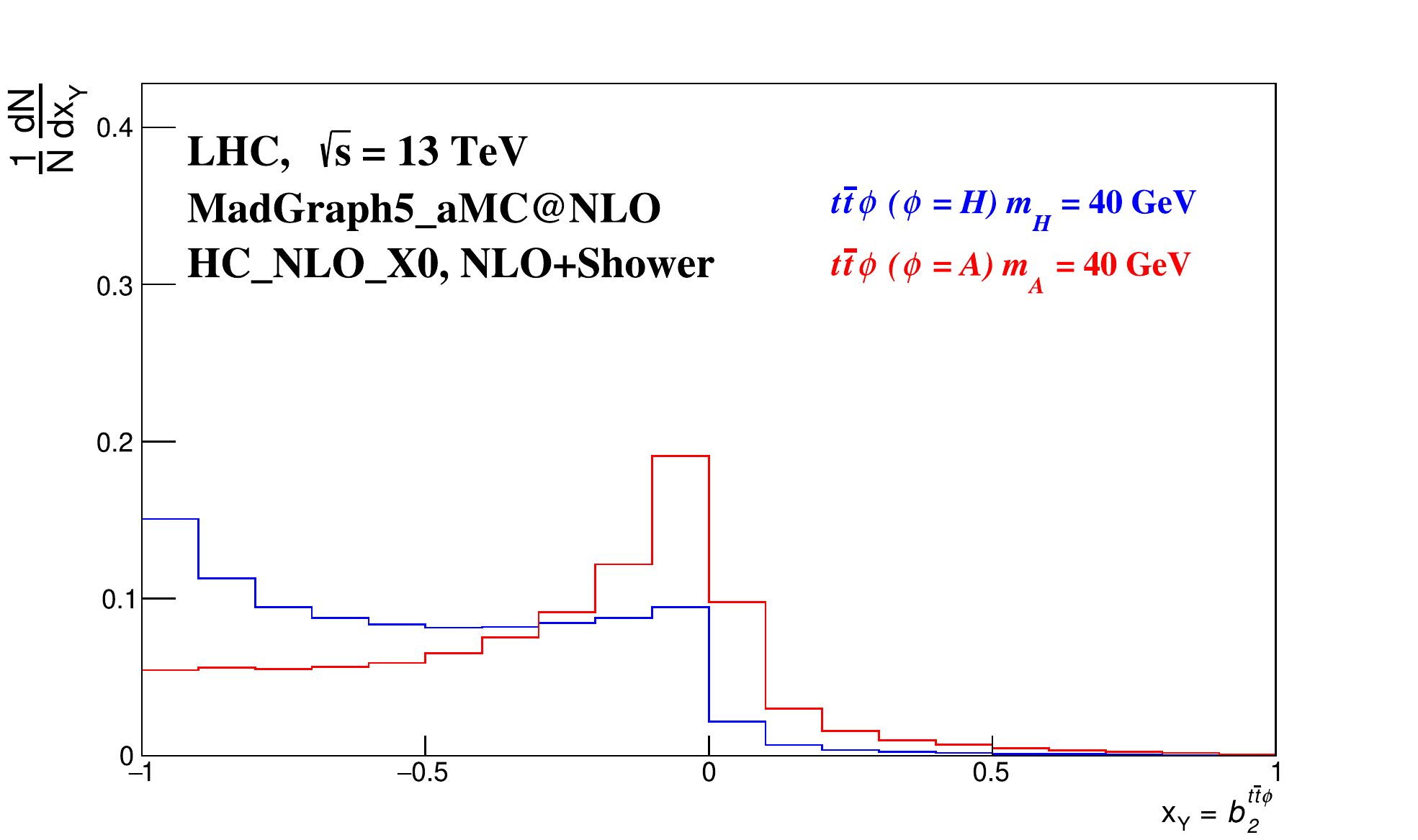}
		\hspace*{-5mm}\includegraphics[height=5.1cm]{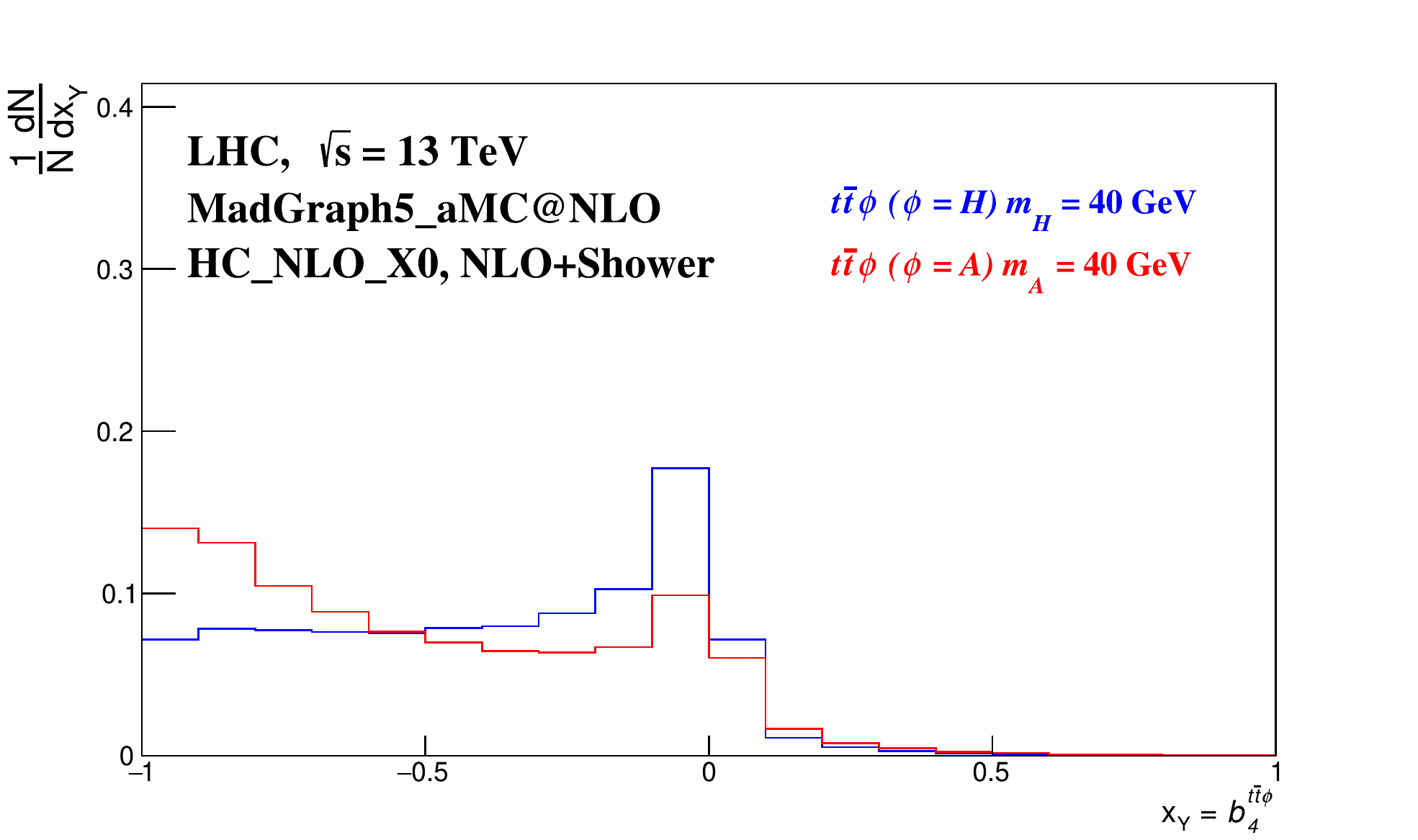}
		\\
		\hspace*{-5mm}\includegraphics[height=5.1cm]{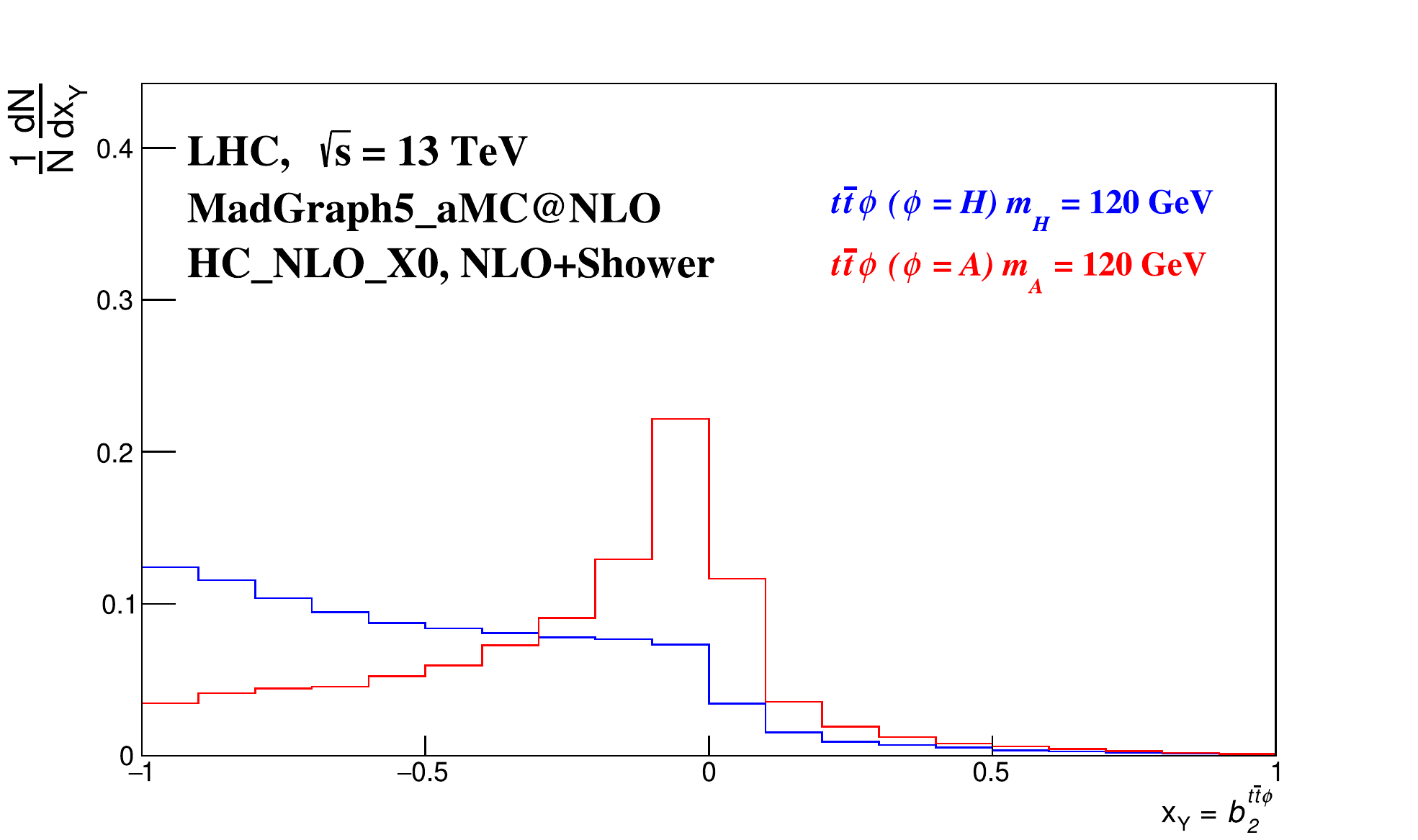}
		\hspace*{-5mm}\includegraphics[height=5.1cm]{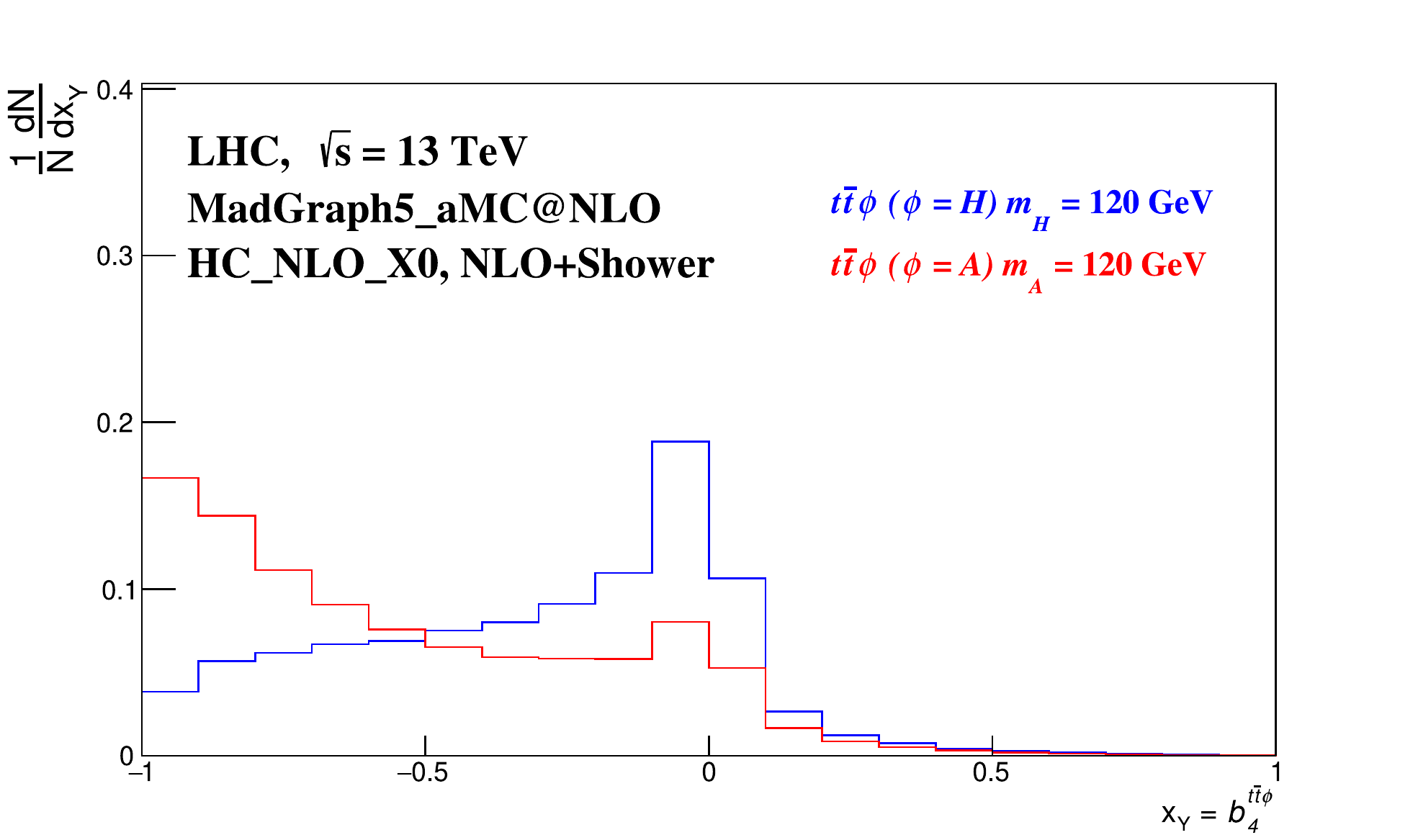}
		\\
		\hspace*{-5mm}\includegraphics[height=5.1cm]{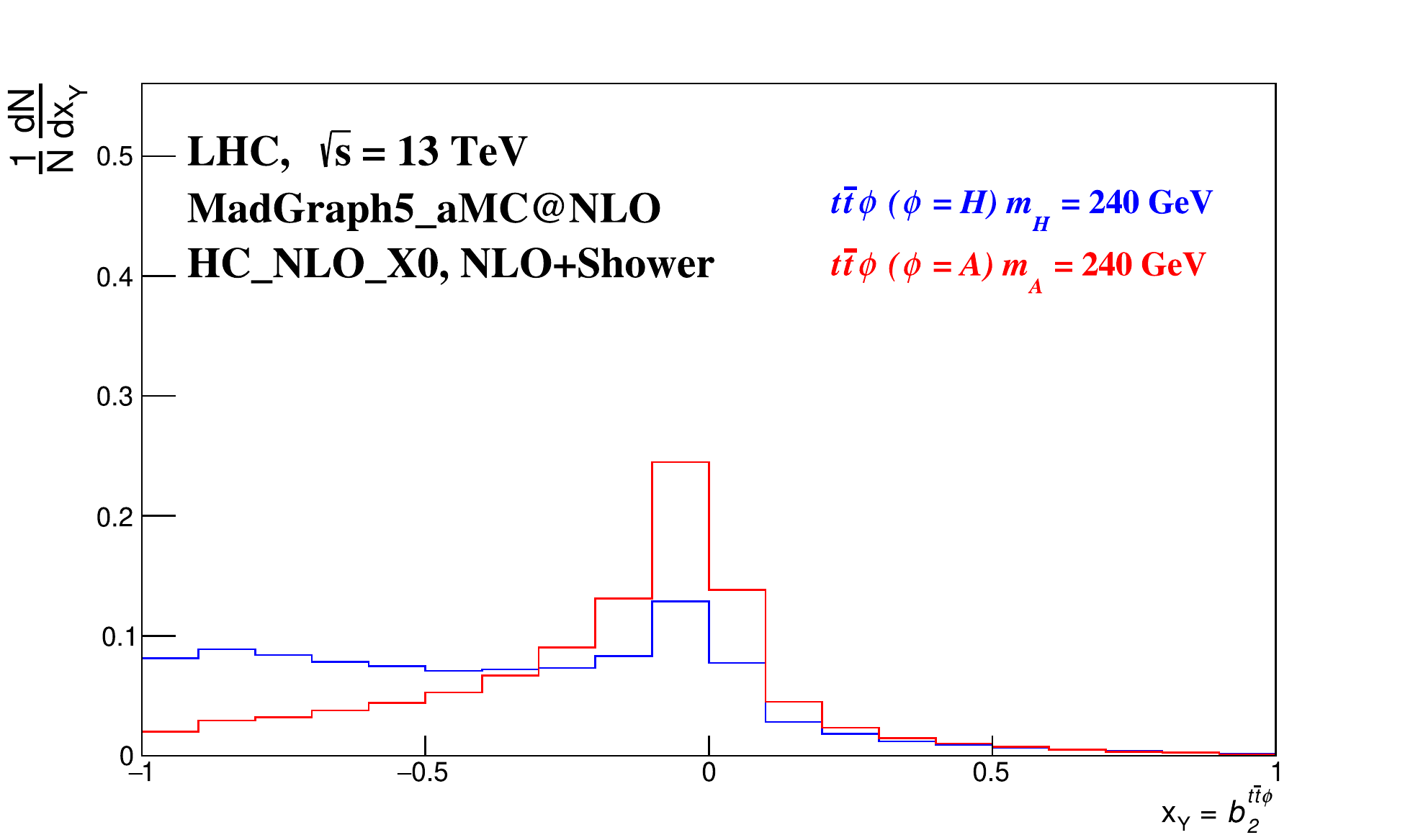}
		\hspace*{-5mm}\includegraphics[height=5.1cm]{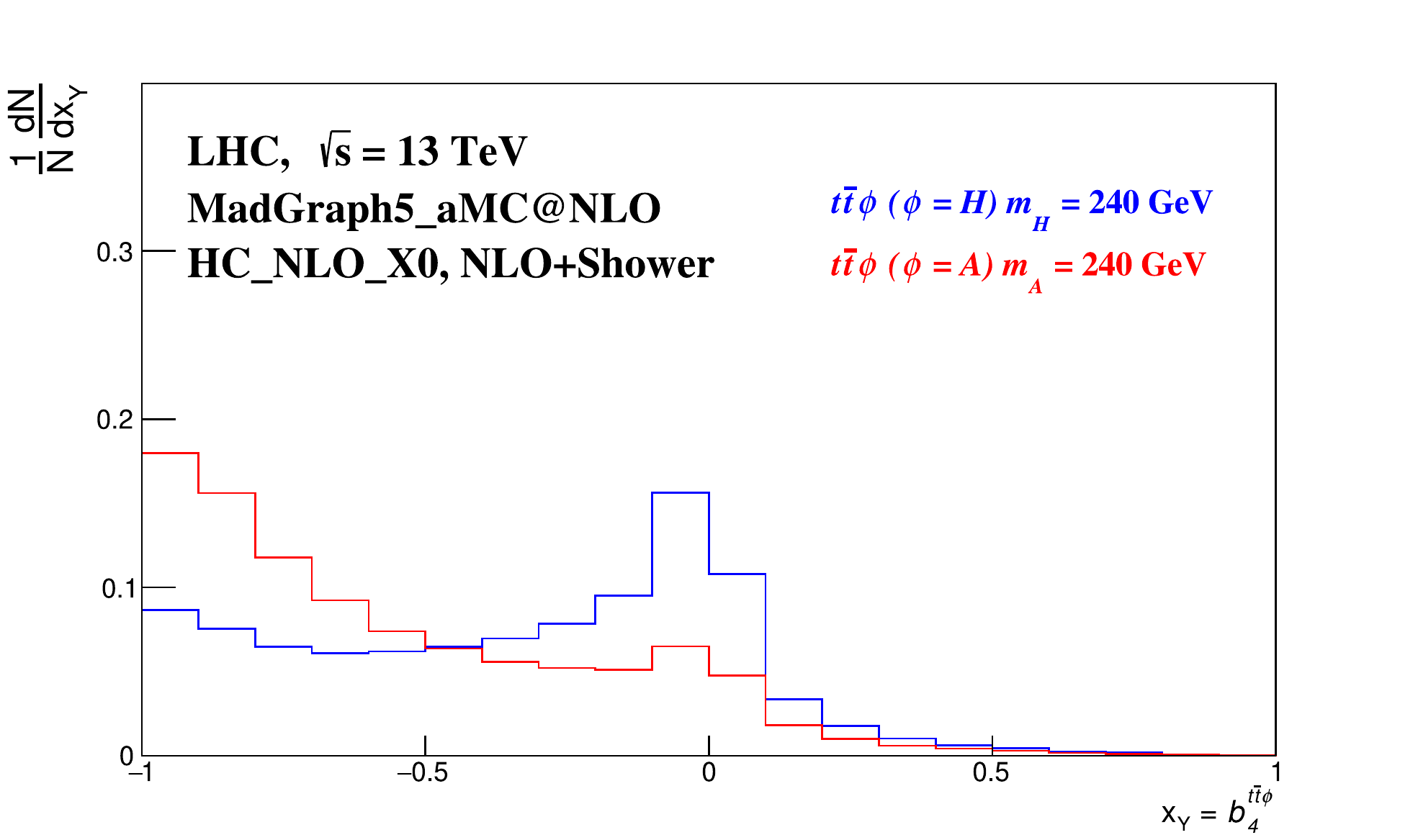}
		
	\end{tabular}
	\caption{Same as in figure \ref{fig:b2b4}, but now the $b_2$ and $b_4$ distributions are computed in the centre-of-mass of the $t\bar{t}\phi$ system.}
	\label{fig:b2b4cm}
\end{figure}

In Figure \ref{fig:mass}, 
we present the total cross section, at NLO, for a centre-of-mass energy of 13 TeV, at the LHC, for the process $pp \to t \bar t H$ (blue) and $pp \to t \bar t  A$ (red) as a function of the $\phi$ boson mass.
The fact that the interference term is much larger compared to the $b\bar{b}\phi$ case means that CP-discrimination between the different CP-components of the Higgs is now possible. Figures \ref{fig:b2b4} and \ref{fig:b2b4cm} show the $b_2$ and $b_4$ distributions for $t\bar{t}H$ and $t\bar{t}A$ events with different $\phi$ boson masses, computed in the LAB and in the centre-of-mass frame of the $t\bar{t}\phi$ system, respectively. They are shown at parton level without any cuts. Next-to-leading order corrections and shower effects (NLO+Shower) are also included. Clear differences are now visible between the scalar and pseudoscalar signals, and also between the distributions computed in the LAB and in the centre-of-mass frames. 

\begin{figure}[h!]
\begin{tabular}{ccc}
\hspace*{-5mm}\includegraphics[height=5.1cm]{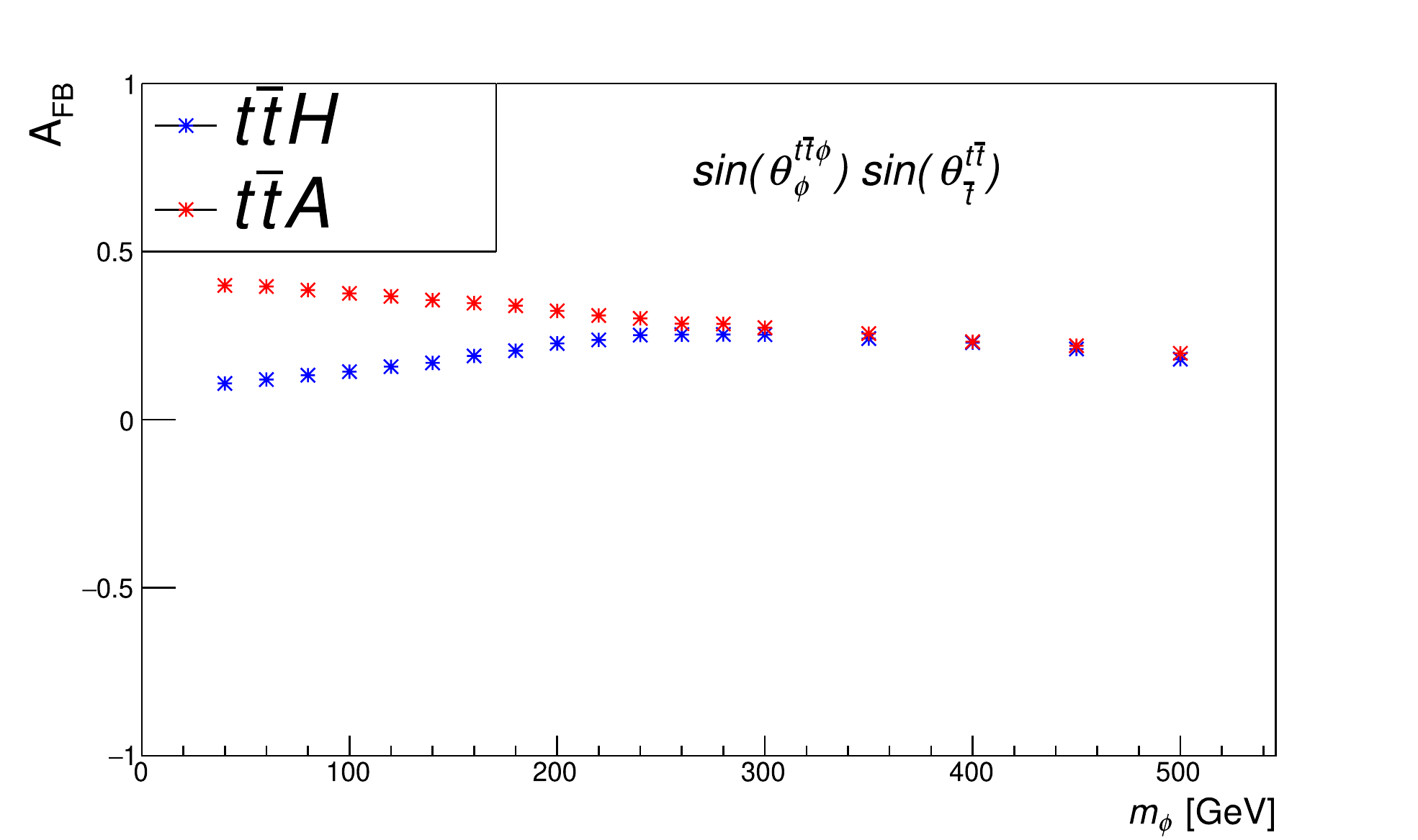}
\hspace*{-5mm}\includegraphics[height=5.1cm]{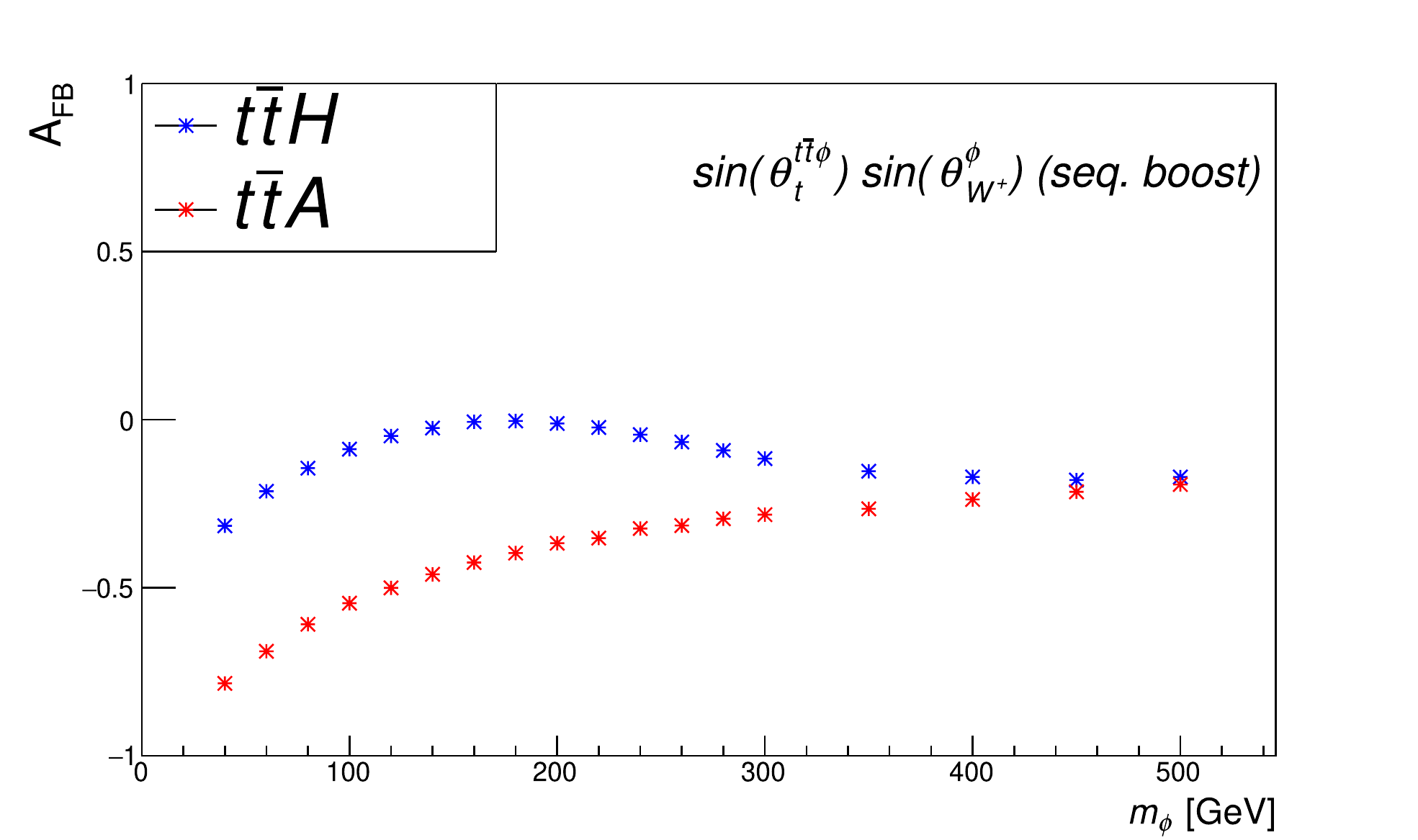}
\\
\hspace*{-5mm}\includegraphics[height=5.1cm]{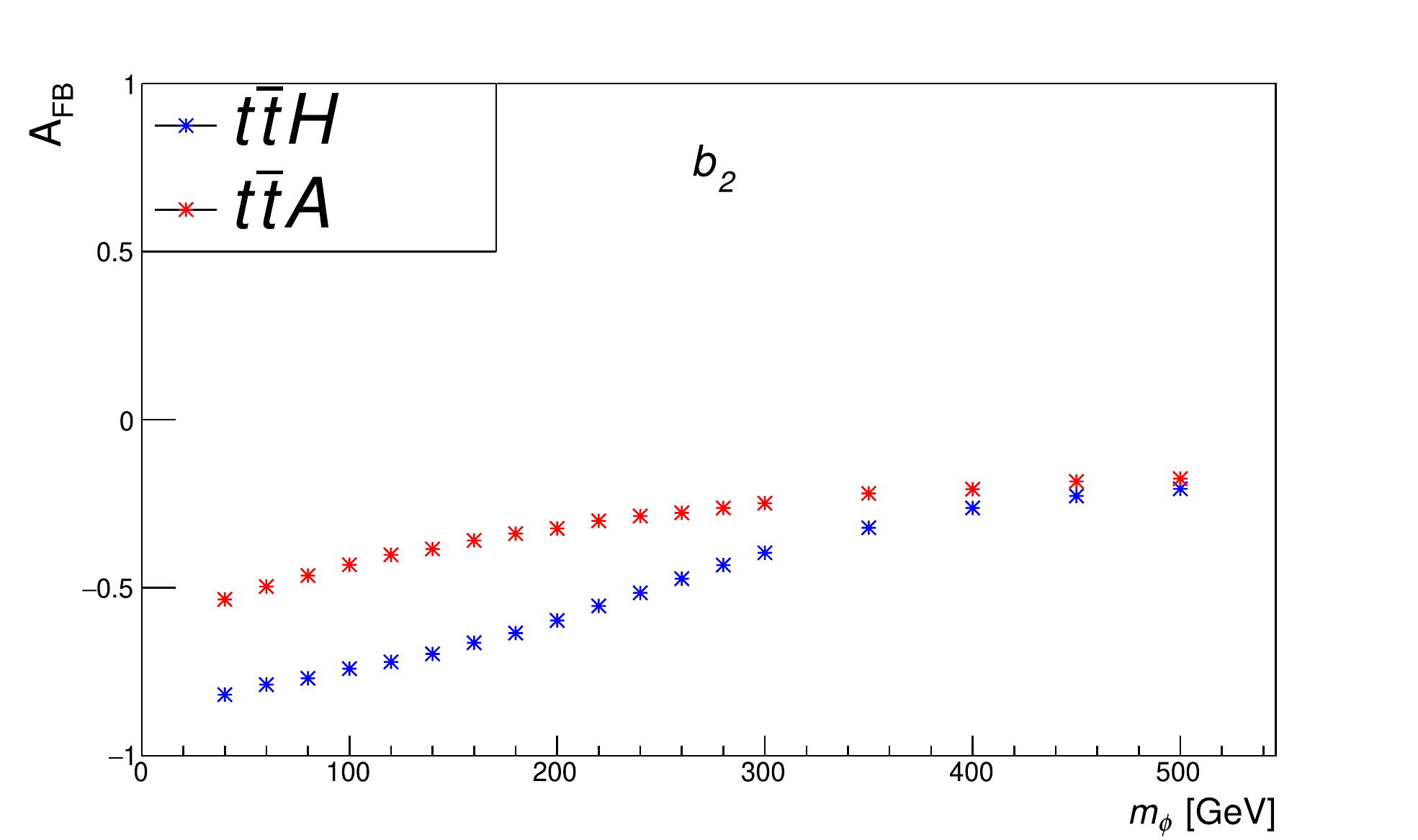}
\hspace*{-5mm}\includegraphics[height=5.1cm]{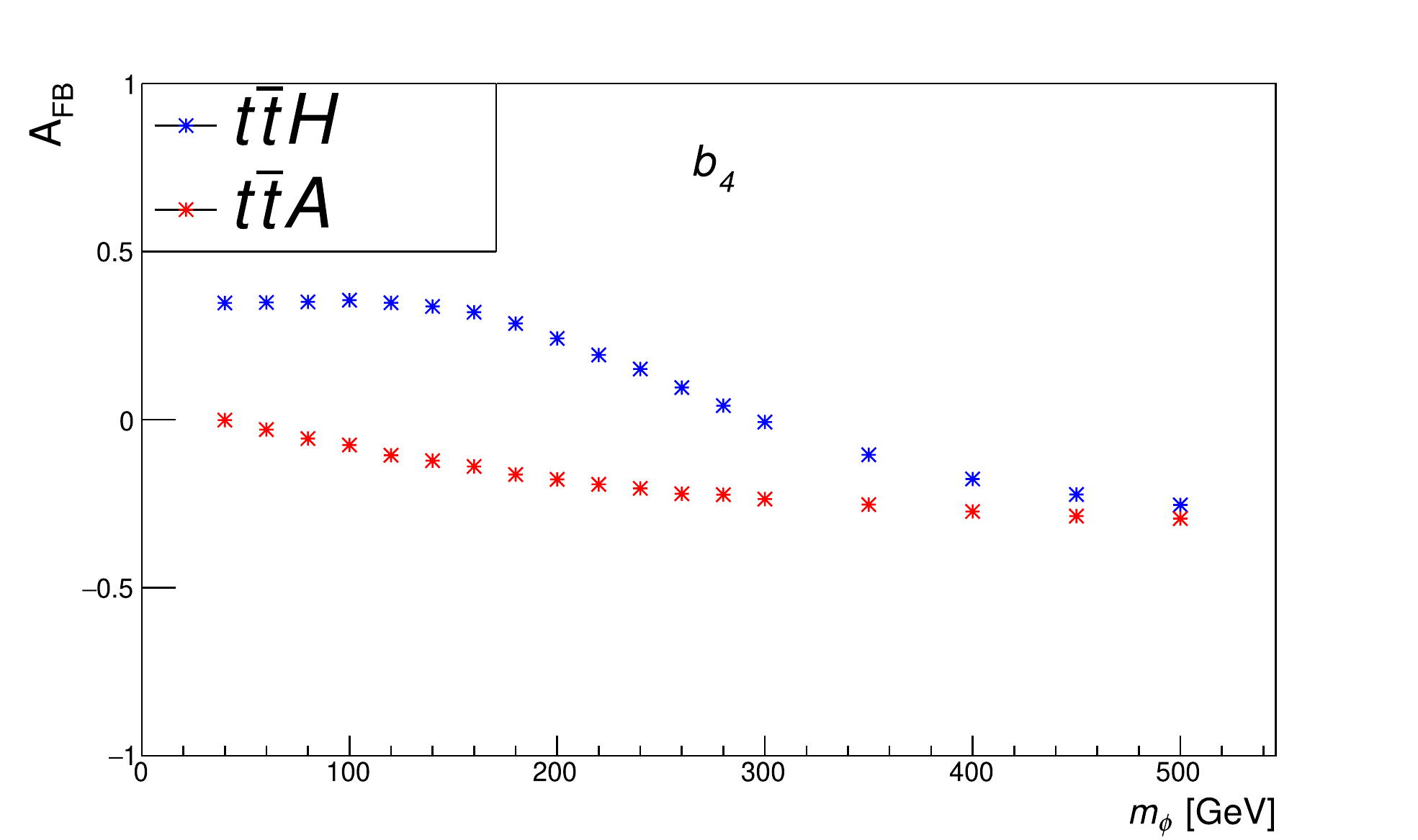}
\\
\hspace*{-5mm}\includegraphics[height=5.1cm]{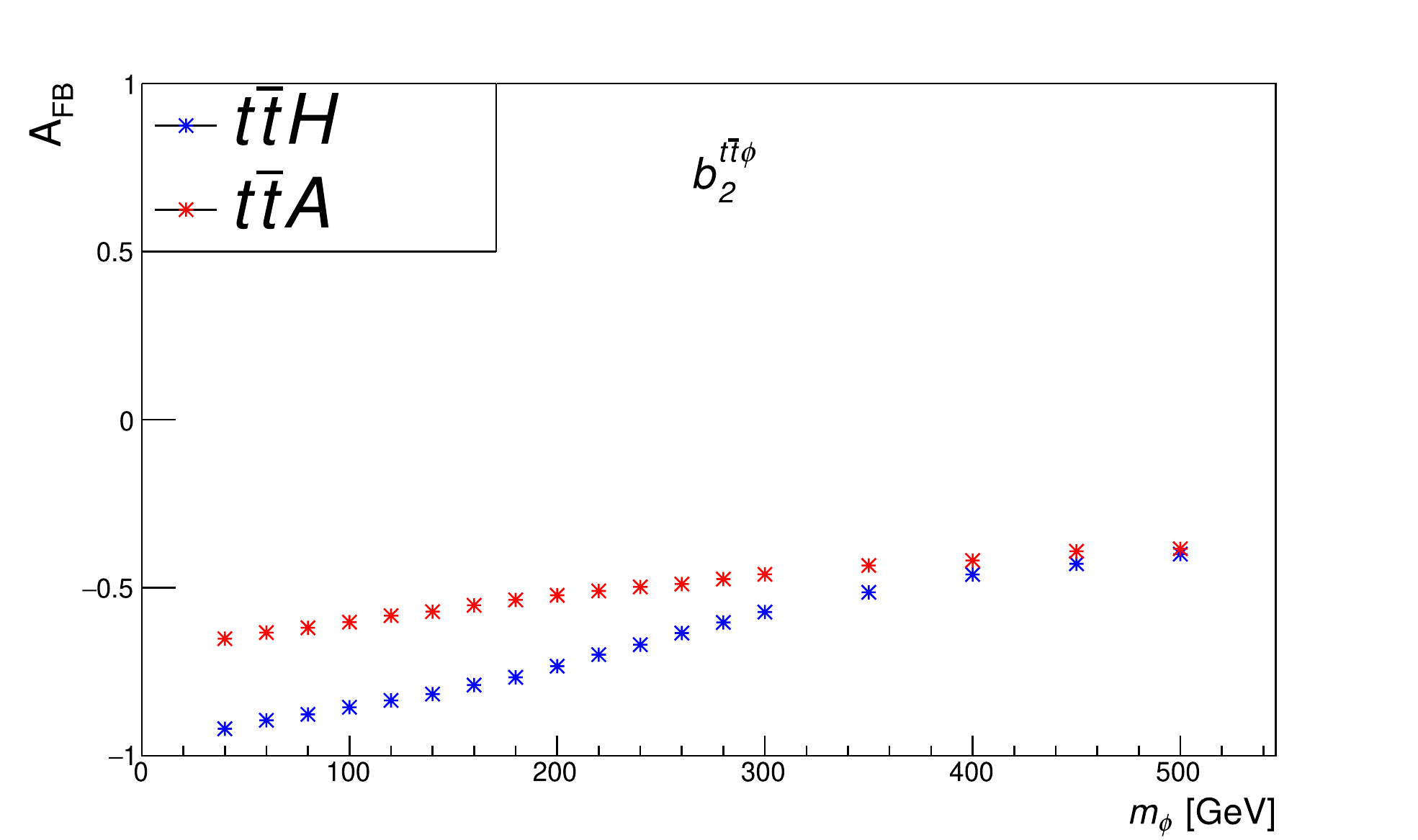}
\hspace*{-5mm}\includegraphics[height=5.1cm]{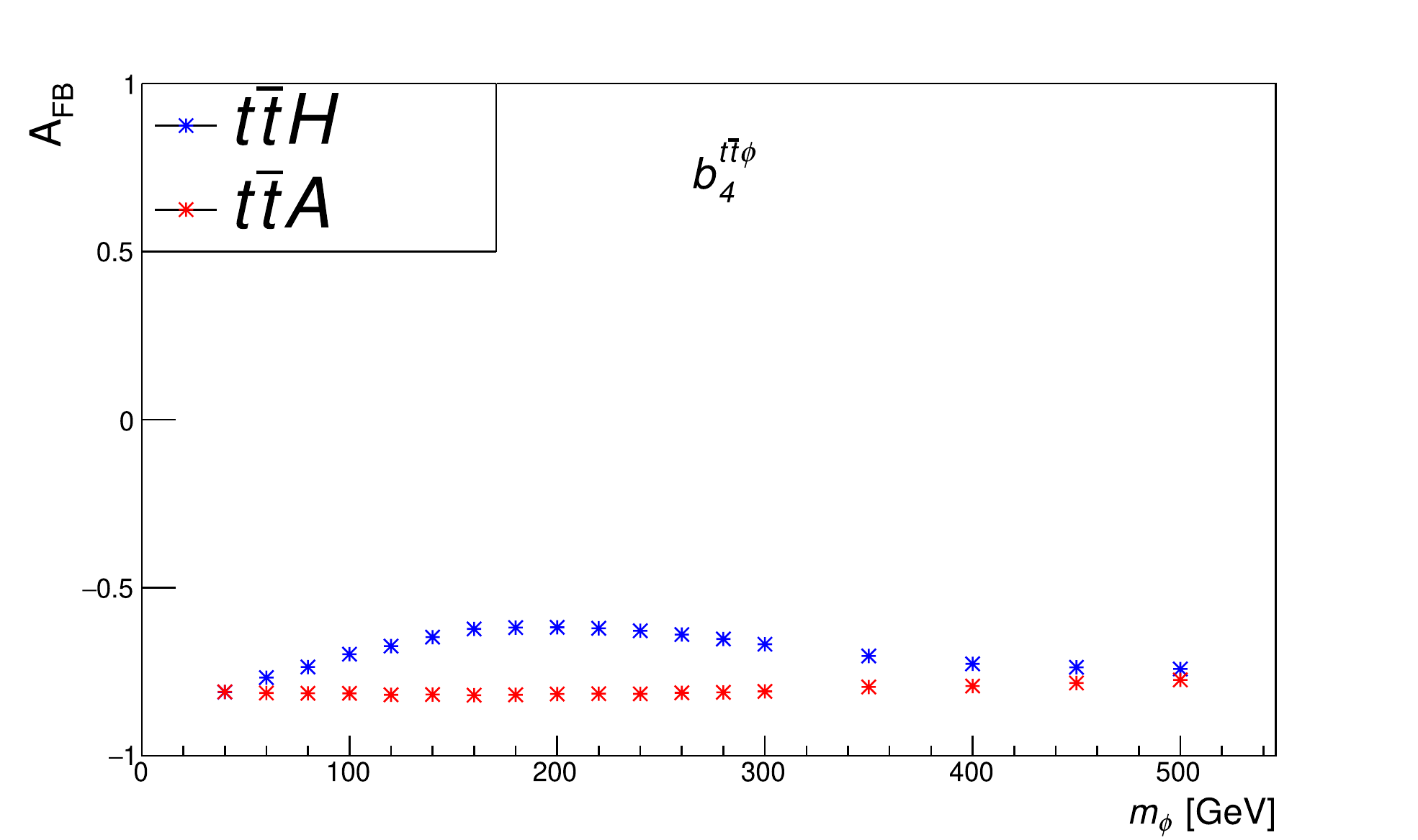}
\end{tabular}
\caption{Forward-backward asymmetries as a function of the $\phi$ boson mass, without cuts, at NLO+Shower. On top, we have two angular distributions, $\sin \theta_\phi^{t \bar t \phi} \, \sin \theta_{\bar t}^{t \bar t}$ (left panel) and $\sin \theta_t^{t \bar t \phi} \, \sin \theta_{W^+}^{\phi}$ with a \textit{sequential} boost (right panel). On the middle and last rows, we show the variables $b_2$ (left) and $b_4$ (right), in the LAB and centre-of-mass frames, respectively.}
\label{fig:afb}
\end{figure}
In order to study the CP-sensitivity as a function of the $\phi$ boson mass, forward-backward asymmetries of some variables were computed for each CP-component of the top quark Yukawa coupling, i.e., CP-even and CP-odd. The variables are,
\begin{itemize}
	\item $X=b_2$,
	\item $X=b_4$,
	\item $X=\sin \theta_\phi^{t \bar{t} \phi} \sin \theta_{\bar{t}}^{t \bar{t}}$,
	\item $X=\sin \theta_t^{t \bar t \phi} \sin \theta_{W^+}^{\phi}$ (with \textit{sequential} boost).
\end{itemize} 

The full normalized distributions, at parton level, are shown in Figure~\ref{fig:afb}. As hinted by the behaviour of the cross sections, for large enough Higgs masses, the difference between CP-even and CP-odd distributions disappears. Although this behaviour was confirmed for all variables, the exact mass value for which the difference becomes negligible depends on the choice of variables. The maximum value of the $\phi$ boson mass for which a meaningful difference between distributions exists is 400 GeV.


\section{Results and Discussion \label{sec:disc}}

\subsection{Generation of events}
\hspace{\parindent} 

Signal events from $pp\to t\bar t\phi$ associated production at the LHC (with $\phi=\{H,A\}$), were generated at NLO with the Higgs Characterization model \texttt{HC\_NLO\_X0}~\cite{Artoisenet:2013puc}, using \texttt{MadGraph5\_aMC@NLO} ~\cite{Alwall:2011uj}. The pure CP-even and the pure CP-odd odd samples were generated by setting the CP-phase to $\cos \alpha=1\text{ or }0$, respectively, following Equation~\ref{eq:higgscharacter}, with $\kappa_t=1$.  Several samples, for both scalar and pseudoscalar signals, were generated with masses $m_\phi$ between 40 and 300~GeV, in steps of 20~GeV, and also the four masses  $m_\phi$ = 350, 400, 450 and 500 GeV. While the CP-even and CP-odd bosons were only allowed to decay to a pair of $b$-quarks ($\phi\to b \bar b$), the $t\bar{t}$ system was assumed to decay to a pair of $b$-quarks and two intermediate $W^\pm$ gauge bosons which, in turn, decay to two charged leptons and two neutrinos $t (\bar t)\to b W^+ (\bar b W^-)\to b \ell^+ \nu_\ell (\bar b \ell^- \bar \nu_\ell )$. Following the decay of all intermediate massive particles, the signal final state is characterized by the presence of two oppositely charged leptons, two neutrinos and two $b\bar{b}$ quark pairs, at parton level. Only $W$ boson decays to electrons ($e$) and muons ($\mu$) were considered as signal. This configuration defines the dileptonic channel.

In addition to the signal samples, backgrounds from \texttt{SM} processes were also generated using \texttt{MadGraph5\_aMC@NLO}. 
The dominant background, a pair of top- and $b$-quarks ($t\bar{t}b\bar{b}$), as well as the associated production of top-quarks with the SM Higgs boson ($t\bar{t}H_{SM}$), were generated at NLO. For the latter, a SM Higgs boson mass of $m_{H_{SM}}=125$ GeV was assumed. These two backgrounds lead to the same partonic final state as the signal. 

The remaining backgrounds considered are:
\begin{itemize}
	\item $t\bar{t}$+3 jets i.e., top-quark pair production with up to three light jets.
	\item $t\bar{t}V$+ jets i.e., top-quark pair production with one gauge boson ($V=Z,W^\pm$), plus up to one light jet.
	\item Single top quark production through the $s$-, $t$-channel (with up to one additional jet) and $Wt$ associated production.
	\item $W$+4 jets, i.e., $W^\pm$ boson production with up to four light jets.
	\item $Wb\bar{b}$+2 jets, i.e., $W^\pm$ boson production with two jets from the hadronization of $b$-quarks ($b$-jets), and up to two additional light jets.
	\item $Z$+4 jets i.e., $Z$ boson production with up to four light jets.
	\item $Zb\bar{b}$+2 jets i.e., $Z$ boson production with a pair of $b$-jets plus up to two light jets.
	\item $WW,WZ,ZZ$+3 jets i.e., diboson production with up to three jets.
\end{itemize}
%


All events were generated assuming proton collisions at the LHC with a centre-of-mass energy of 13 TeV. The masses of the top quarks ($m_t$) and the $W$ bosons ($m_W$), were set to 173~GeV and 80.4~GeV, respectively. For all samples, the \texttt{NNPDF2.3}~\cite{Ball:2012cx} parton distribution functions (PDFs), were used.
The renormalization and factorisation scales were fixed to the sum of the transverse masses of all final state particles and partons.
 The decay of particles was performed by \texttt{MadSpin}~\cite{Artoisenet:2012st} for signal and background events in order to preserve spin correlations among the decay products and with the respective heavy parent resonances. Parton shower and hadronization was performed by \texttt{Pythia6}~\cite{Sjostrand:2006za}. The matching between the generator and the parton shower used the MLM scheme~\cite{Alwall:2014hca} for the LO samples and the MC@NLO matching~\cite{Frixione:2010wd} for the NLO events. For a fast, parametrised detector simulation of a LHC-like experiment, we used \texttt{Delphes}~\cite{deFavereau:2013fsa} with the default ATLAS parameter card. For jet reconstruction of the signal and background events, \texttt{FastJet}~\cite{Cacciari:2011ma} is employed with the anti-$k_t$ algorithm~\cite{Cacciari:2008gp} with a cone size of $\Delta R=0.7$~\footnote{$\Delta R\equiv\sqrt{\Delta \phi^2+\Delta \eta^2}$, where $\Delta \phi \, (\Delta \eta)$ correspond to the difference in the azimuthal angle (pseudo-rapidity) of two objects.}. Transverse momentum ($p_T$) cuts are applied to jets and photons such that, in any events, these objects are kept if the following conditions are met
\begin{equation}
\begin{aligned}
p_T^\text{jet}&\geq 10 \text{ GeV} , \qquad &p_T^\text{photon}\geq 20 \text{ GeV} . \\
\end{aligned}
\end{equation}

No additional cuts were applied to the transverse momentum of leptons nor to the pseudo-rapidity ($\eta$) of jets, leptons and photons.

\subsection{Kinematic reconstruction}
\hspace{\parindent} 

After event generation, hadronization and detector simulation, we use a kinematic reconstruction to assign detector level jets to partons from the hard-scattering process and, using the detected charged leptons, reconstruct the massive intermediate particles i.e., the top quarks, the $W$ and $\phi$ bosons. This, unavoidably, requires the reconstruction of the undetected neutrinos, which is performed on an event by event basis, using the \texttt{MadAnalysis5}~\cite{Conte:2012fm} framework.

Only events with at least two charged leptons of opposite charge and four or more jets are selected and reconstructed. Both leptons and jets were required to have $p_T\geq 20$ GeV and $|\eta| \leq 2.5$, which leads to signal selection efficiencies that vary from 9\% (12\%) to 18\% (19\%) for masses of the scalar (pseudoscalar) from 40~GeV to 200~GeV, respectively. The uncertainties on these numbers are smaller than 0.2\%. It should be stressed at this point, that no attempt to optimize the selection was applied by looking for instance, to boosted jets, which is outside the scope of this paper. 

\begin{figure}[h!]
	\begin{tabular}{ccc}
		\hspace*{-3mm}\includegraphics[height = 4.5cm]{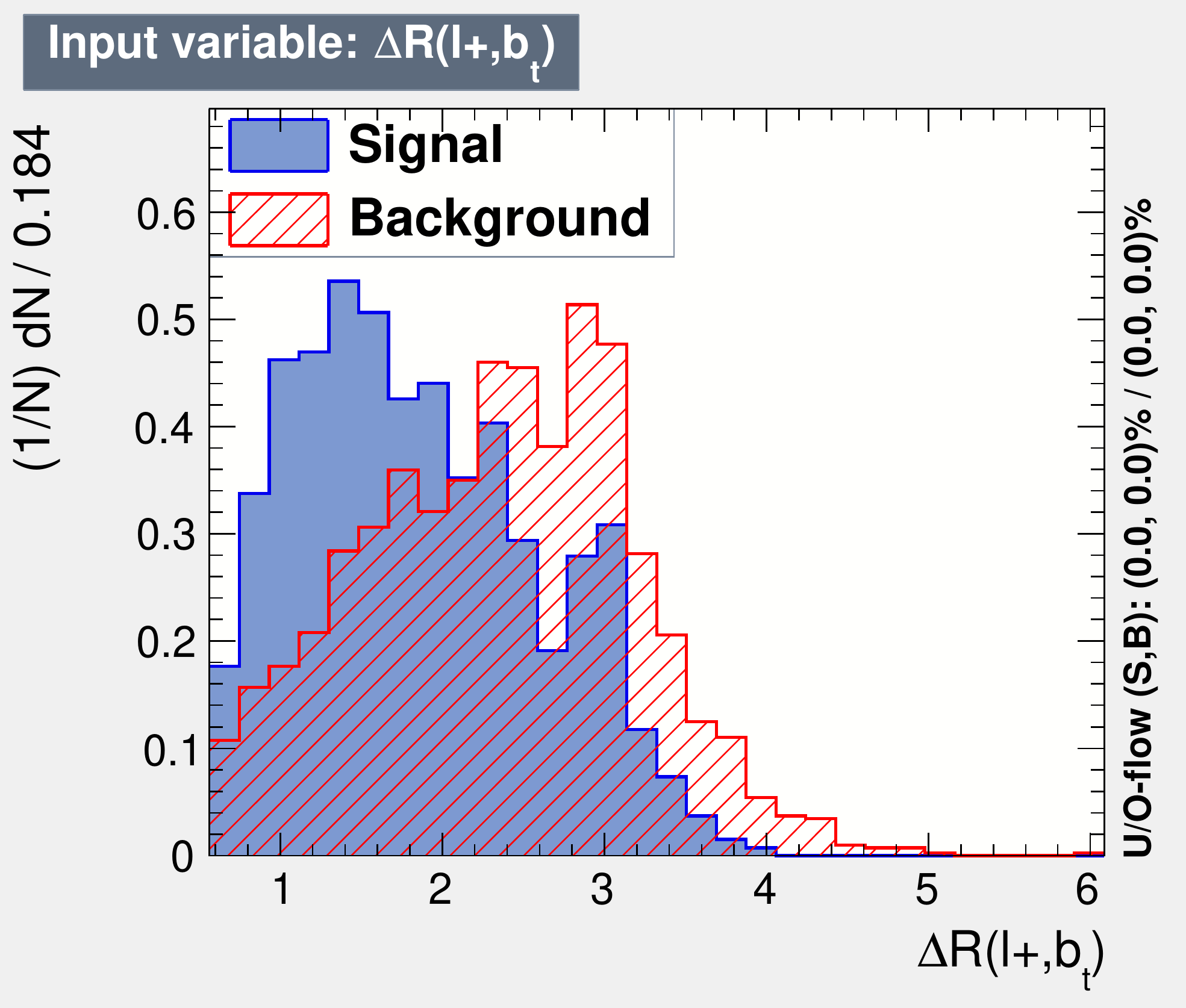}
		\includegraphics[height = 4.5cm]{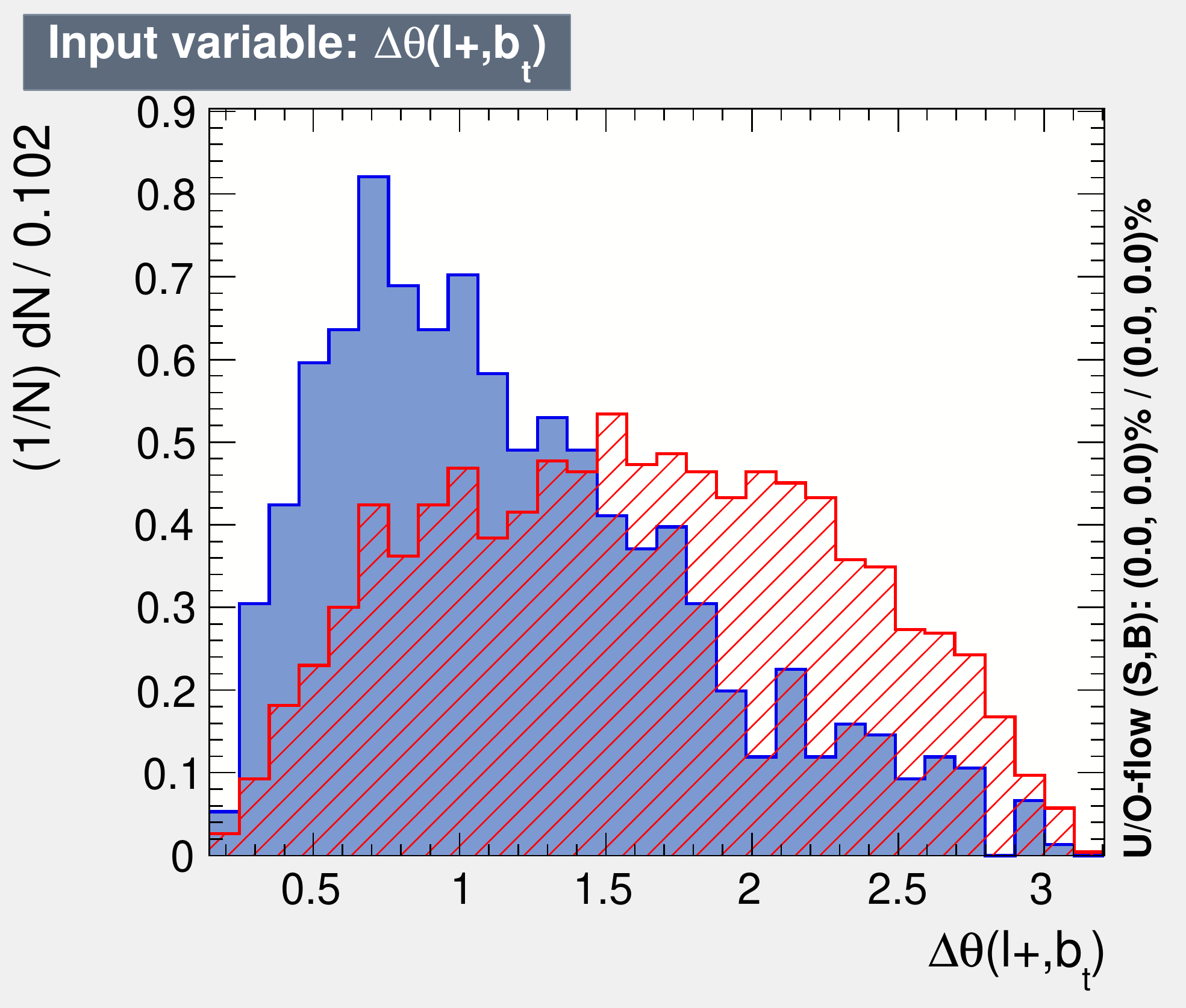}
		\includegraphics[height = 4.5cm]{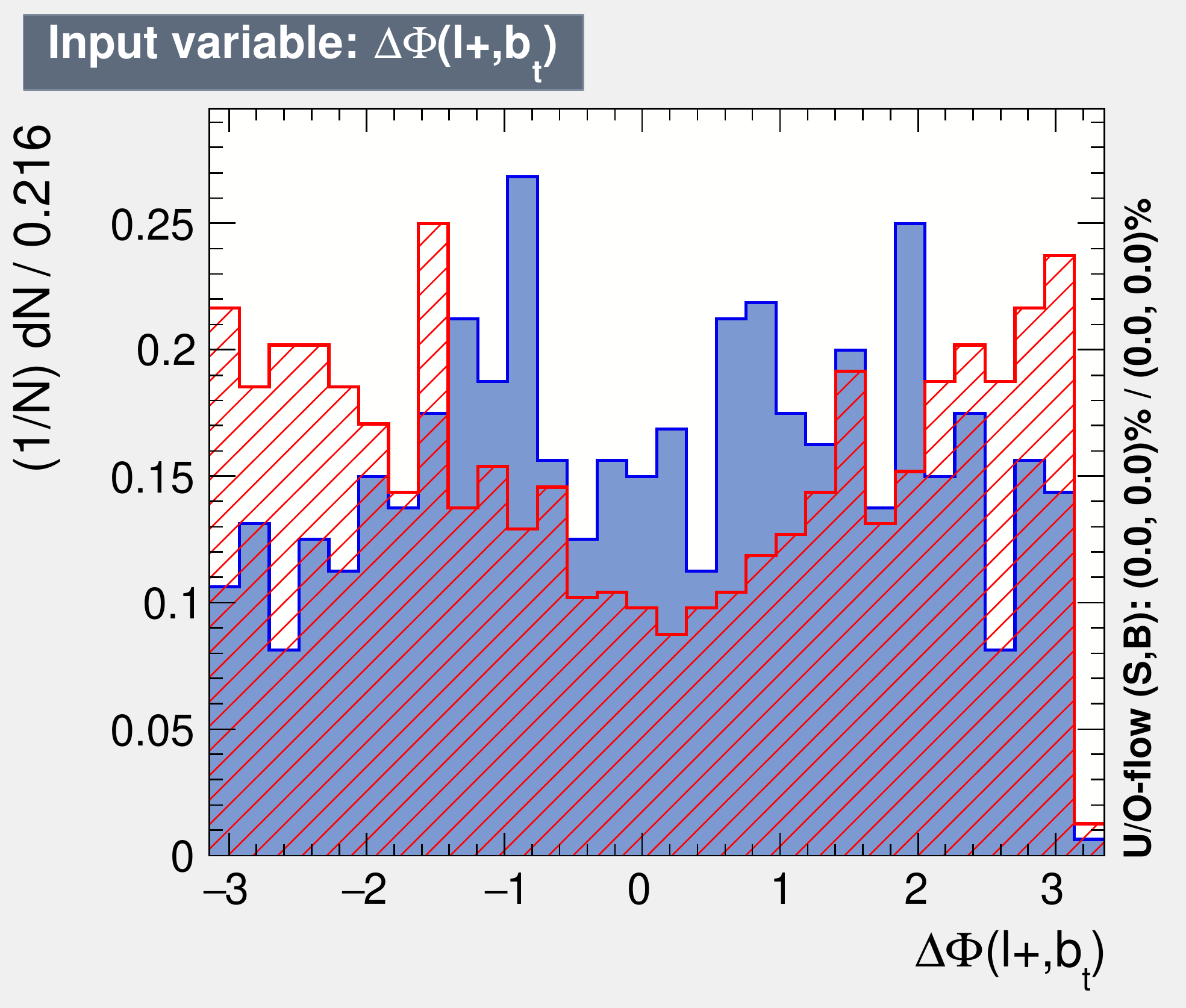}
		\\		
		\hspace*{-2mm}\includegraphics[height = 4.5cm]{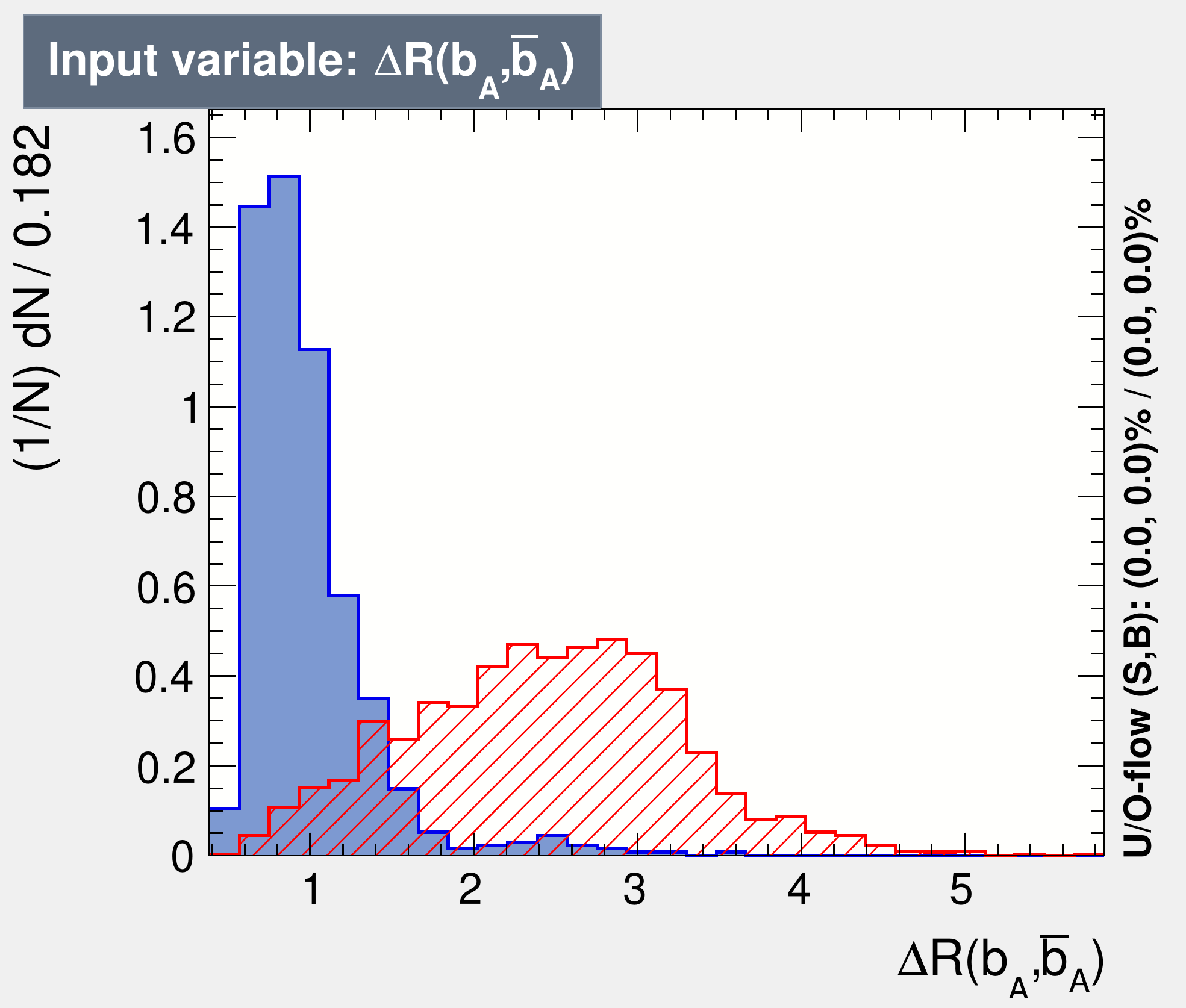}
		\includegraphics[height = 4.5cm]{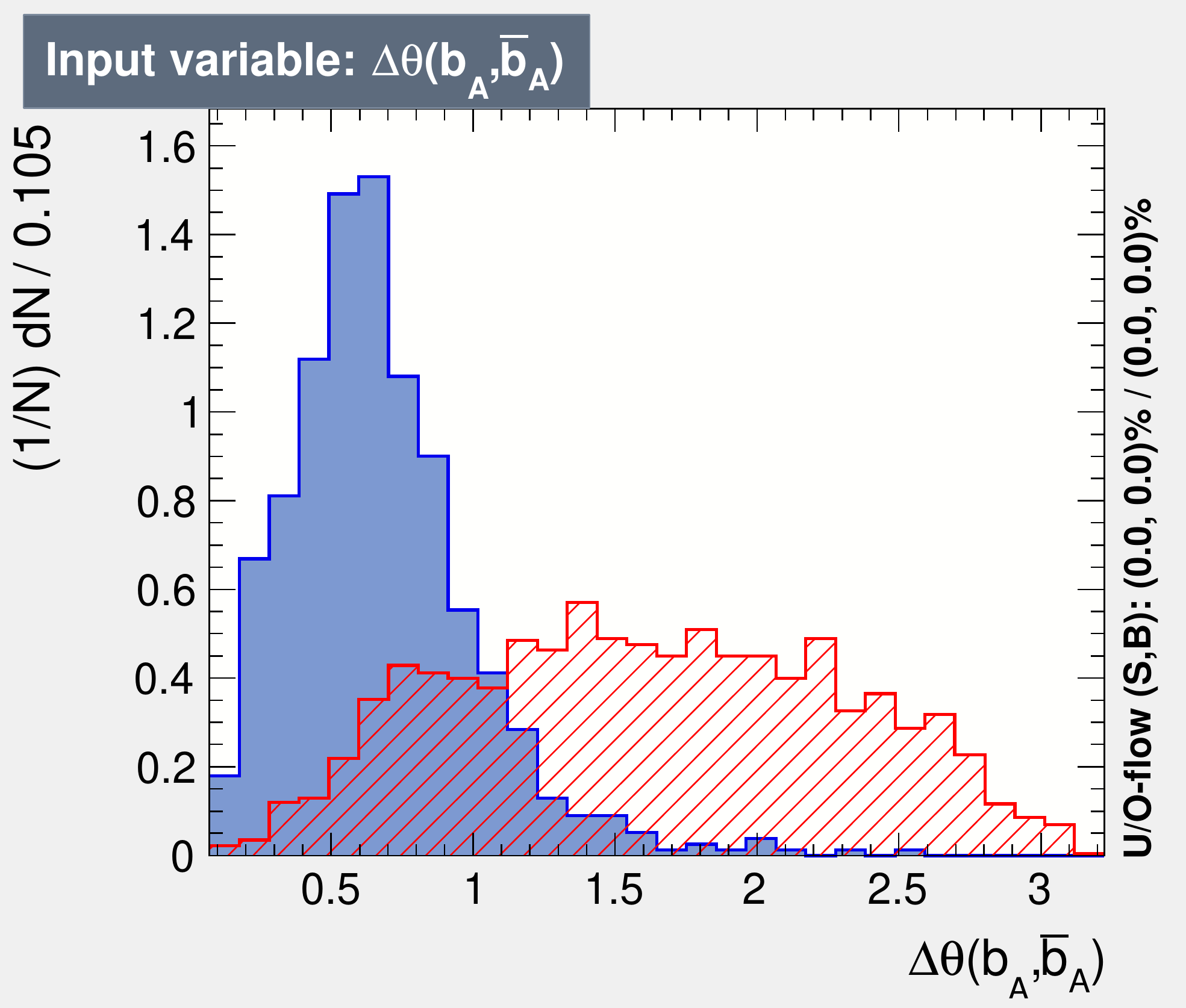}
		\includegraphics[height = 4.5cm]{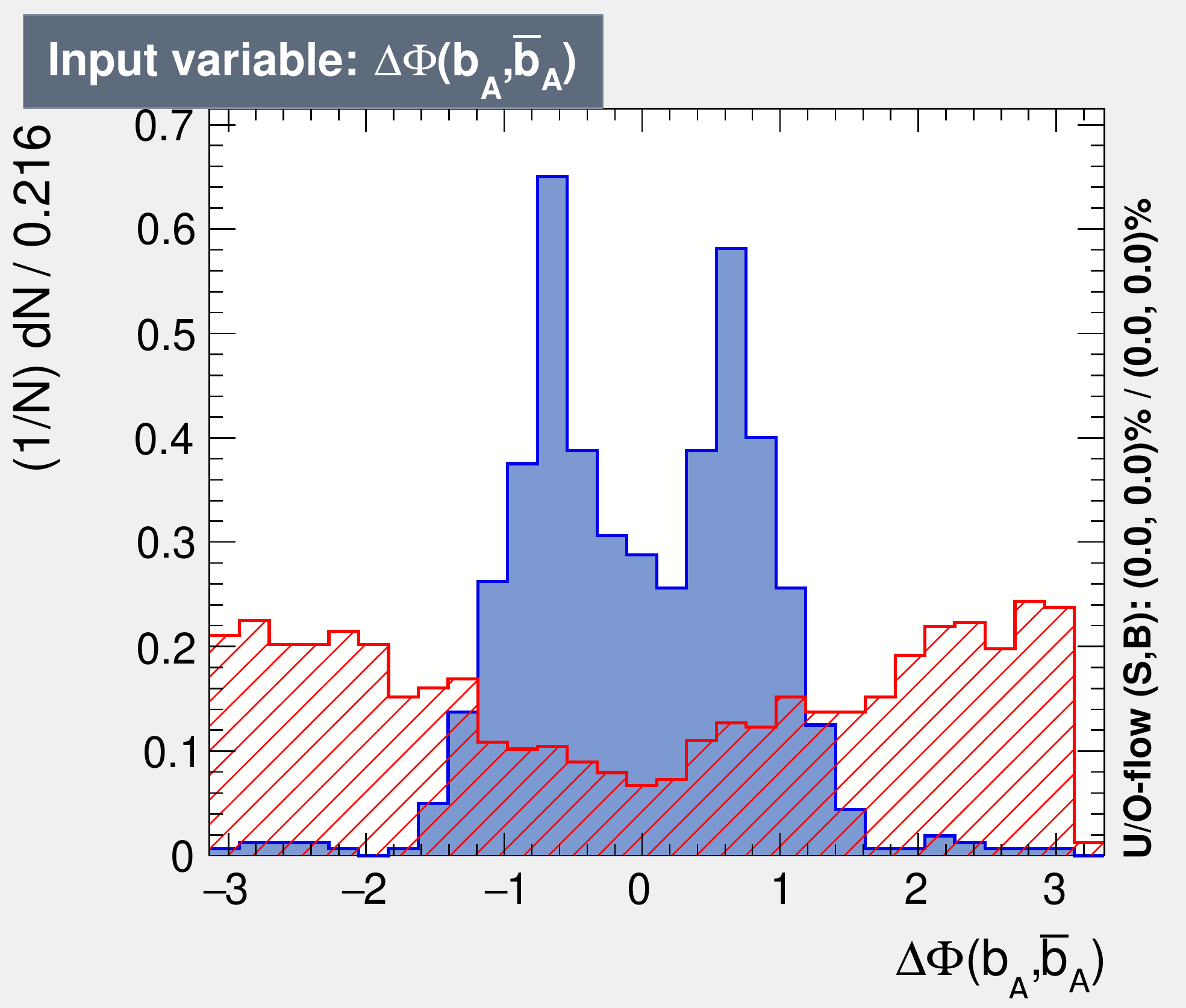}
	\end{tabular}
	\caption{Distributions of the TMVA input variables for the signal (blue) and background (red) samples for $t\bar{t}A$ events with $m_A = 40$ GeV. The angular variables $\Delta R$, $\Delta \theta$ and $\Delta \Phi$ for the pairs $(l^+, b_t)$ (top) and $(b_A, \bar{b}_A)$ (bottom) are computed at parton level.}
	\label{fig:tmva1_A40}
\end{figure}

\begin{figure}[!htbp]
	\begin{center}
		\begin{tabular}{ccc}
			\includegraphics[height=5.8cm]{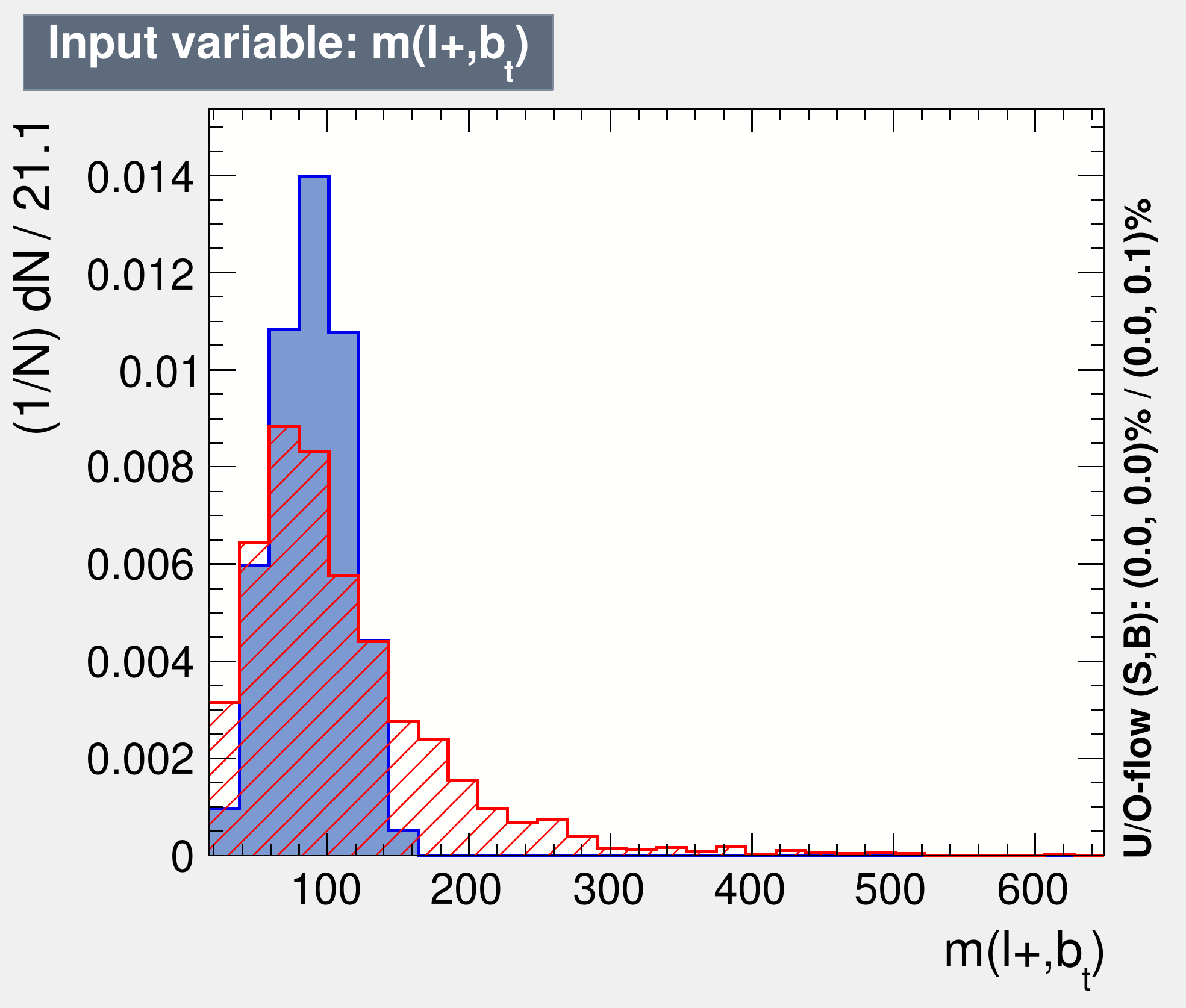} \quad \quad \quad
			\includegraphics[height=5.8cm]{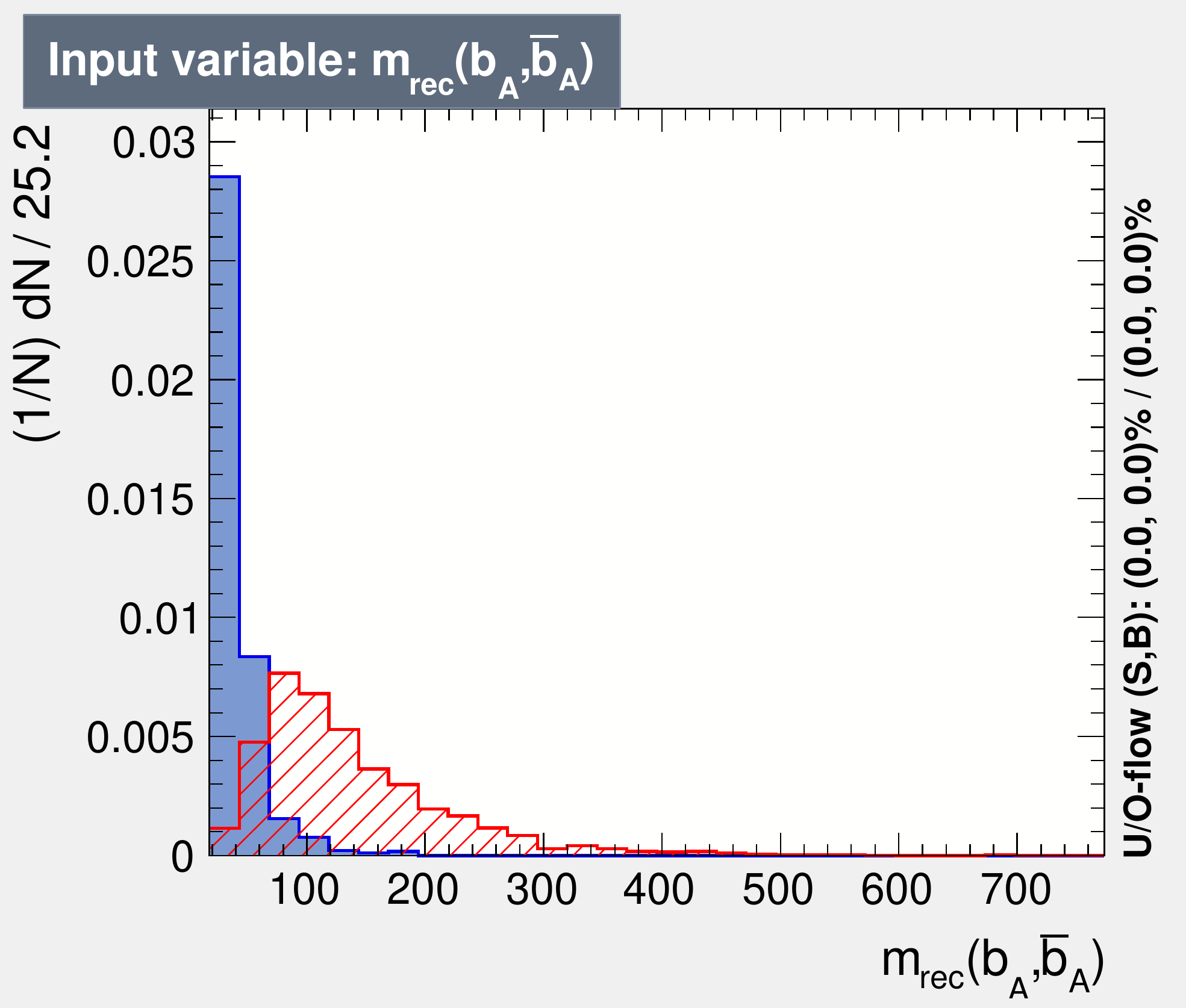}
			\\
			\includegraphics[height=6.5cm]{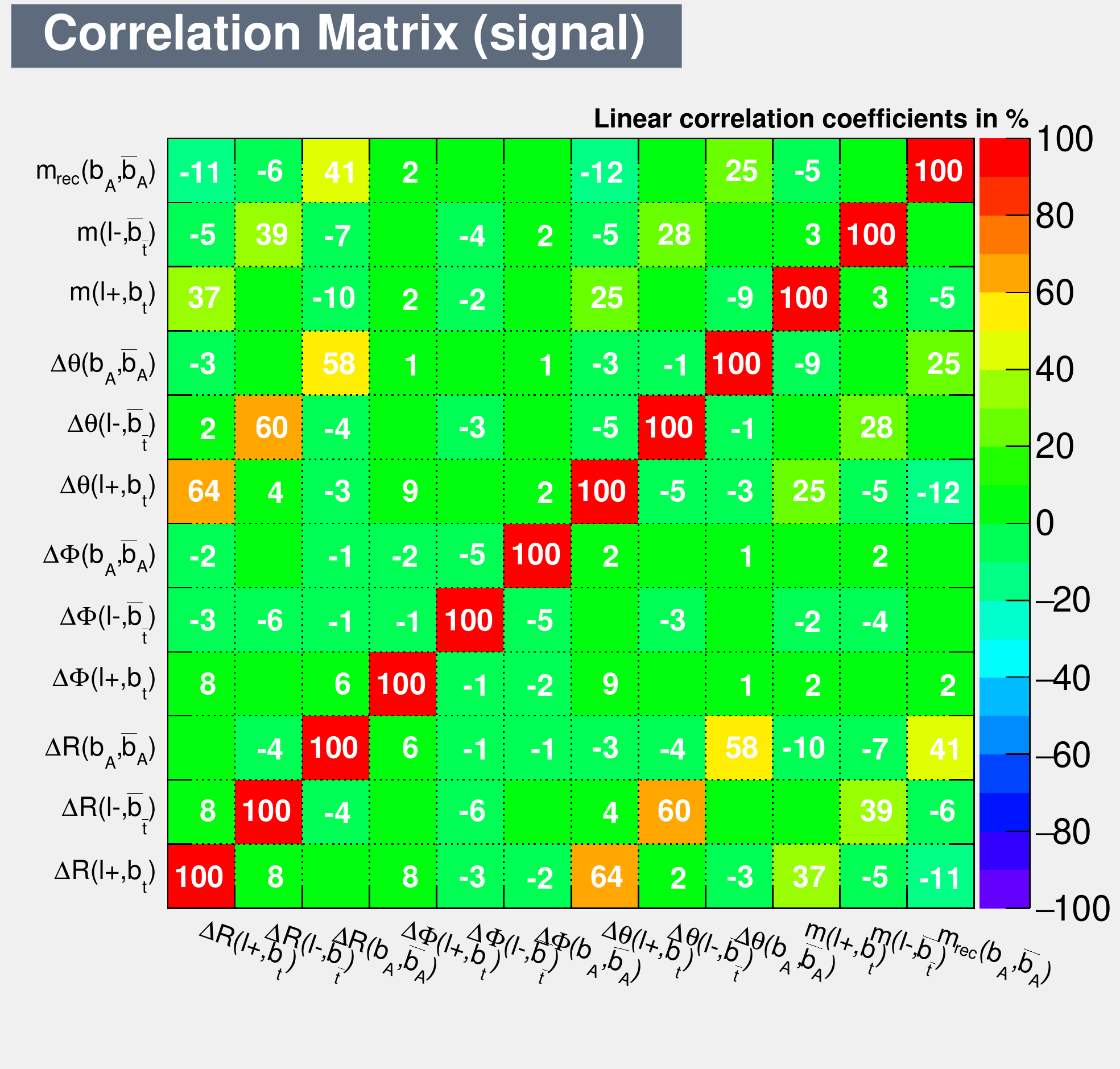} \quad \quad \quad	 	
			\includegraphics[height=6.5cm]{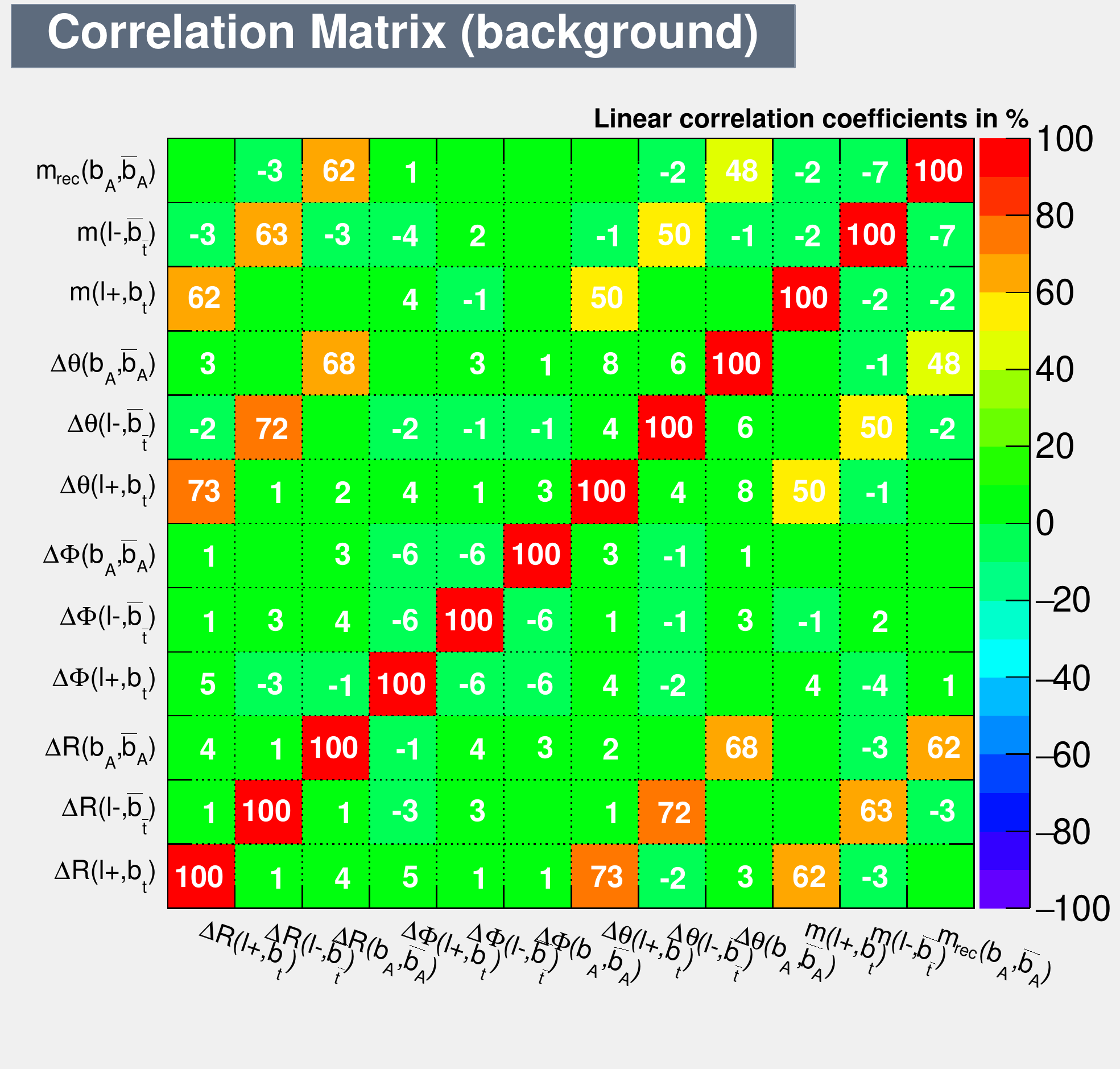}
		\end{tabular}
		\caption{Invariant masses for the systems $(l^+, b_t)$ at parton level (top left) and $(b_A, \bar{b}_A)$ obtained with truth-matching (top right), for $t\bar{t}A$ events with $m_A = 40$ GeV. Below, we have the matrix correlations between the TMVA input variables for the signal (left) and background (right).}
		\label{fig:tmva2_A40}
	\end{center}
\end{figure}

\begin{figure}[!htbp]
	\begin{center}
			\begin{tabular}{ccc}
			\includegraphics[height = 5.5cm]{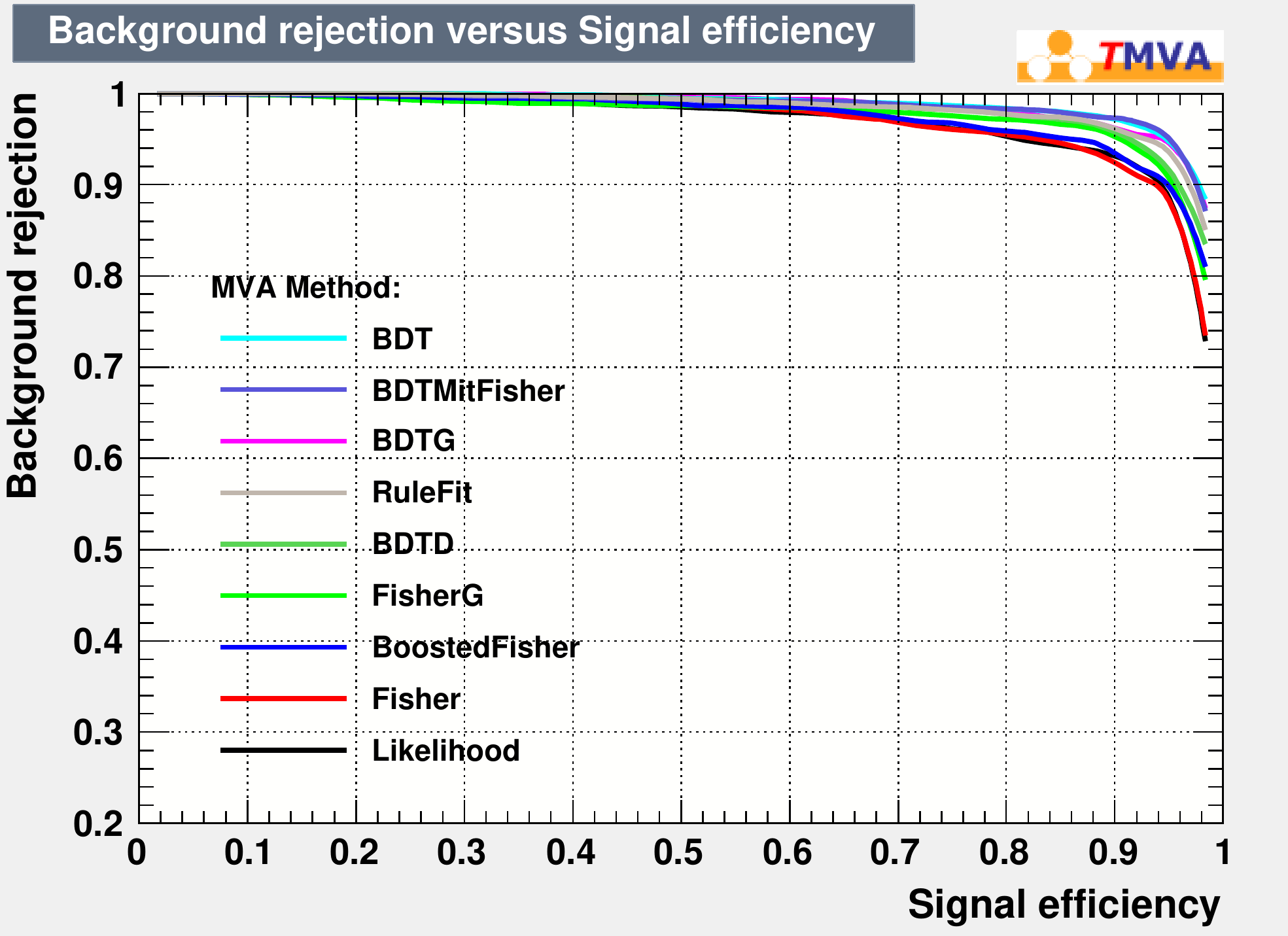}
		\end{tabular}
		\caption{Most performant methods in terms of correct jet assignment for $t\bar{t}A$ events with $m_A = 40$ GeV. We found that typically BDT or BDTG are the best for all $\phi$ boson masses.}
		\label{fig:tmva3_A40}
	\end{center}
\end{figure}

\begin{figure}[!htbp]
	\begin{center}
			\begin{tabular}{ccc}
			\hspace*{-2mm}\includegraphics[height = 5.5cm]{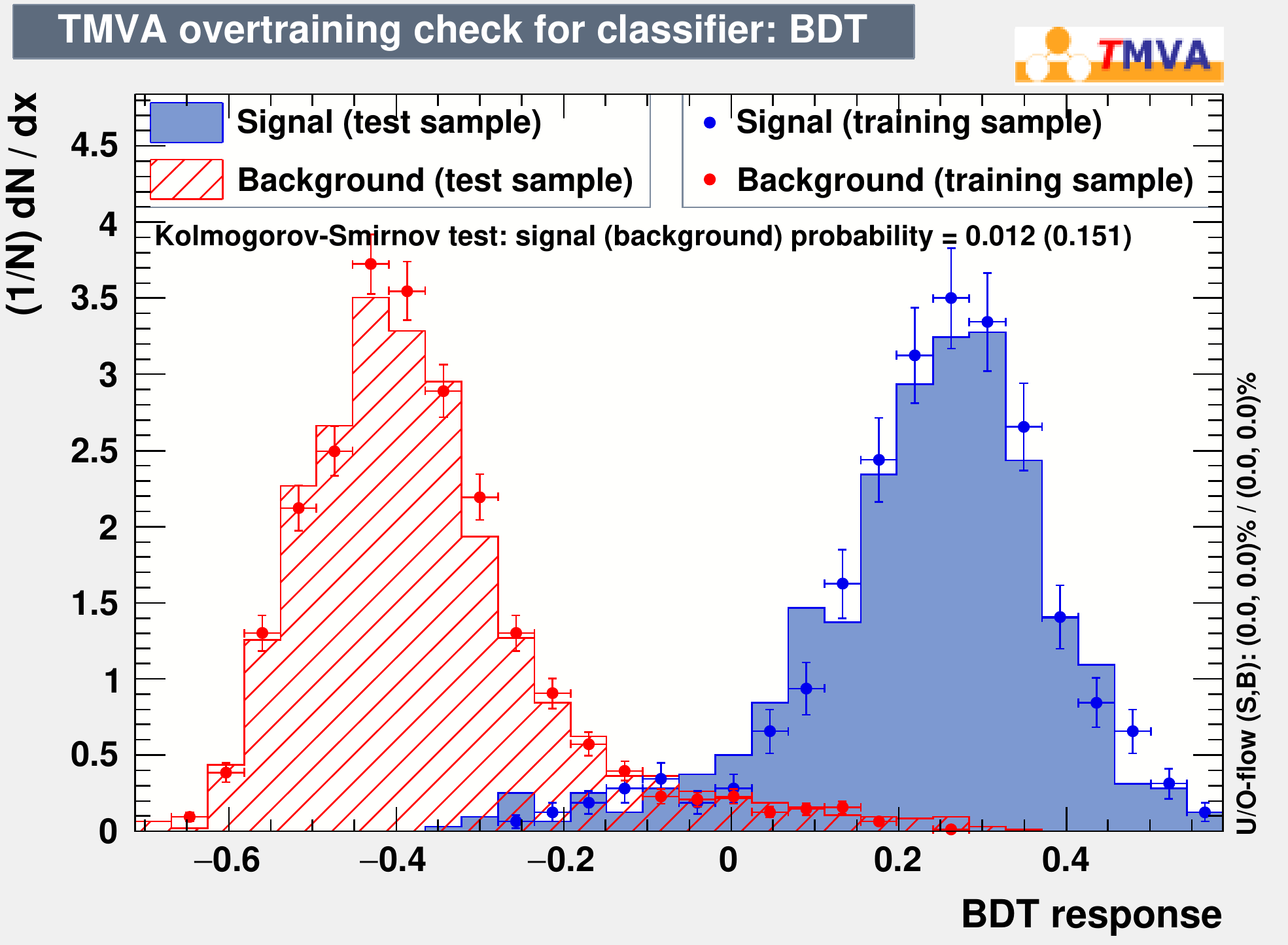} \quad \quad \includegraphics[height = 5.5cm]{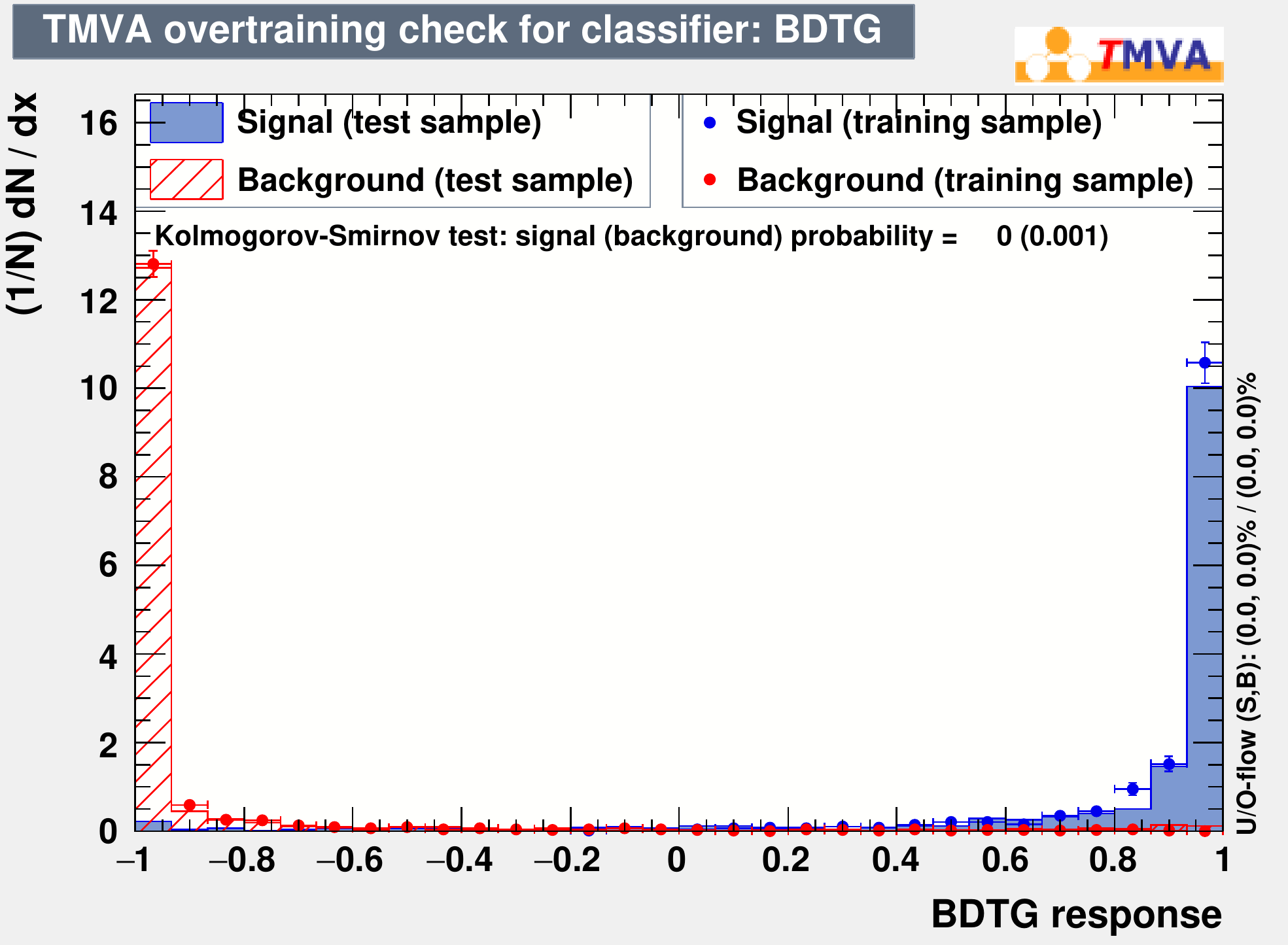}
		\end{tabular}
		\caption{Distributions of the BDT (left) and BDTG (right) discriminants for the signal and background in training and test samples, for $t\bar{t}A$ events with $m_A = 40$ GeV.}
		\label{fig:tmva4_A40}
	\end{center}
\end{figure}

One of the main challenges of the kinematic reconstruction, is the assignment of jets to the reconstructed parton level objects, that match correctly the decay particles of the top quarks, the $W$ and the $\phi$ bosons. In order to check the performance of the kinematic reconstruction a \textit{truth-match} approach was used for the assignment, by finding the four jets with smallest $\Delta R$ distance to the parton level $b$-quarks. As we expect a one-to-one correspondence, a wrong assignment leads, necessarily, to combinatorial background. As when dealing with real data no possible \textit{truth-match} information is available, an association criteria needs to be applied to the events. This relies on a multivariate analysis method tailored for each case, CP-even and CP-odd, using \texttt{TMVA}~\cite{2007physics...3039H}. In both cases, two samples labelled as signal and background were created from simulated $t \bar{t} \phi$ signal events and used for training and testing. While signal samples contain kinematic distributions only from the correct association, background samples contain equivalent kinematic distributions from wrong associations. The following variables were used for training the methods: $\Delta R$, $\Delta \Phi$, $\Delta \theta$ for the pairs $(b_t,l^+)$, $(\bar{b}_{\bar{t}} , l^-)$ and $(b_\phi,\bar{b}_\phi)$, where $b_t$ ($\bar{b}_{\bar{t}}$) represents the bottom (anti-bottom) quark from the top (anti-top) decay and $b_\phi$ ($\bar{b}_\phi$) represents the bottom (anti-bottom) quark from the Higgs decay. The invariant mass of the first two pairs, at parton level, and the invariant mass of the system $(b_\phi,\bar{b}_\phi)$ at the detector level, were also considered. These variables and their correlations are shown in Figures~\ref{fig:tmva1_A40} and \ref{fig:tmva2_A40} for the CP-odd case with $m_A=40$~GeV. 
We have found that, for all the mass values of the CP-even and CP-odd signals, the methods with best performance are the Boosted Decision Tree (BDT) and the Gradient Boosted Decision Tree (BDTG). The latter is the method used in the kinematic reconstruction. During the testing phase, the jet combination chosen is the one returning the highest value of the BDTG discriminant. The {\it Receiver Operating Characteristic} (ROC) curve and the BDT and BDTG discriminant distributions are shown in Figures~\ref{fig:tmva3_A40} and \ref{fig:tmva4_A40}, respectively, for $t\bar{t}A$ with $m_A=40$~GeV.

In events with jet multiplicity above six, only the six highest $p_T$ jets are considered. The reason for this choice relates to the fact that, in about 95\% of all signal events, the jets corresponding to the hadronization of parton level $b$-quarks are among the six with highest $p_T$. Jet combinations also need to verify $m_{l^+ b_t} (m_{l^- \bar{b}_{\bar{t}}})< 150$~GeV and $20$~GeV $< m_{b_\phi \bar{b}_\phi} < 300$~GeV. 

Following the pairing of jets and leptons, the reconstruction of the undetected neutrinos 3-momentum is performed by solving the following set of equations,
\begin{equation}
	\begin{split}
	&(p_{l^+}+p_\nu)^2=m_{W}^2 , \\
	&(p_{l^-}+p_{\bar{\nu}})^2=m_{W}^2 , \\
	&(p_{W^+}+p_b)^2=m_t^2 , \\
	&(p_{W^-}+p_{\bar{b}})^2=m_t^2 , \\
	&p^x_\nu+p^x_{\bar{\nu}}=\slashed{E}^x , \\
	&p^y_\nu+p^y_{\bar{\nu}}=\slashed{E}^y .
	\end{split}
\end{equation}
In the first four, relativistic mass constraints are imposed to signal and background events, by assuming the four-momentum of the $W$ bosons ($p_{W^{\pm}}$), with masses set to $m_W$, are reconstructed using the charged leptons and neutrinos four-momentum $p_{\ell^{\pm}}$ and $p_{\nu(\bar \nu)}$, respectively. The top quarks, $t$ and $\bar t$, with masses set to $m_{t}$, are reconstructed with the $b$-quarks four-momentum, $p_b$ and $p_{\bar b}$, correctly paired to the respective $W^+$ and $W^-$. In the last two equations, the 3-momentum $x$ and $y$ components of the undetected neutrinos (anti-neutrinos),  $p^x_\nu$ ($p^x_{\bar\nu}$) and $p^y_\nu$ ($p^y_{\bar\nu}$), respectively, fully account for the $x$ and $y$ components of the missing transverse energy ($\slashed{E}$). 

Since top quarks and $W$ bosons have non-zero widths, their mass distributions follow Breit-Wigner probability distribution functions ({\it p.d.f.}s), with pole masses fixed to $m_t$ and $m_W$, respectively. In order to reconstruct the neutrino and anti-neutrino four-momenta, we generate random top and anti-top quark masses from 1-dimensional parton level {\it p.d.f.}s, and generate, consistently, random $W^\pm$ masses, following 2-dimensional mass {\it p.d.f.}s of $(m_{W^+},m_t)$ and $(m_{W^-},m_{\bar{t}})$. This ensures kinematic correlations are preserved when generating the top quark and $W$ boson masses. We then solve the equations for all momentum components of the neutrinos. If no solution is found, the mass generation is repeated up-to a maximum of 500 trials. If there is still no solution, the event is discarded. 
Additionally, as the mass equations are of quadratic form, several solutions may exist for a single event. In order to choose the best one, a likelihood function is constructed using {\it p.d.f.}s from the transverse momenta of the neutrinos, the top quarks and the $t\bar{t}$ system, respectively $P(p_{T_\nu})$, $P(p_{T_{\bar{\nu}} })$, $P(p_{T_t })$, $P(p_{T_{\bar{t}} })$, $P(p_{T_{t\bar{t}} })$, all obtained from parton level distributions. Furthermore, we consider the two dimensional mass {\it p.d.f.}  of the $t\bar{t}$ pair, $P(m_{t},m_{\bar{t}})$, and the mass of the reconstructed Higgs, $P(m_\phi)$, obtained with truth-matching. The likelihood is defined according to
\begin{equation}
	L_{t\bar{t}\phi}\propto \frac{1}{p_{T_\nu} p_{T_{\bar{\nu}}} } P(p_{T_\nu})P(p_{T_{\bar{\nu}} })P(p_{T_t })P(p_{T_{\bar{t}} })P(p_{T_{t\bar{t}} })P(m_t, m_{\bar{t}})P(m_\phi) .
\end{equation}

A normalization factor $1/(p_{T_\nu} p_{T_{\bar{\nu}}})$ is applied in the likelihood because energy losses due to radiation emission and effects from detector resolutions will tend to increase the reconstructed neutrino four-momentum. This factor compensates for too extreme values of the neutrinos $p_T$, giving less weight to solutions of that type. We have checked, after event selection and considering only truth-matched signal events, that 66\% to 73\% of the total number of events are correctly reconstructed, corresponding to $\phi$ masses in the range 40~GeV to 300~GeV (for both scalar and pseudoscalar $t\bar{t}\phi$ signals). If truth-match is not applied, the reconstruction efficiency varies from 49\% (51\%) to 63\% (62\%), for scalars (pseudoscalars), in the same mass range. In this case, the number of times the reconstruction results in the same jet configuration as the one found with truth-match varies from 29\% (31\%) to 49\% (55\%) for the same mass range of scalar (pseudoscalar) signals. It is worth mentioning here that the current kinematic reconstruction nicely extends the one discussed in \cite{AmorDosSantos:2017ayi} to a wider mass range of scalar and pseudoscalar bosons with very similar performance numbers, if not better.

\begin{figure*}
\begin{center}
\begin{tabular}{ccc}
\hspace*{-3mm}\epsfig{file=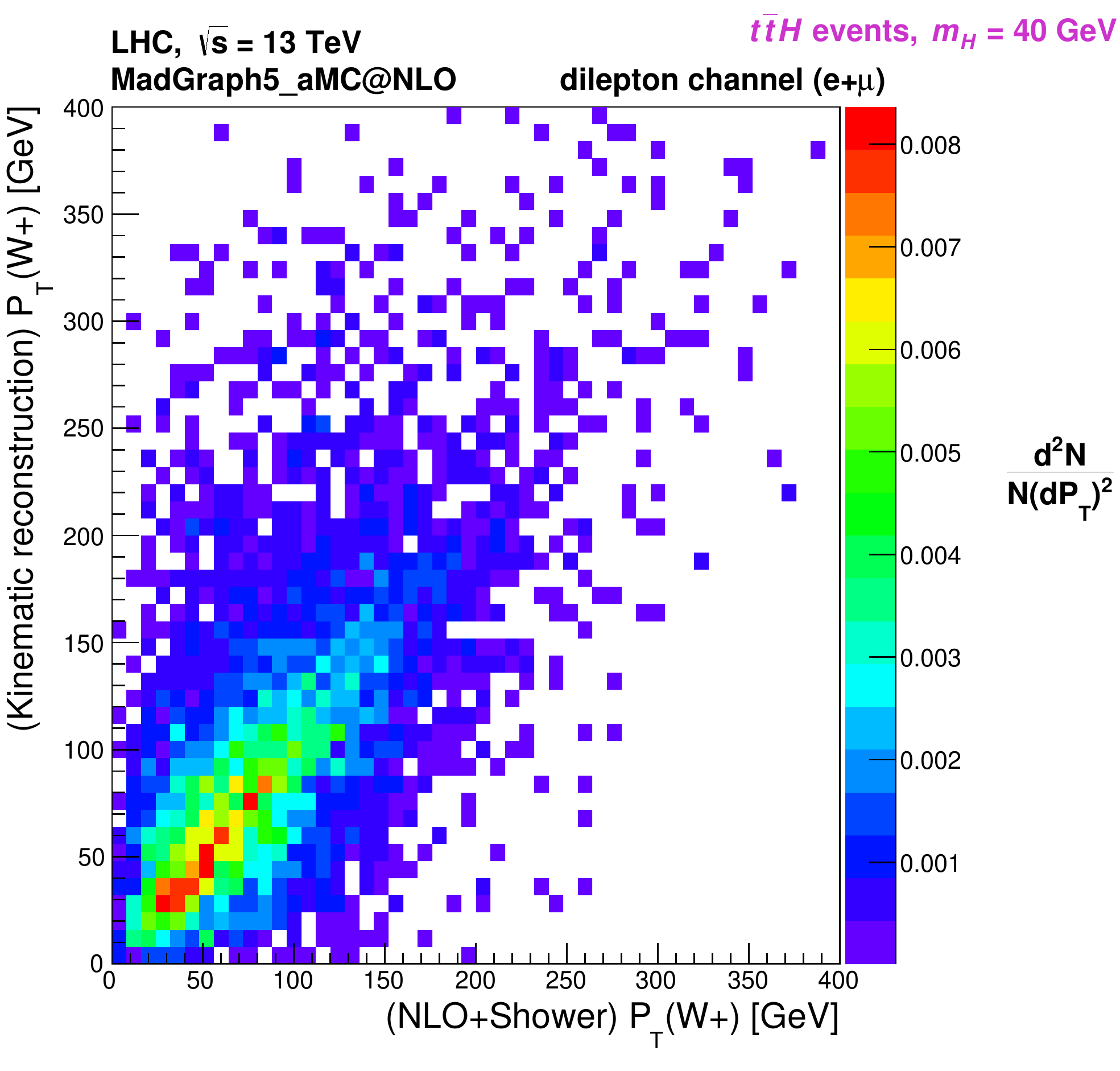,height=5.5cm,width=8cm,clip=} \epsfig{file=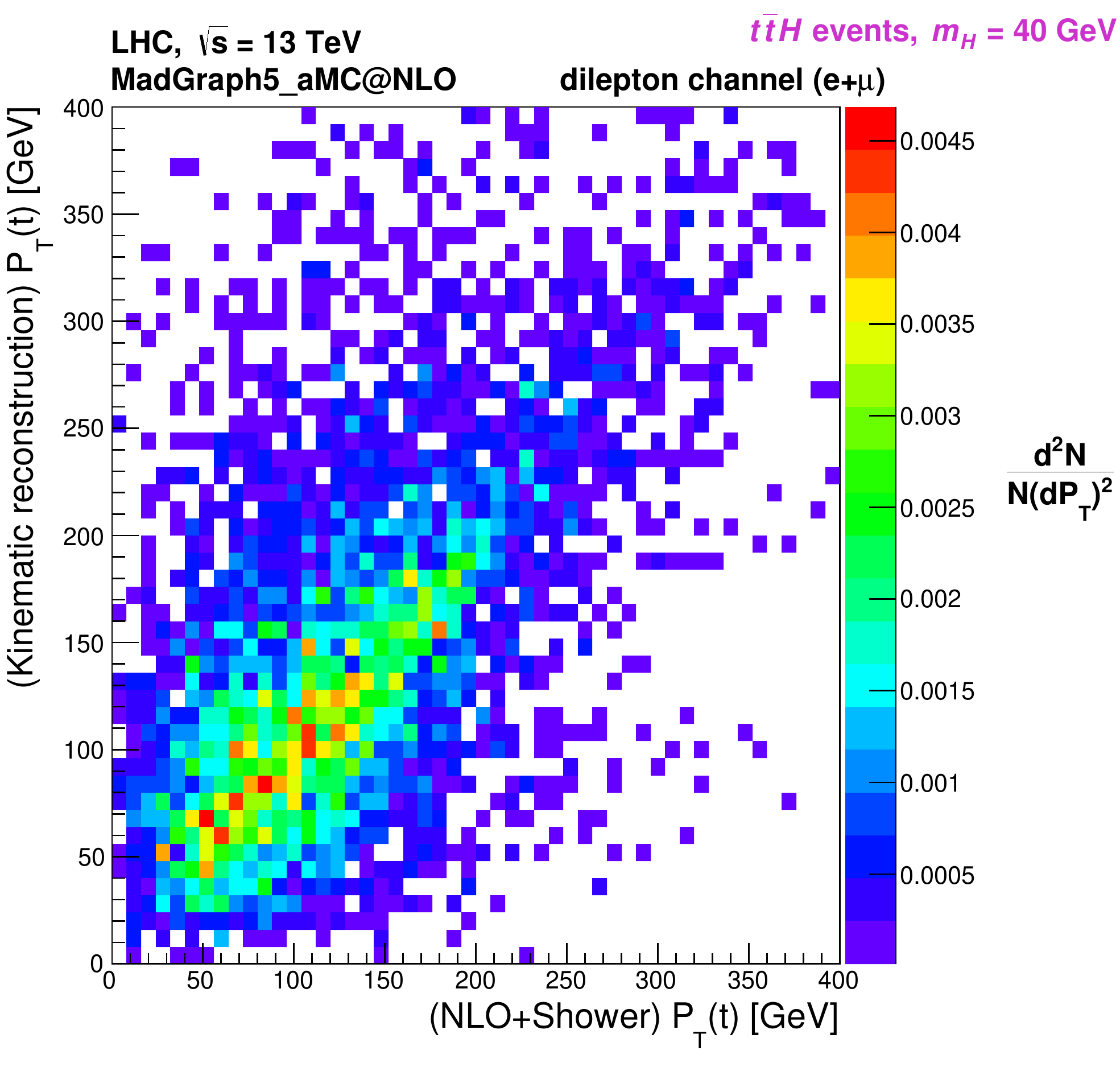,height=5.5cm,width=8cm,clip=} \\[-2mm]
\hspace*{-3mm}\epsfig{file=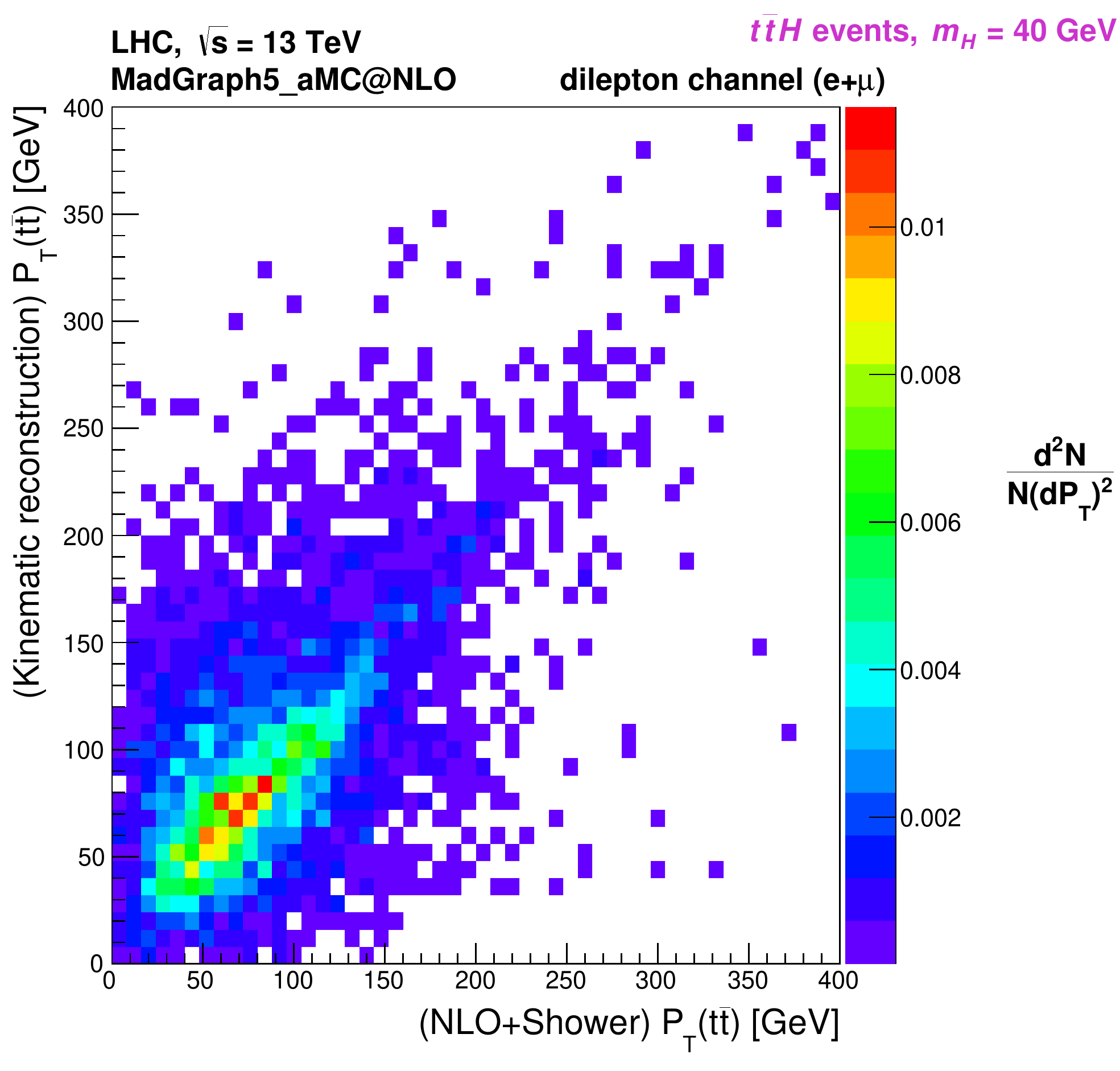,height=5.5cm,width=8cm,clip=} \epsfig{file=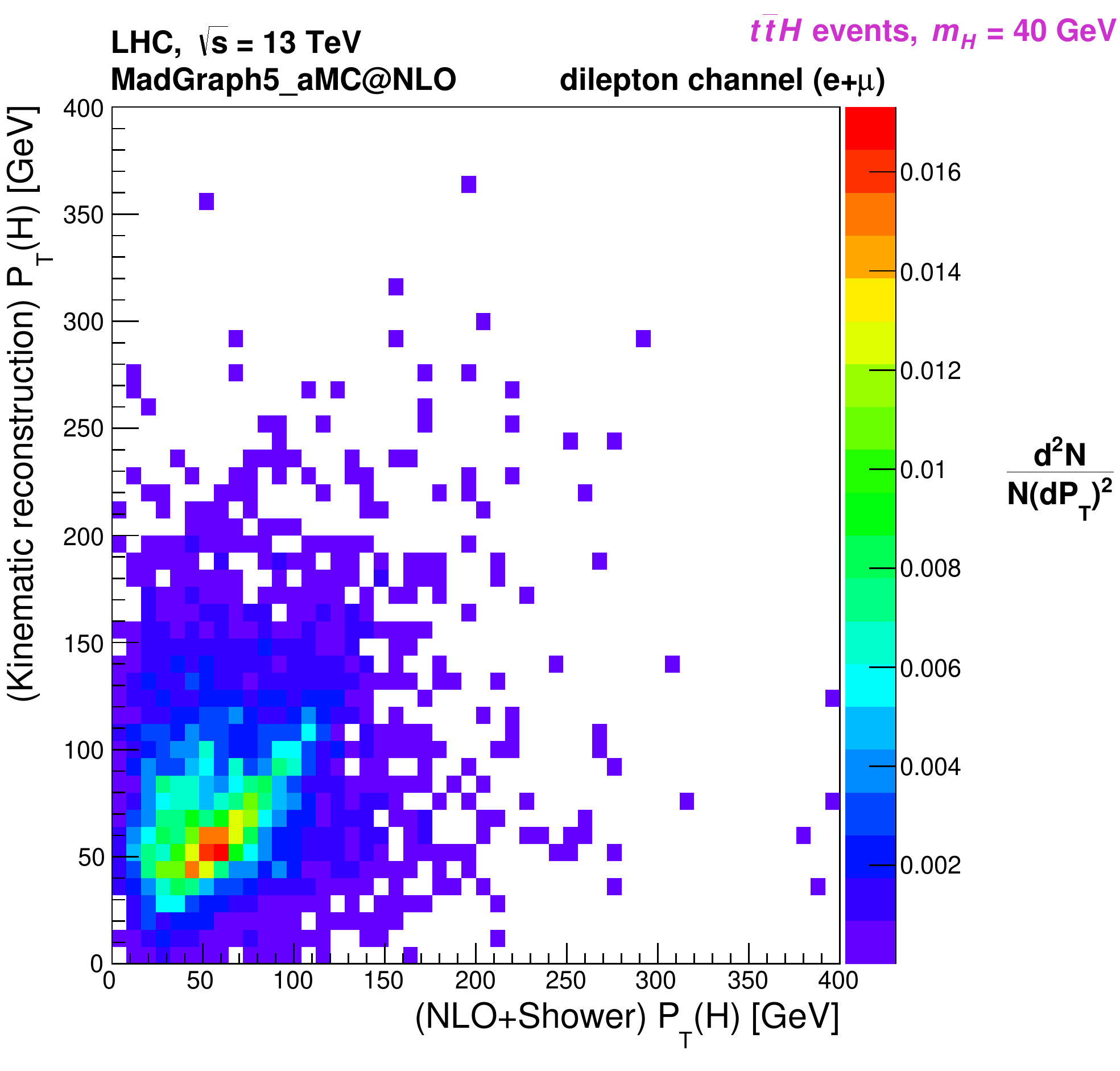,height=5.5cm,width=8cm,clip=} \\[-2mm]
\end{tabular}
\caption{Two-dimensional distributions of $p_T$ in $t\bar{t}H$ events (similar distributions can be found for the $t\bar{t}A$). The horizontal axes represent variables recorded at NLO+Shower, and the vertical axes represent the corresponding variables after kinematic reconstruction. Upper-left: distribution for $W^+$. A similar distribution is obtained for $W^-$, but is not shown here. Upper-right: distribution for $t$. A similar distribution is obtained for $\bar{t}$, but is not shown here. Lower-left: distribution for $t\bar{t}$. Lower-right: distribution for $H$. All distributions are shown for a Higgs mass of 40 GeV.}
\label{fig:genexp2D}
\end{center}
\end{figure*}

Figure \ref{fig:genexp2D} shows two-dimensional $p_T$ distributions of the $W^+$ (top-left), the top quark (top-right), the $t\bar{t}$ system (bottom-left) and the Higgs boson (bottom-right) after kinematic reconstruction of $t\bar{t}H$ events, for $m_H=40$ GeV. The correlation between the parton level ($x$-axis) and reconstructed ($y$-axis) $p_T$ distributions, is clearly visible. The same behaviour is observed for the $t\bar{t}A$ signals.

\begin{figure*}
\begin{center}
\begin{tabular}{ccc}
\hspace*{-3mm}\epsfig{file=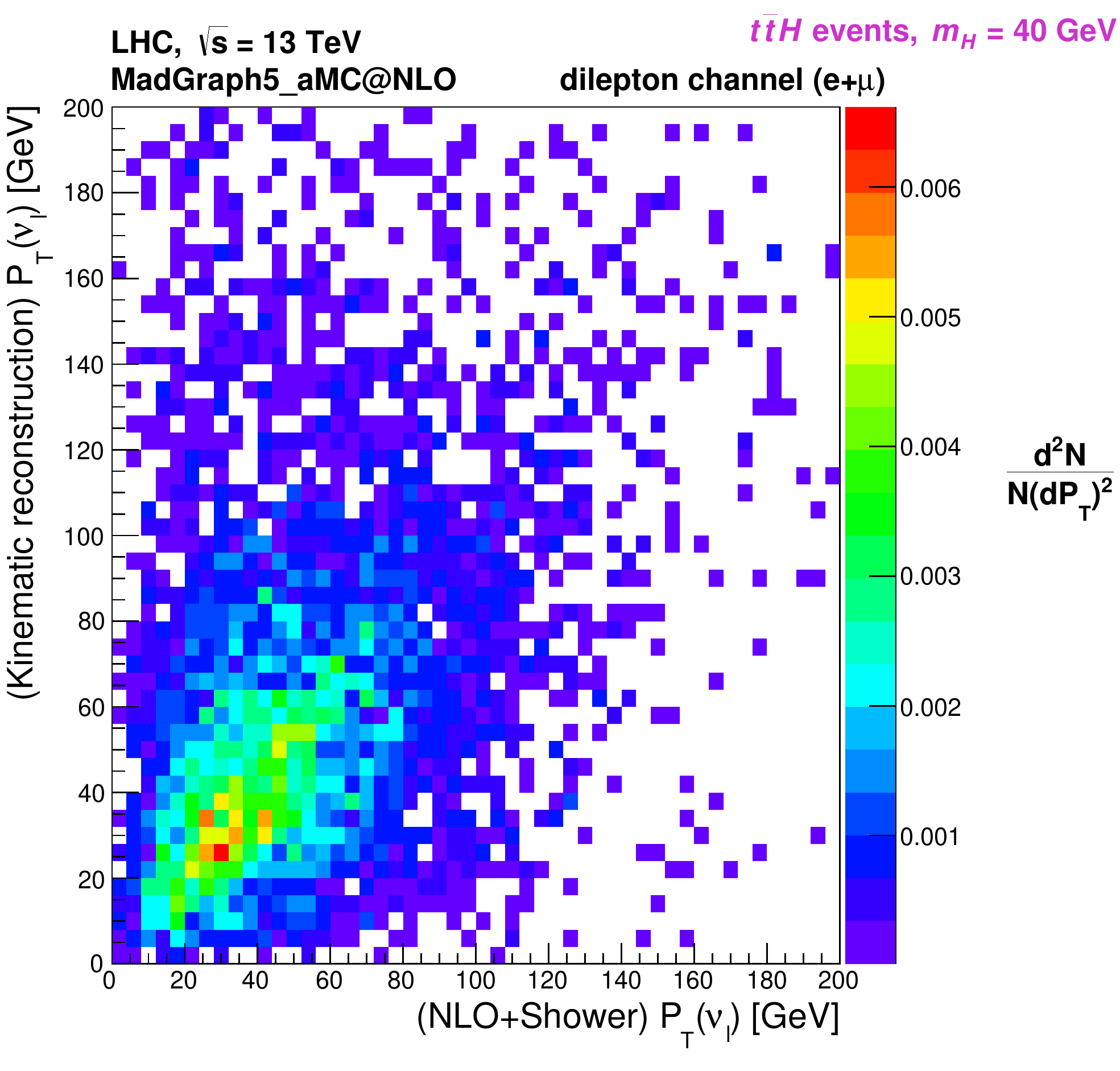,height=5.5cm,width=8cm,clip=} 
\epsfig{file=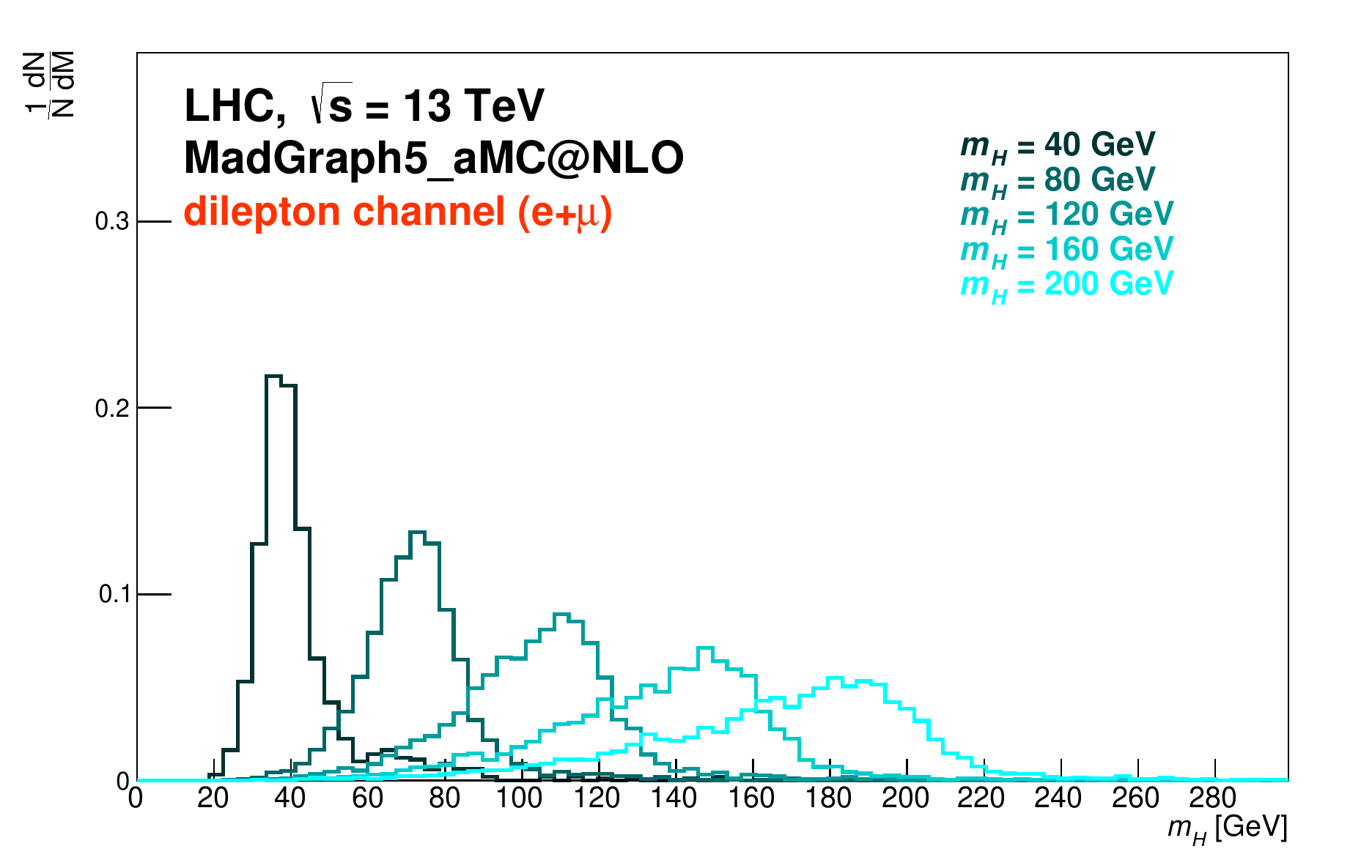,height=5.3cm,clip=} \\[-2mm]
\end{tabular}
\vspace*{-0.2cm}
\caption{Two-dimensional distribution of the neutrino $p_T$ in $t\bar{t}H$ events (left): the NLO+Shower $p_T$ (x-axis) against the reconstructed $p_T$ (y-axis) is shown. Distribution of the reconstructed Higgs boson mass with truth-matched jets in $t\bar{t}H$ events (right).}
\label{fig:genexp}
\end{center}
\end{figure*}

In Figure~\ref{fig:genexp}, we show the neutrino reconstructed $p_T$ versus the parton level value (left) and the distribution of the Higgs boson reconstructed masses, obtained with truth-matching, for several masses of the scalar boson (right).  In spite of the wider spread of values in the neutrino $p_T$ distribution, a clear correlation between the parton level and reconstructed $p_T$ is observed.

\begin{figure}[h!]
\begin{center}
	\begin{tabular}{ccc}
		\hspace*{-3mm}
		\includegraphics[height=7.5cm]{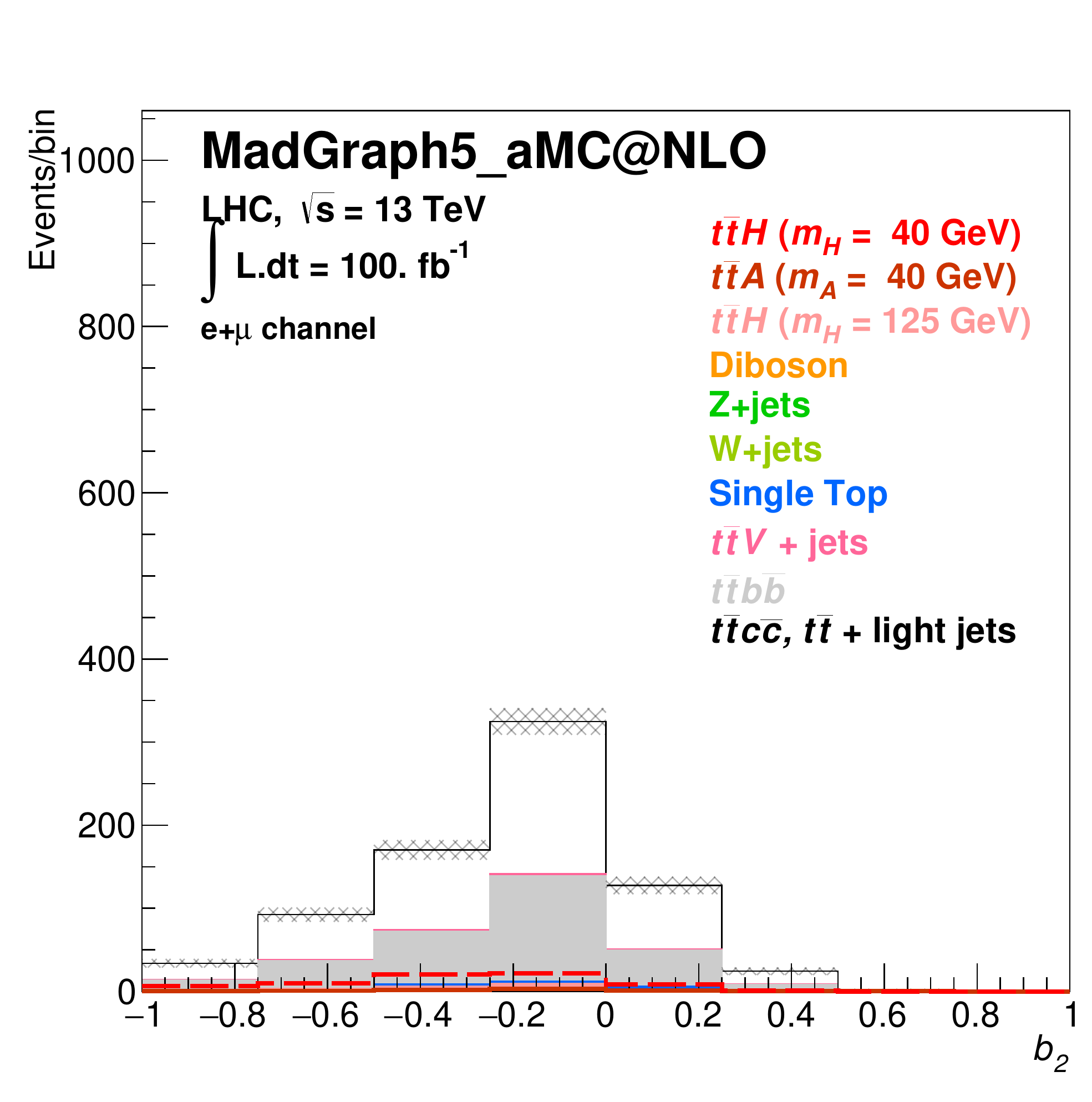}
		\includegraphics[height=7.5cm]{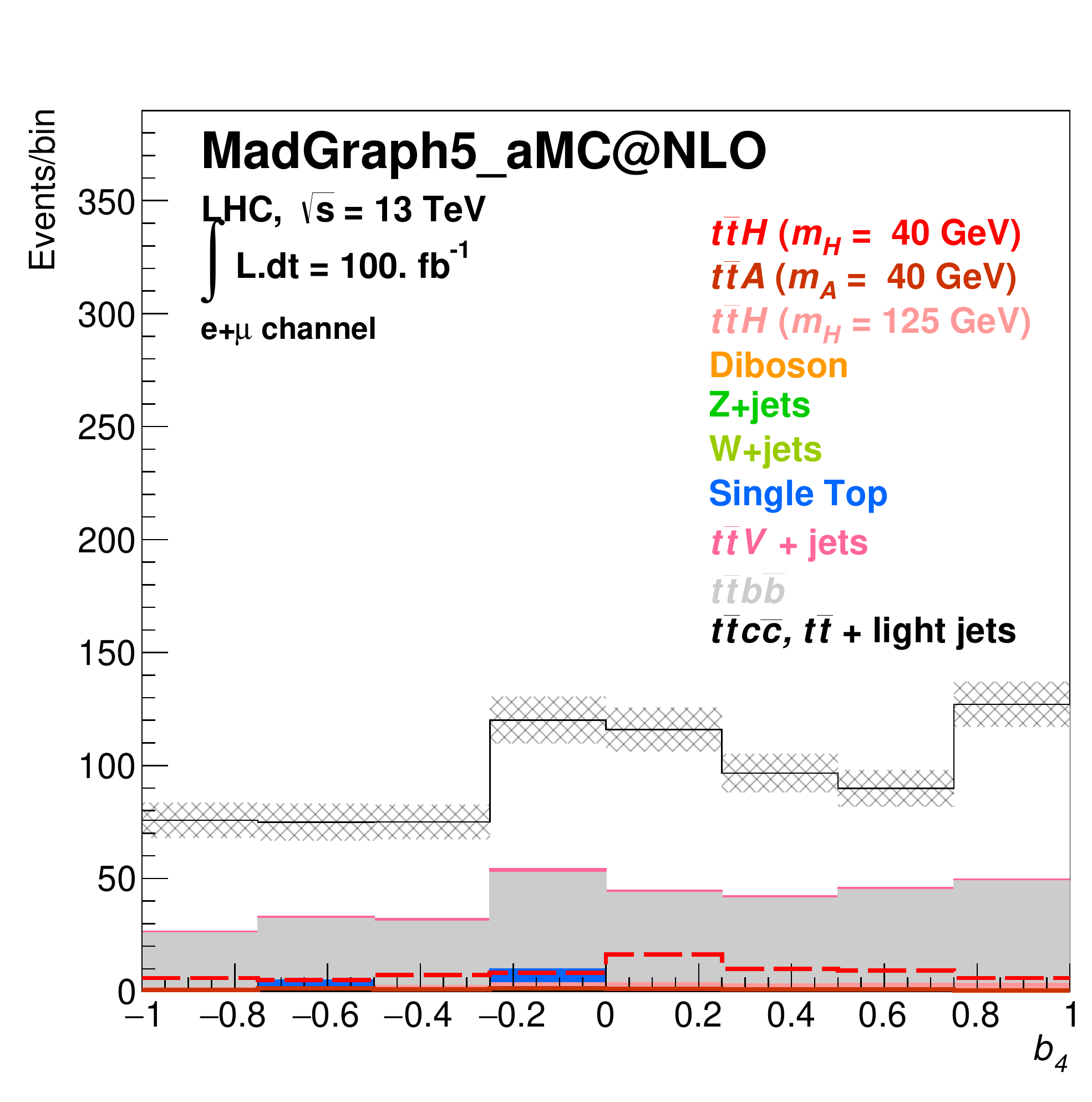}
		\\
		\includegraphics[height=7.5cm]{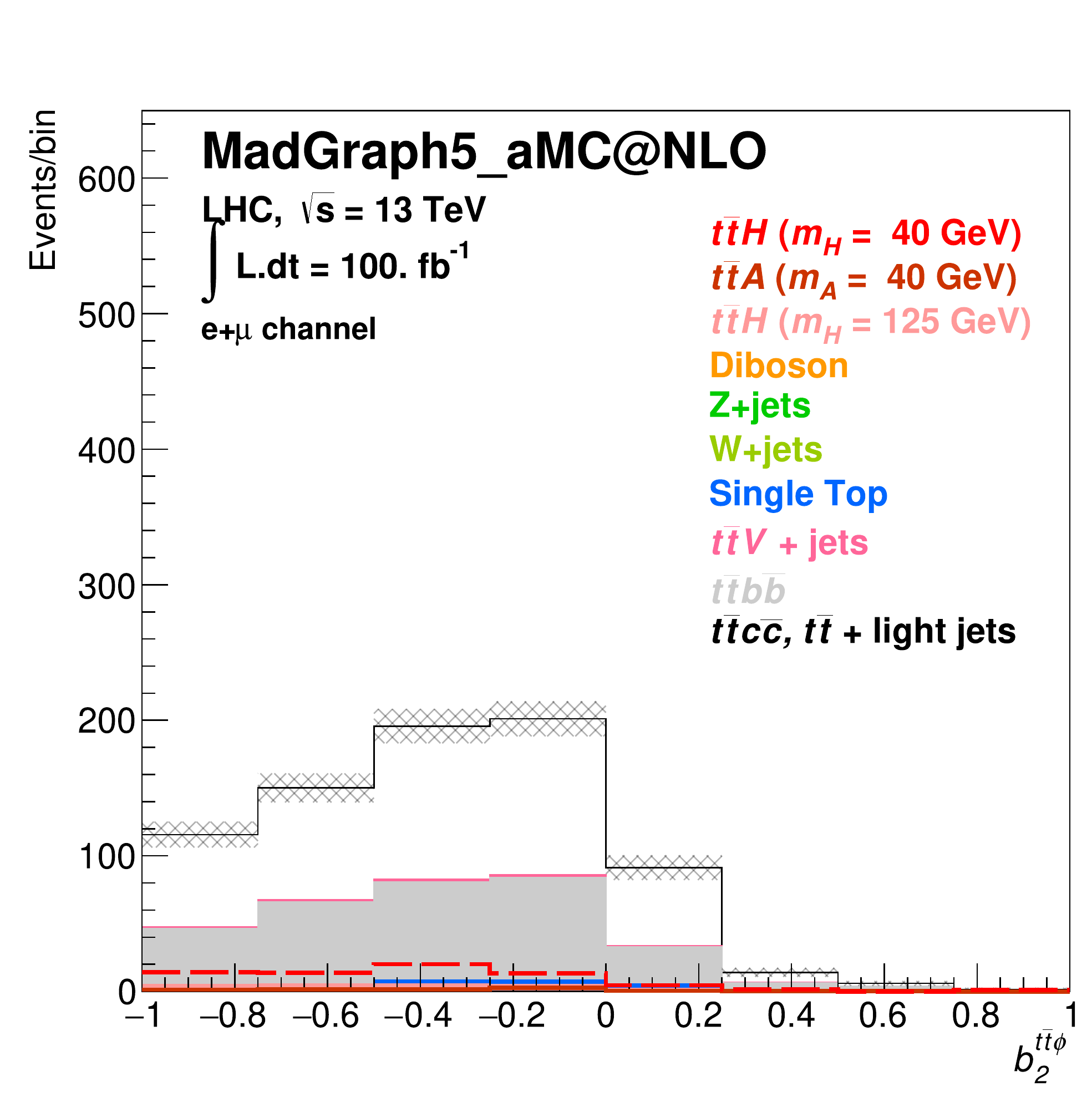}
		\includegraphics[height=7.5cm]{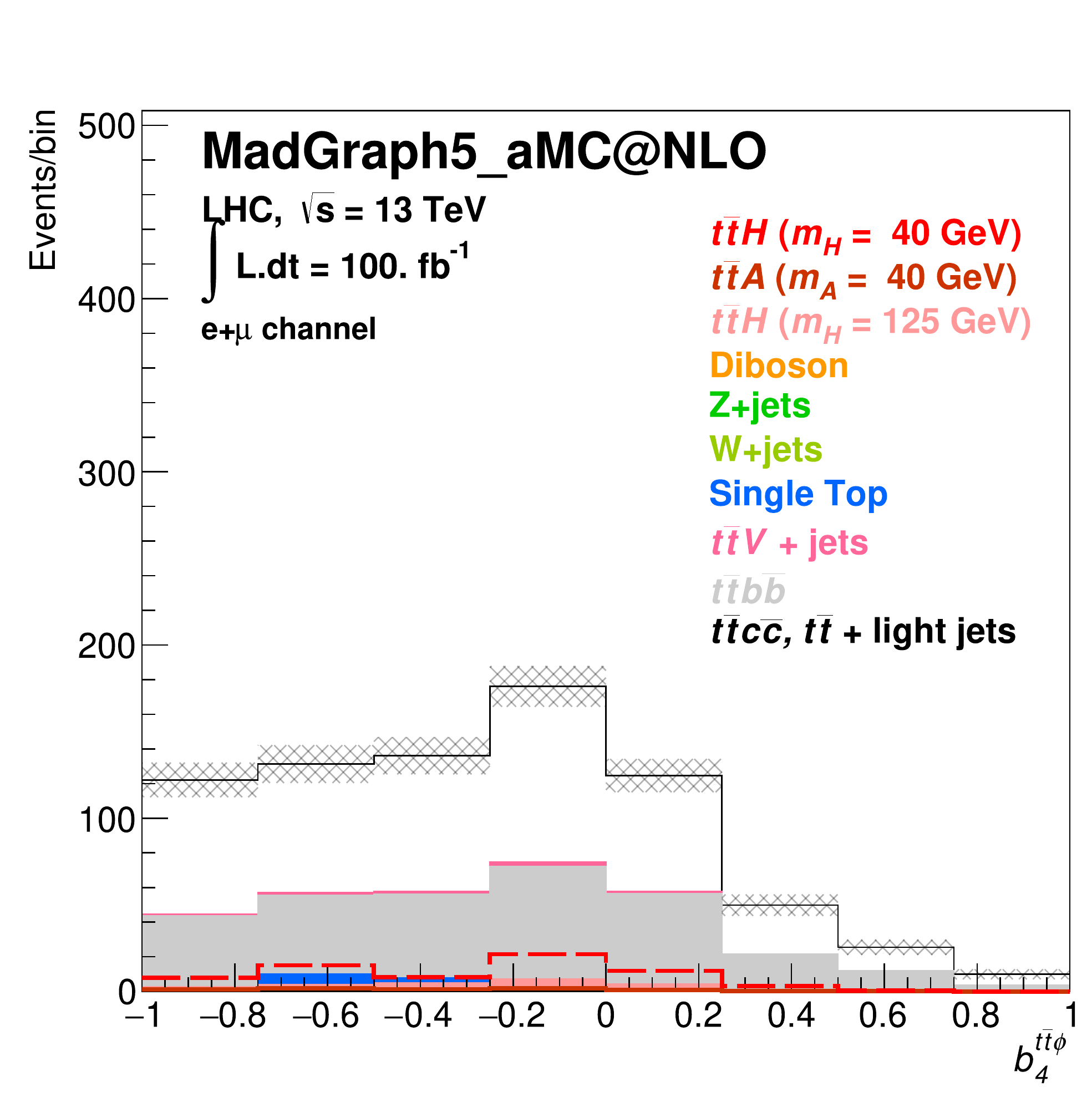}
	\end{tabular}
	\caption{Expected number of background and signal events for the distributions $b_2$ (top left), $b_4$ (top right), $b_2^{t\bar{t}\phi}$ (bottom left) and $b_4^{t\bar{t}\phi}$ (bottom right), for a luminosity of 100 fb$^{-1}$. Kinematic reconstruction and final selection cuts are considered.}
	\label{fig:stackb2b4}
\end{center}
\end{figure}

Additional selection criteria were applied to events following the kinematical reconstruction (final selection cuts), to further increase the signal to background ratio. The depletion of $Z+4$~jets and $Zb\bar{b}+2$~jets backgrounds is accomplished by selecting events with a dilepton invariant mass ($m_{\ell^+ \ell^-}$) outside a window around the $Z$ boson mass ($m_Z=91$~GeV). That is defined by $|m_{\ell^+ \ell^-} - m_Z| > 10$ GeV. Most backgrounds, notably the $t\bar{t}+3$ jets, are mitigated by selecting events with at least 3 $b$-jets. In Figure \ref{fig:stackb2b4}, the expected number of events that survive the full selection criteria, for the different SM backgrounds is shown at the LHC and for an integrated luminosity of 100~fb$^{-1}$. The distributions are compared to the CP-even and CP-odd signals, with $m_\phi = 40$ GeV, for different observables. The $Z$+jets includes the $Z+4$ jets and the $Zb\bar{b}+2$ jets contributions. The $W$+jets, includes the contributions from $W+4$ jets and $Wb\bar{b}+2$ jets. Diboson events are composed of the $WW+3$~jets, $WZ+3$~jets and $ZZ+3$~jets backgrounds, $t\bar{t}c\bar{c}$, $t\bar{t}$ + light jets is the $t\bar{t}+3$ jets process and $t\bar{t}H$ ($m_H = 125$ GeV) is the $t\bar{t}H_{SM}$ process.

\clearpage

\subsection{CLs results for different exclusion scenarios}
\hspace{\parindent} 

In this section, CLs on the exclusion of scalar and pseudoscalar signals $t\bar{t}\phi$ ($\phi=H,A$) evaluated for different scenarios, are computed as a function of the LHC luminosity, up to the High-Luminosity Phase (HL-LHC). Several mass values of the $\phi$ boson are considered, in the range 40-200~GeV.

The $b_2$ and $b_4$ distributions are used to set the CLs evaluated in both the LAB and $t\bar t \phi$ centre-of-mass systems, for comparison. The contribution of all SM backgrounds is taken into account, normalized to the LHC luminosity, as well as the different signal hypotheses. For each scenario under study, one million pseudo experiments are generated, using bin-by-bin Poisson fluctuations around a mean value, which is set to the number of events in each individual bin of the distributions. The probability that a $H_0$ and an alternative $H_1$ hypothesis can describe the pseudo experiment is evaluated for each of them. The likelihood ratio of the $H_1$ and $H_0$ probabilities is used as test statistics, to compute the CLs with which hypothesis $H_1$ can be excluded assuming $H_0$ is true. The expected CLs for exclusion were calculated as a function of the integrated luminosity, from 100 to 3000~fb$^{-1}$, using the $b_2$ and $b_4$ observables. The calculation of the CLs follows the prescription set by \cite{Read:2002hq, Junk:1999kv}. The different scenarios under consideration are: 

\begin{itemize}
	\item Scenario 1: Exclusion of the SM plus a new CP-even scalar particle, assuming the SM. In this case, $H_0$ is the SM only hypothesis~\footnote{Consisting of diboson, Z + jets, W + jets, single top, $t\bar{t}$V + jets, $t\bar{t}b\bar{b}$, $t\bar{t}c\bar{c}$, $t\bar{t}$ + light jets and $t\bar{t}H$ ($m_H=125$ GeV) events.}, while $H_1$ is the SM plus a new CP-even signal;
	\item Scenario 2: Exclusion of the SM plus a new CP-odd scalar particle, assuming the SM. In this case, $H_0$ is the SM only hypothesis, while $H_1$ is the SM plus a new CP-odd signal;
	\item Scenario 3: Exclusion of the SM plus a new CP-odd scalar particle, assuming the SM plus a new CP-even scalar particle of the same mass. In this case, $H_0$ is the SM plus a new CP-even signal hypothesis, while $H_1$ is the SM plus a new CP-odd signal;
	\item Scenario 4: SM exclusion, assuming the SM plus a new CP-even scalar particle. In this case, $H_0$ is the SM plus a new CP-even signal hypothesis, while $H_1$ is the SM only.
\end{itemize} 

\begin{figure}[h!]
	\begin{center}
		\begin{tabular}{ccc}
			\hspace*{-3mm}
			\includegraphics[height=7.cm]{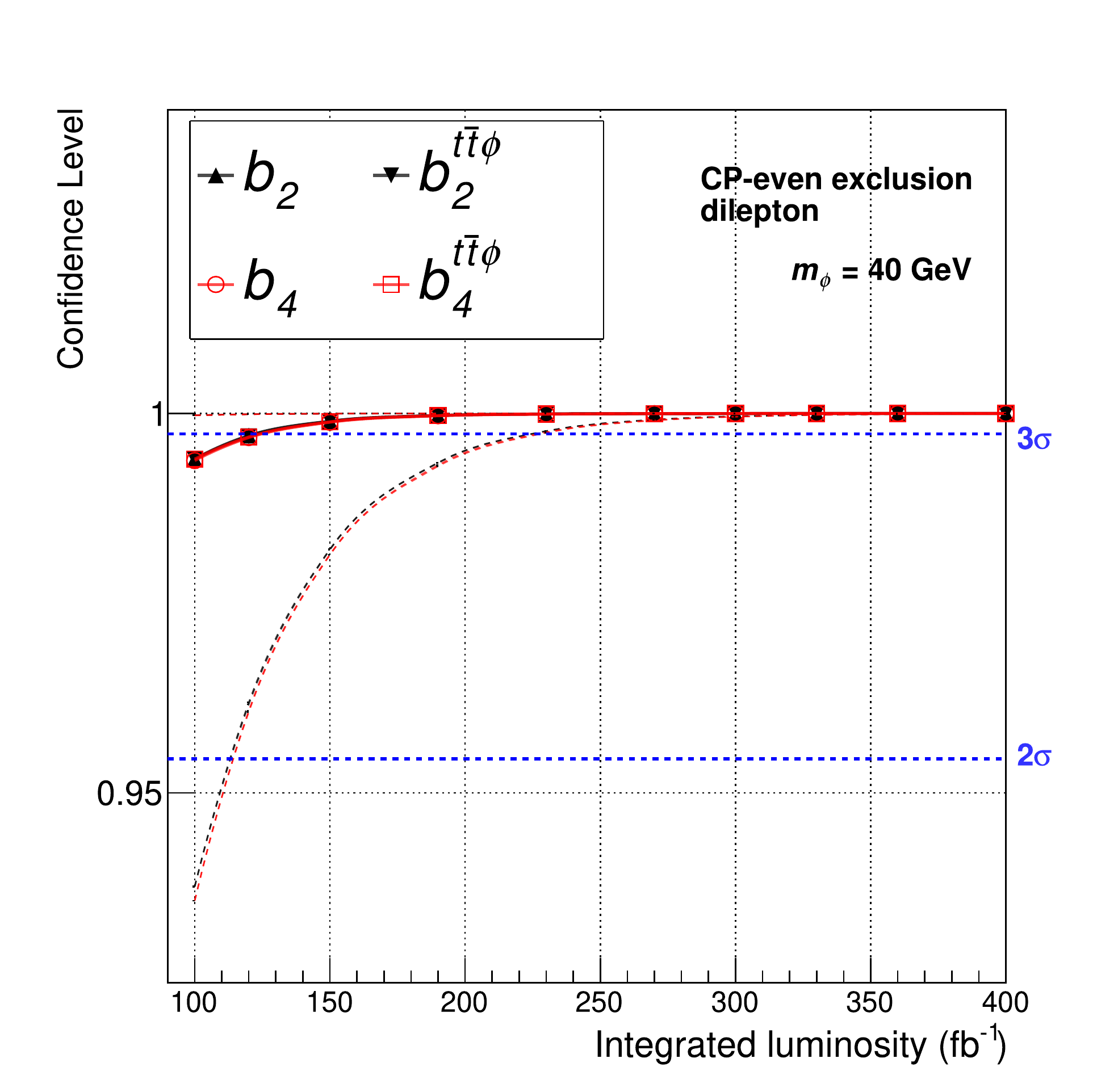}
			\includegraphics[height=7.cm]{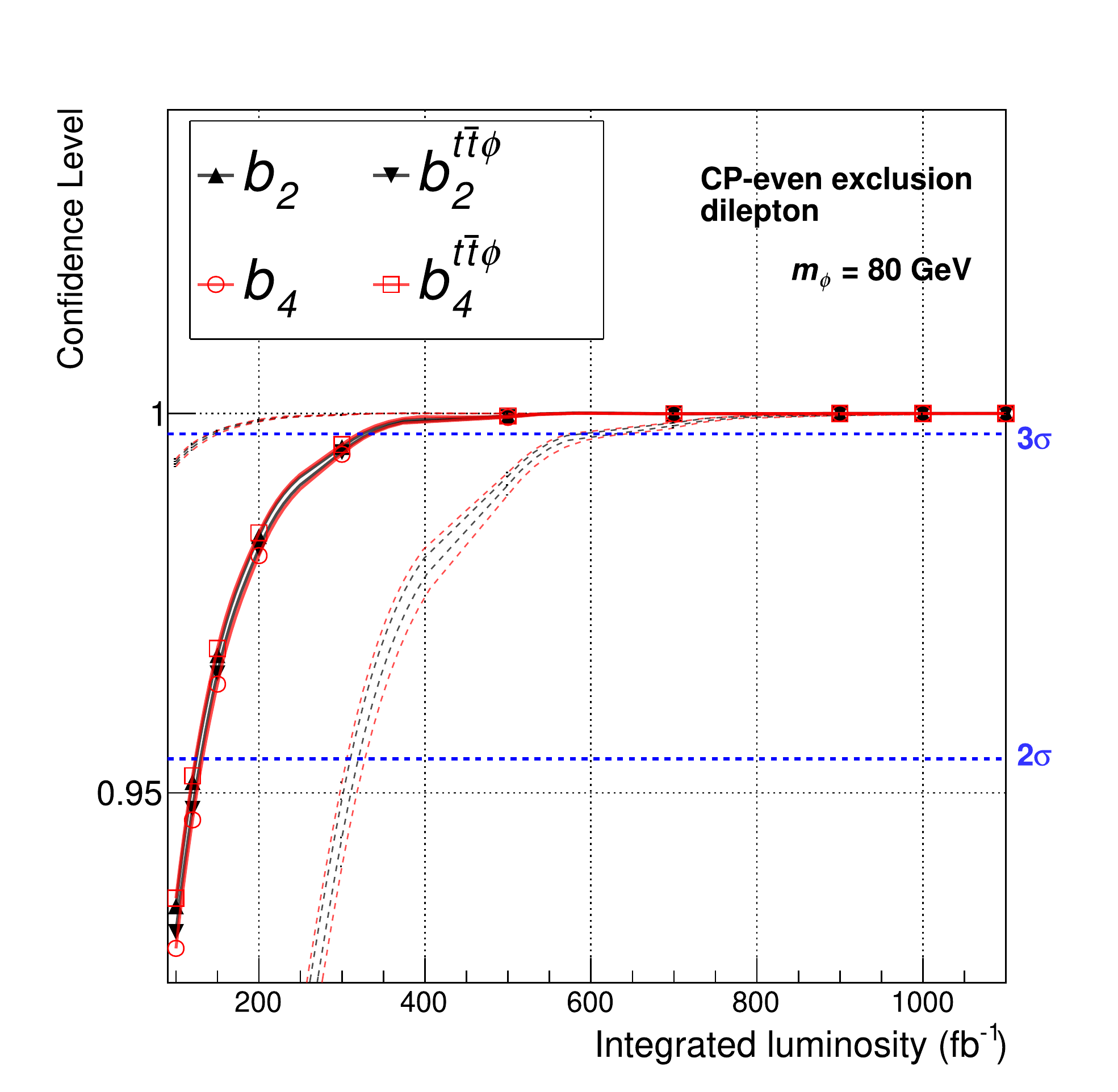}
			\\
			\includegraphics[height=7.cm]{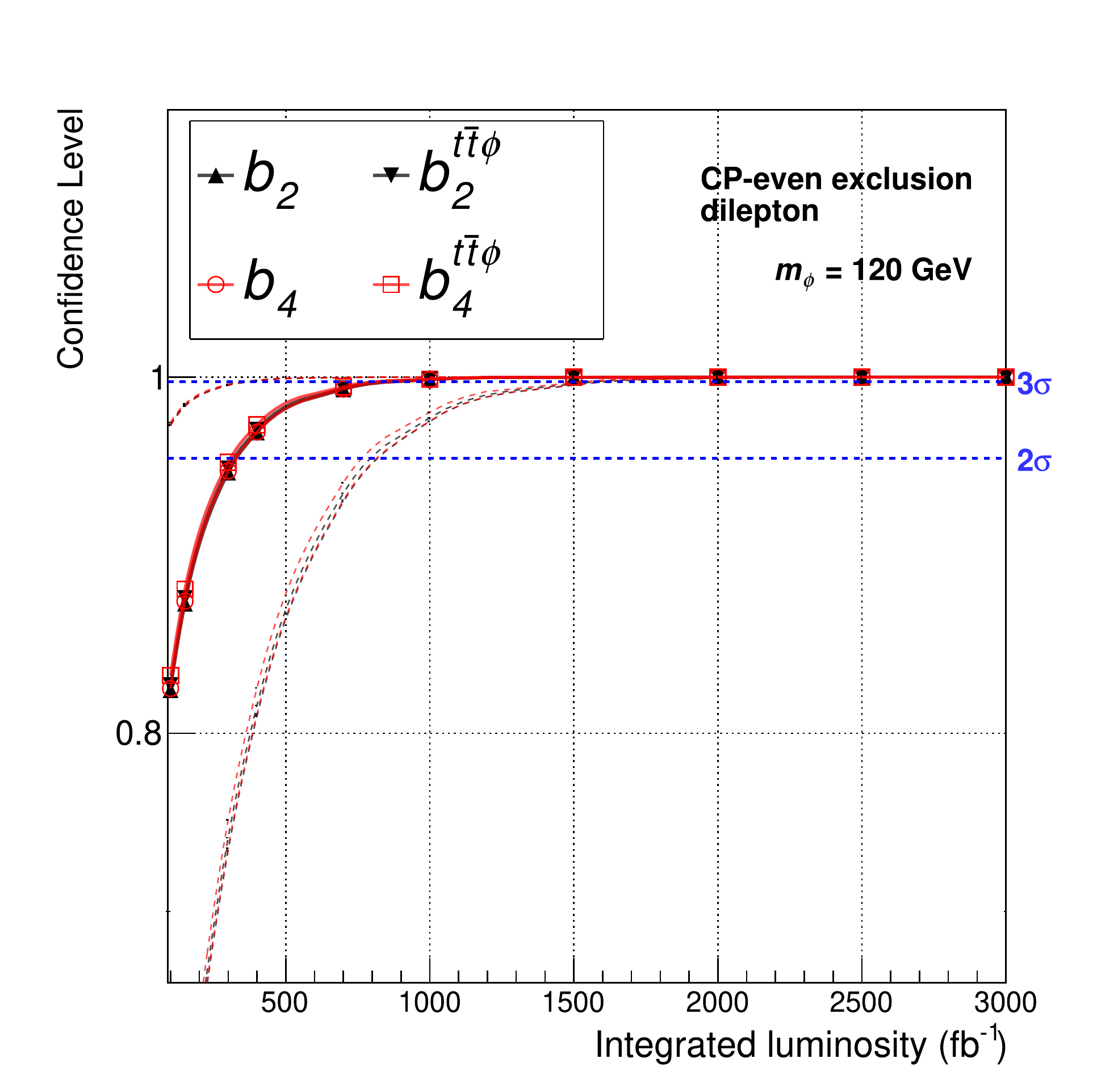}
			\includegraphics[height=7.cm]{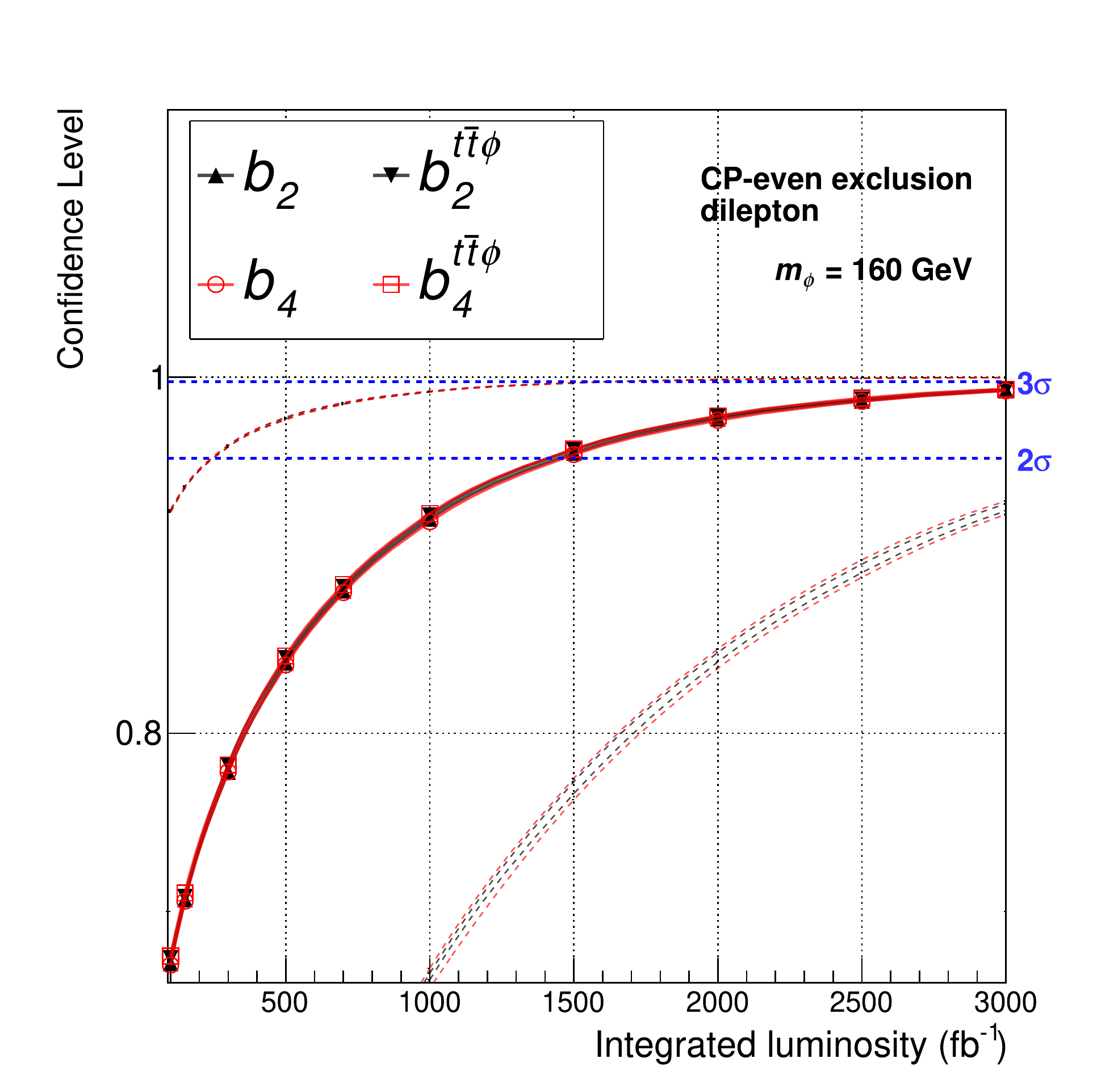}
		\end{tabular}
		\caption{Expected CLs for CP-even exclusion assuming the SM (scenario 1), as a function of the integrated luminosity. The luminosity is increased until CL$ = 1$ or $\mathcal{L} = 3000$ fb$^{-1}$. Each plot represents a different $\phi$ boson mass. The vertical axis scale also changes, for visualization purposes. The red and black dashed lines represent the $\pm 1\sigma$ bands for the variables considered. The blue dashed lines represent the $2\sigma$ or $3\sigma$ CL lines.}
		\label{case1}
	\end{center}
\end{figure}

\begin{figure}[h!]
	\begin{center}
		\begin{tabular}{ccc}
			\hspace*{-3mm}
			\includegraphics[height=7.cm]{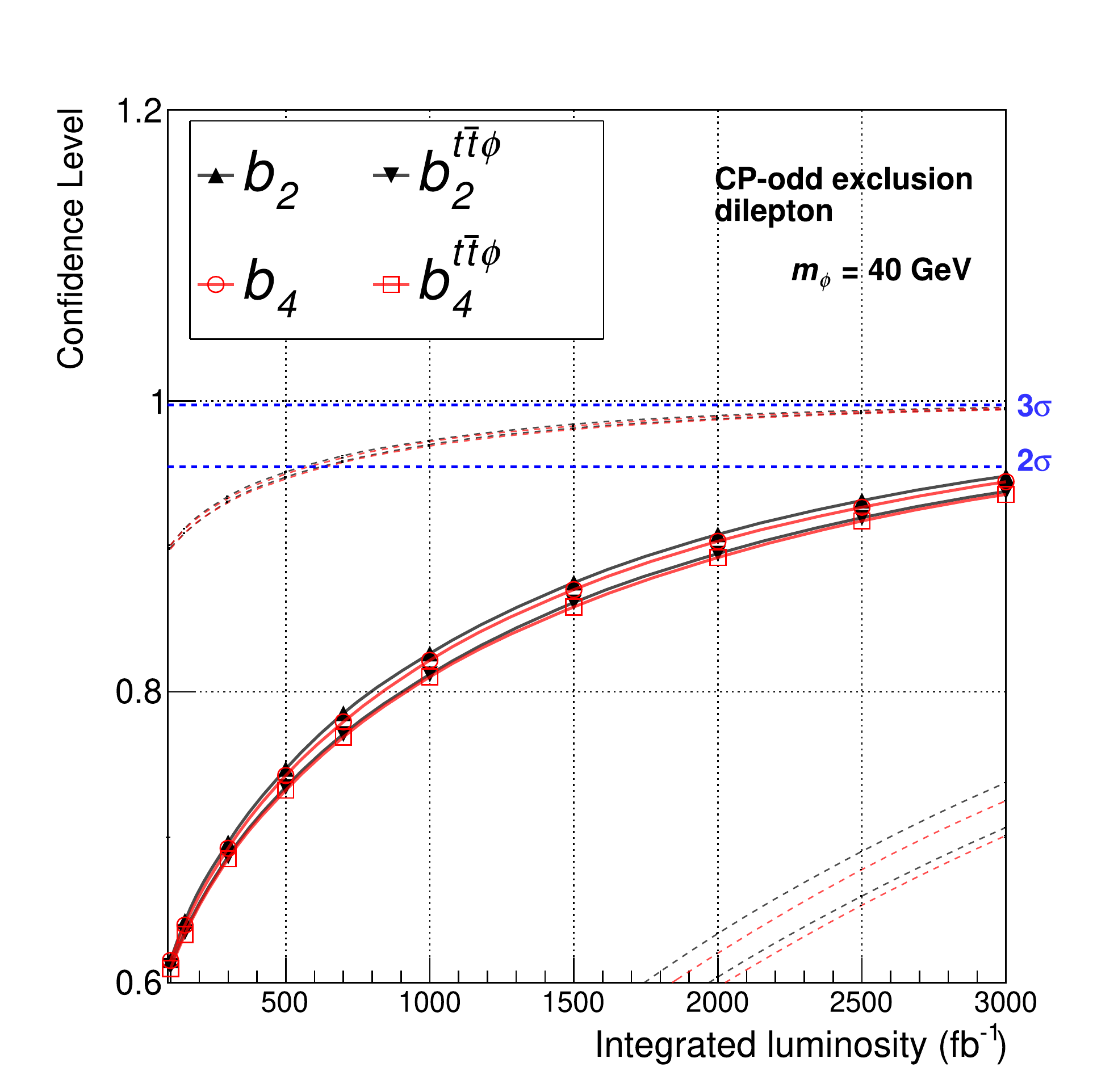}
			\includegraphics[height=7.cm]{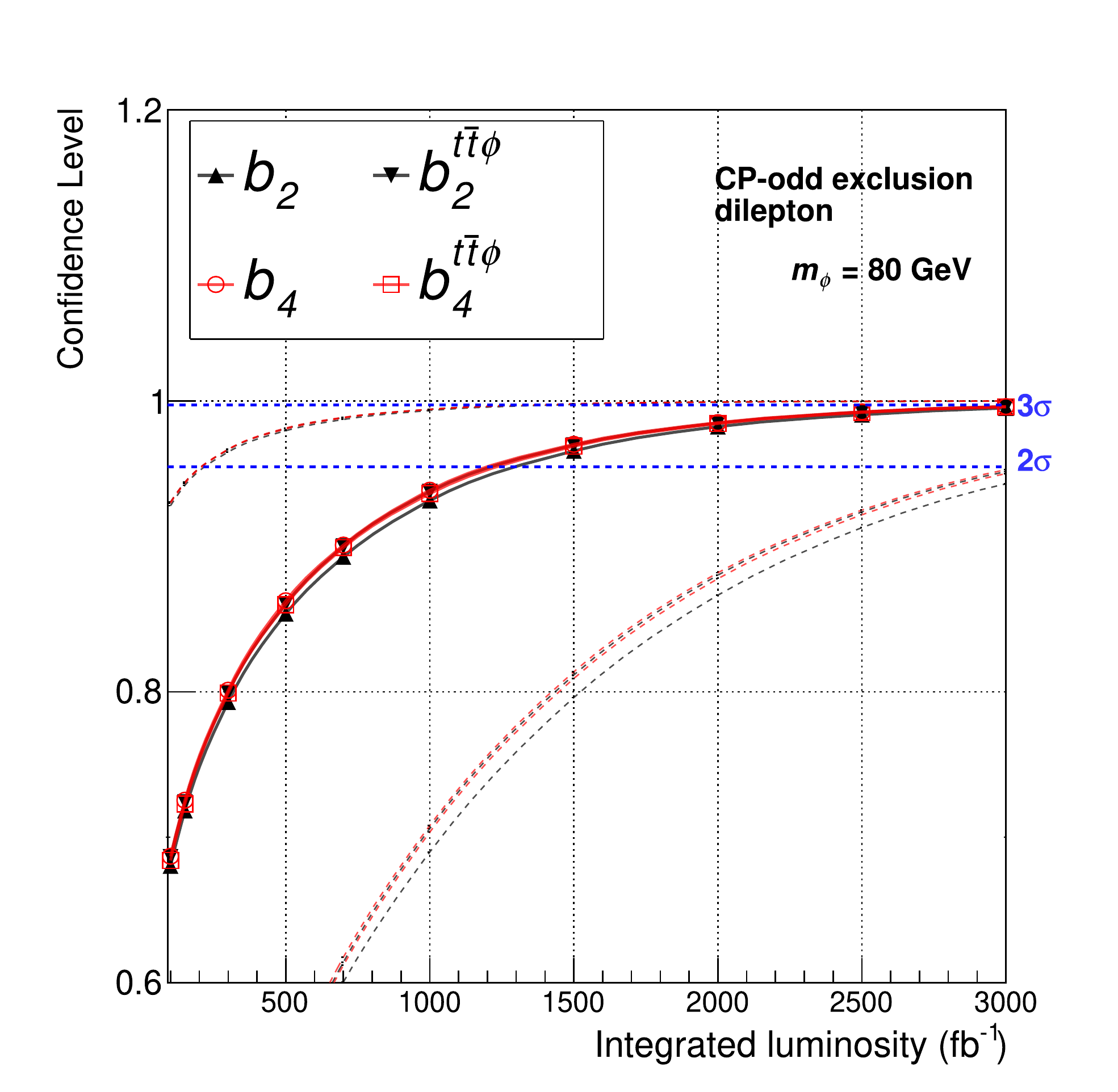}
			\\
			\includegraphics[height=7.cm]{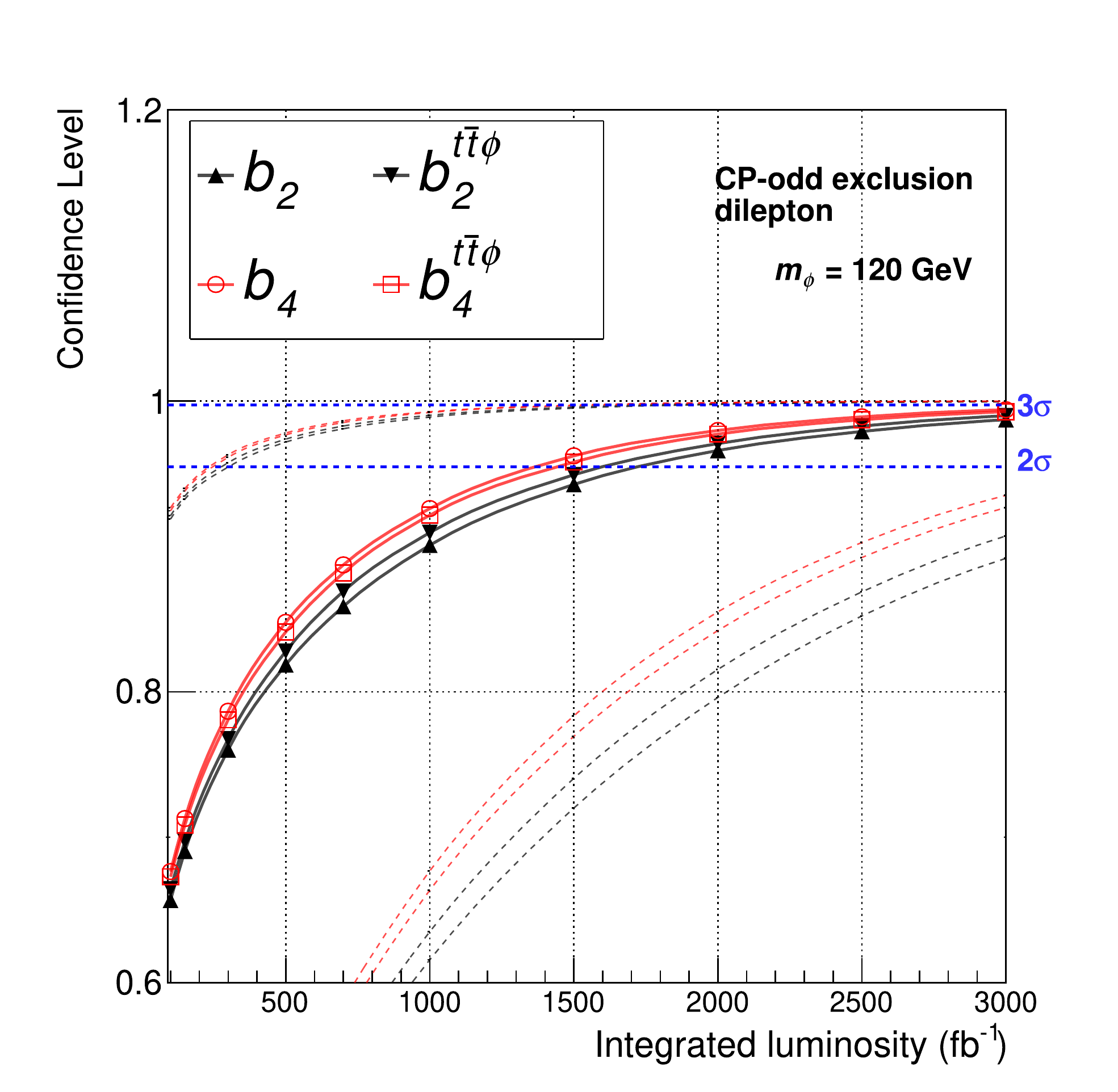}
			\includegraphics[height=7.cm]{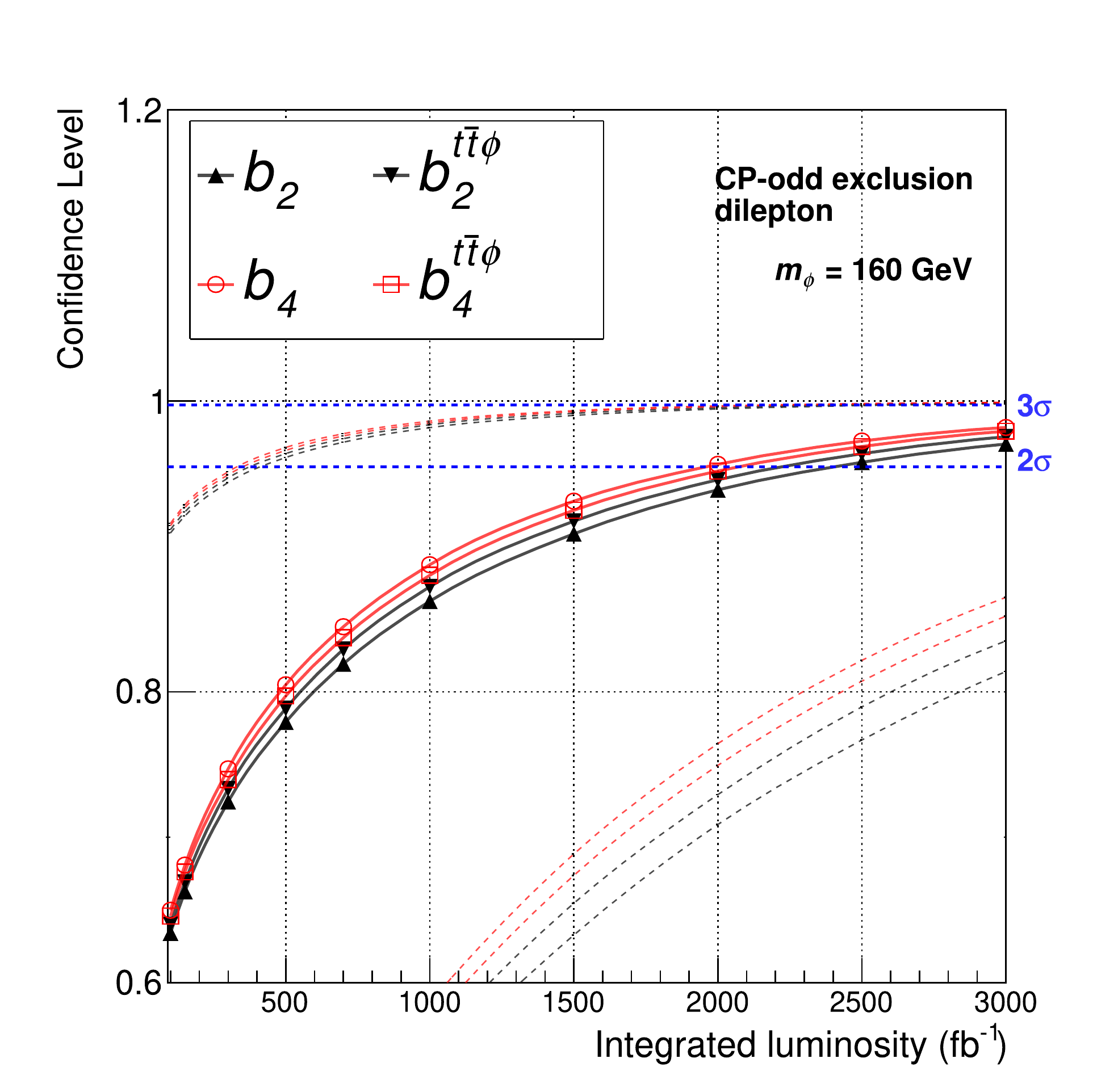}
			\\
			\includegraphics[height=7.cm]{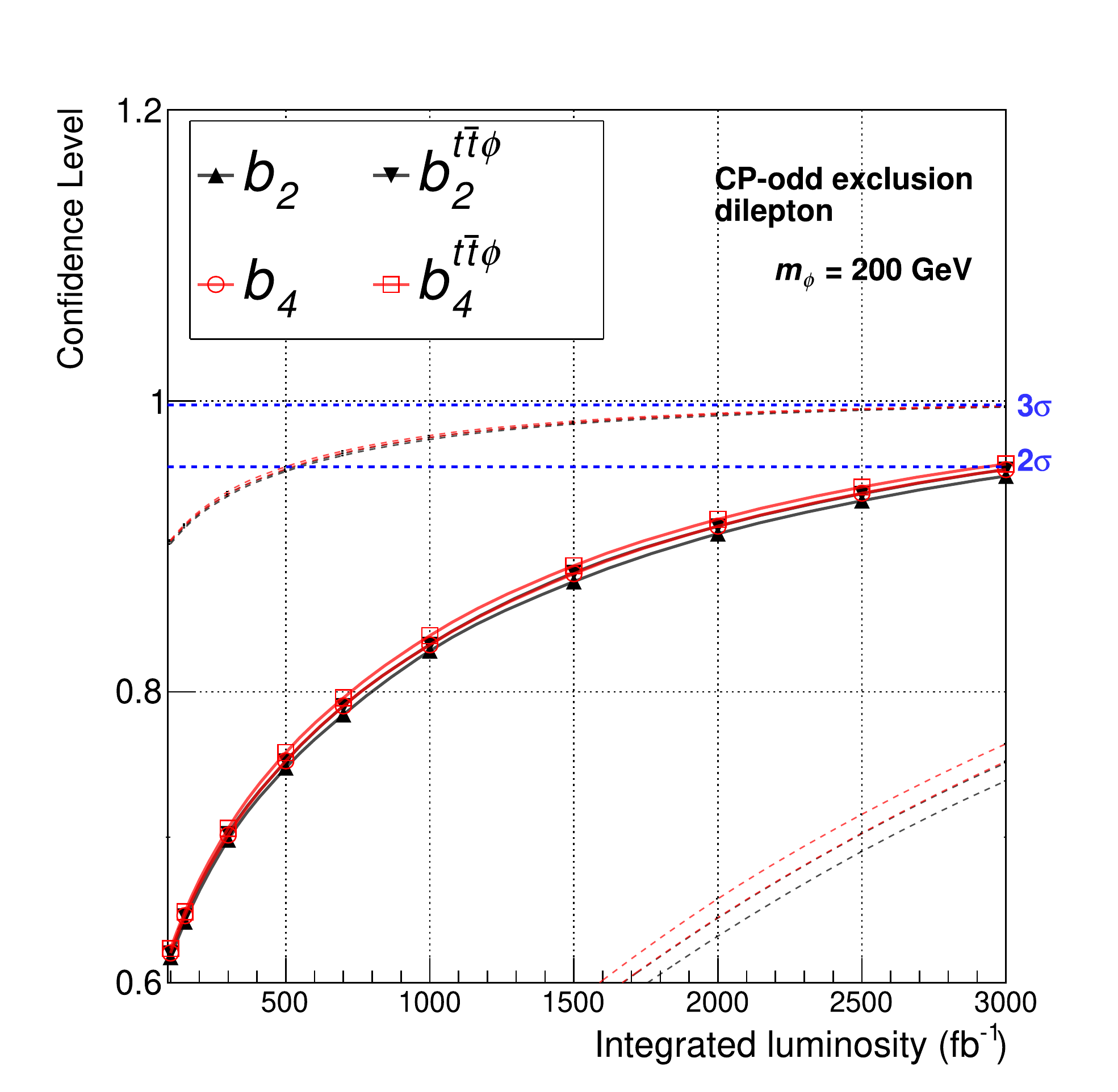}
		\end{tabular}
		\caption{Expected CLs for CP-odd exclusion assuming the SM (scenario 2), as a function of the integrated luminosity.}
		\label{case2}
	\end{center}
\end{figure}

\begin{figure}[h!]
	\begin{center}
		\begin{tabular}{ccc}
			\hspace*{-3mm}
			\includegraphics[height=7.cm]{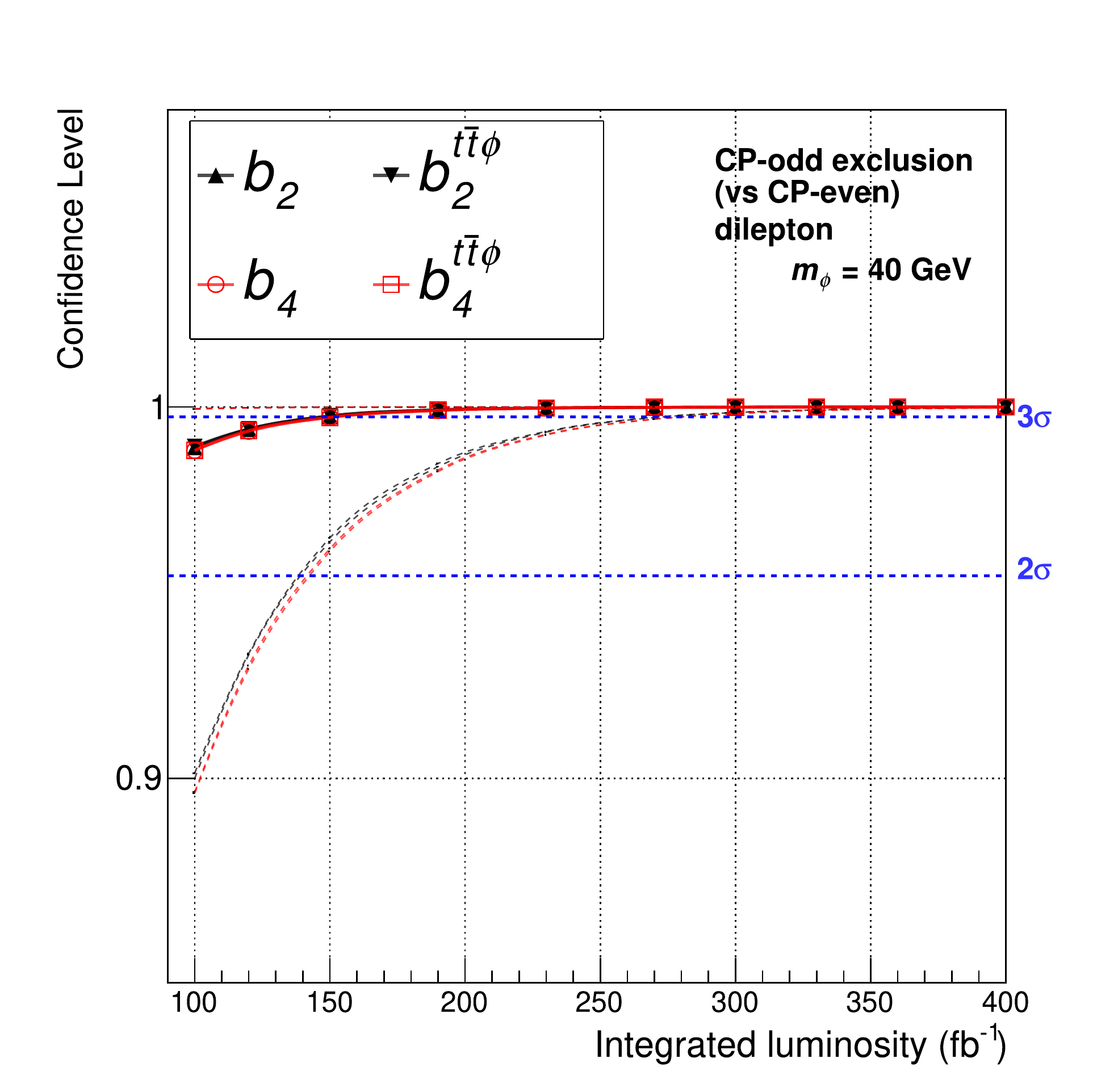}
			\includegraphics[height=7.cm]{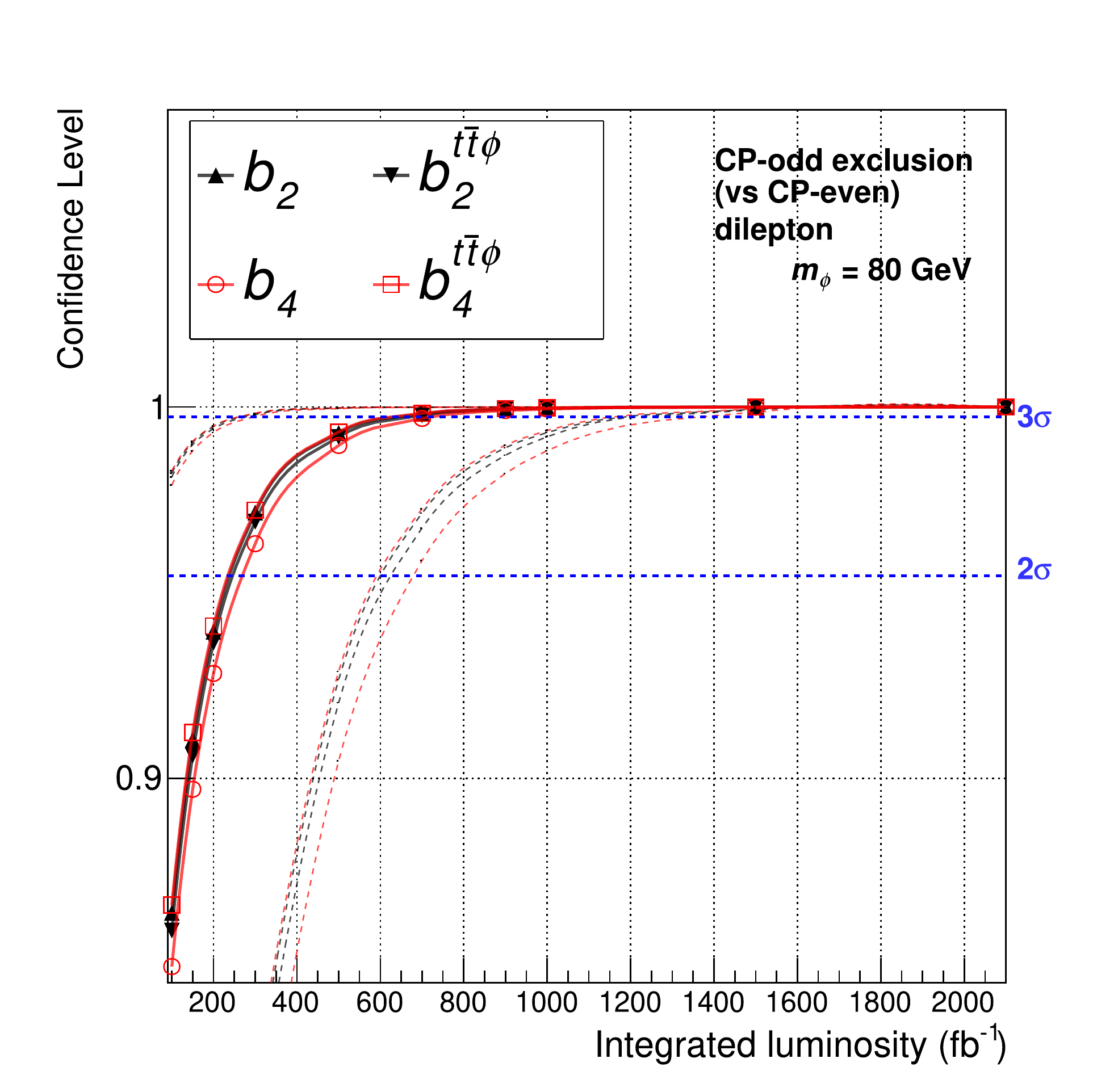}
			\\
			\includegraphics[height=7.cm]{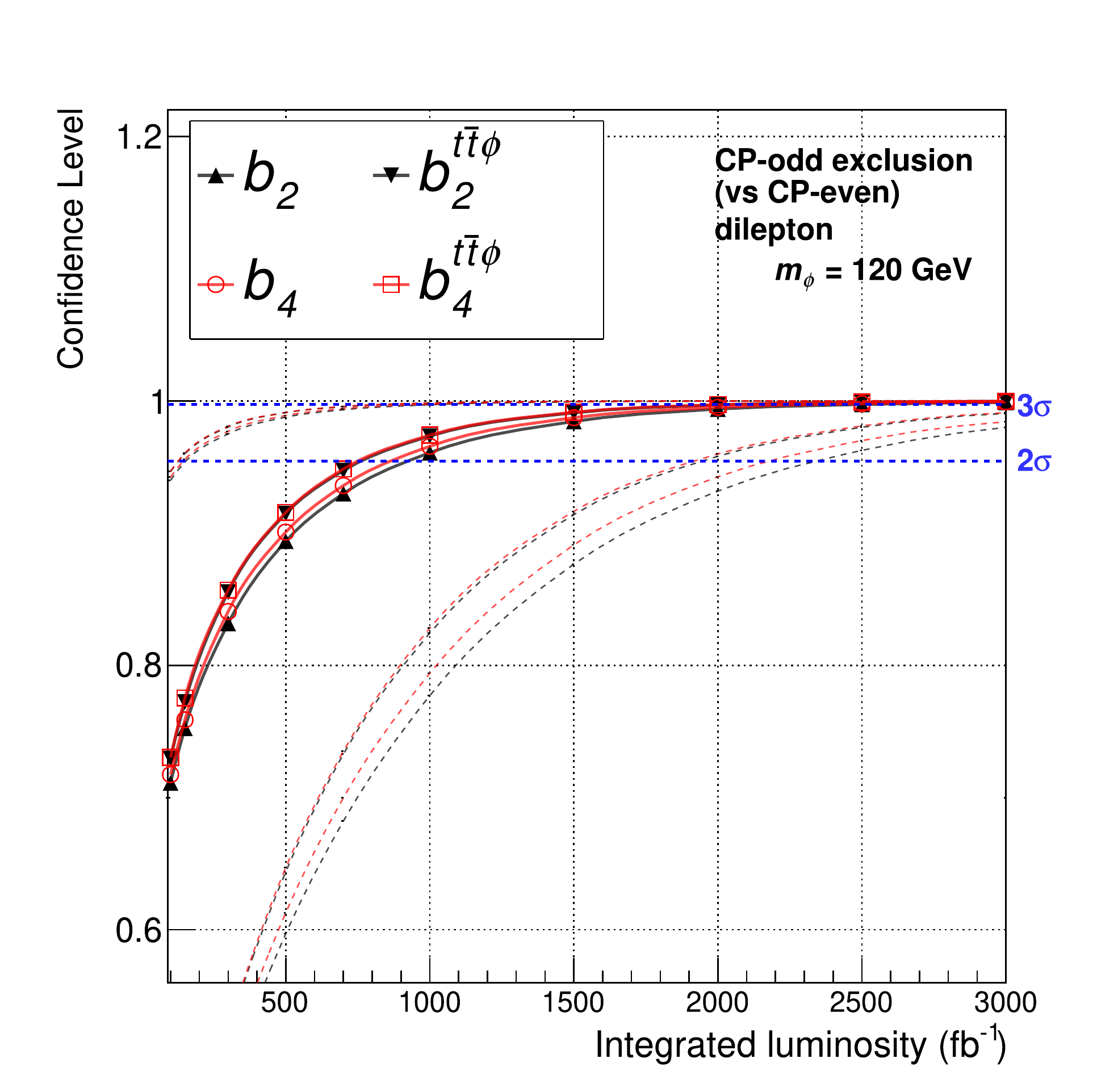}
			\includegraphics[height=7.cm]{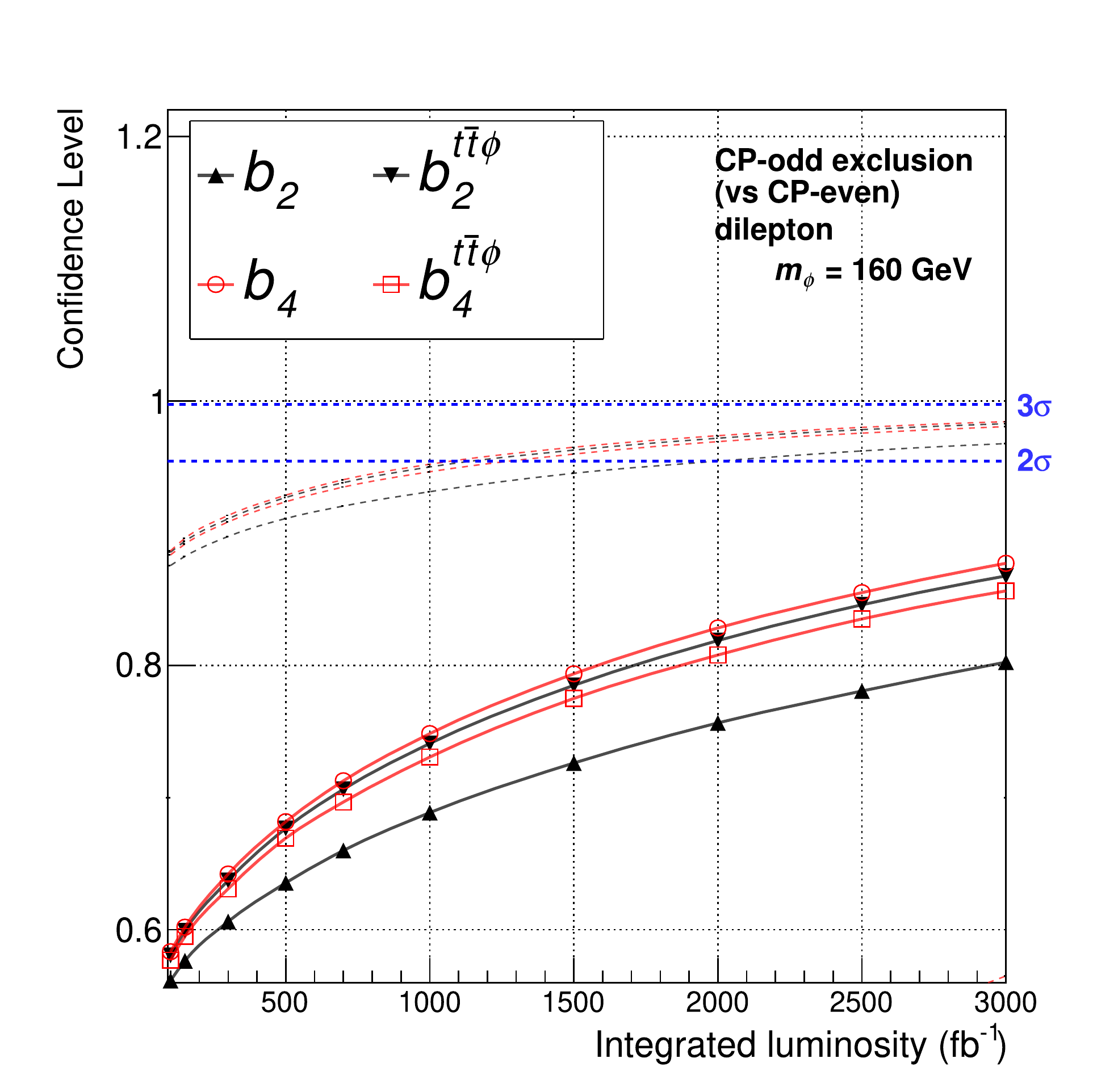}
			\\
			\includegraphics[height=7.cm]{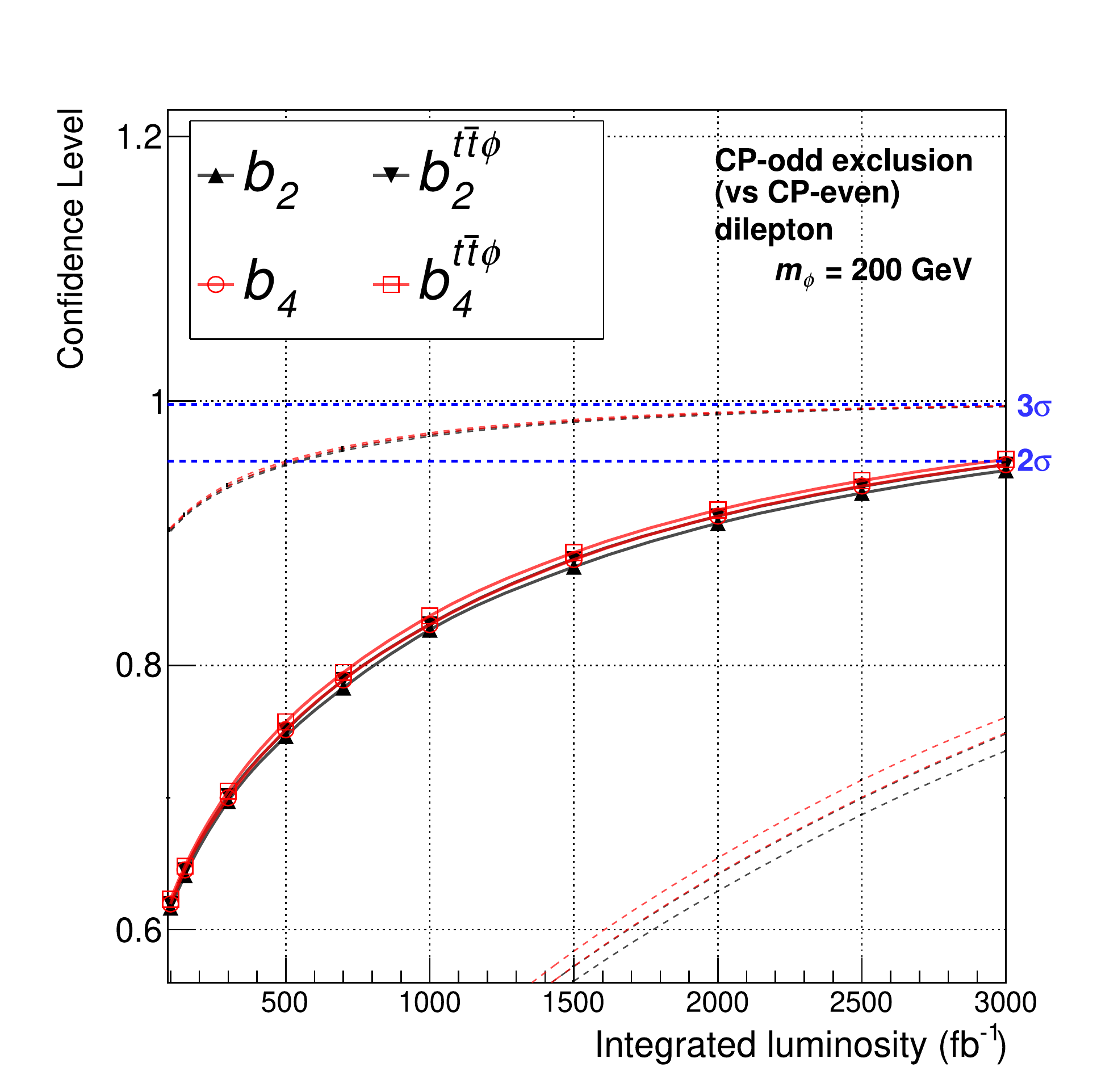}
		\end{tabular}
		\caption{Expected CLs for CP-odd exclusion assuming the SM plus a new CP-even scalar particle (scenario 3), as a function of the integrated luminosity.}
		\label{case3}
	\end{center}
\end{figure}

\begin{figure}[h!]
	\begin{center}
		\begin{tabular}{ccc}
			\hspace*{-3mm}
			\includegraphics[height=7.cm]{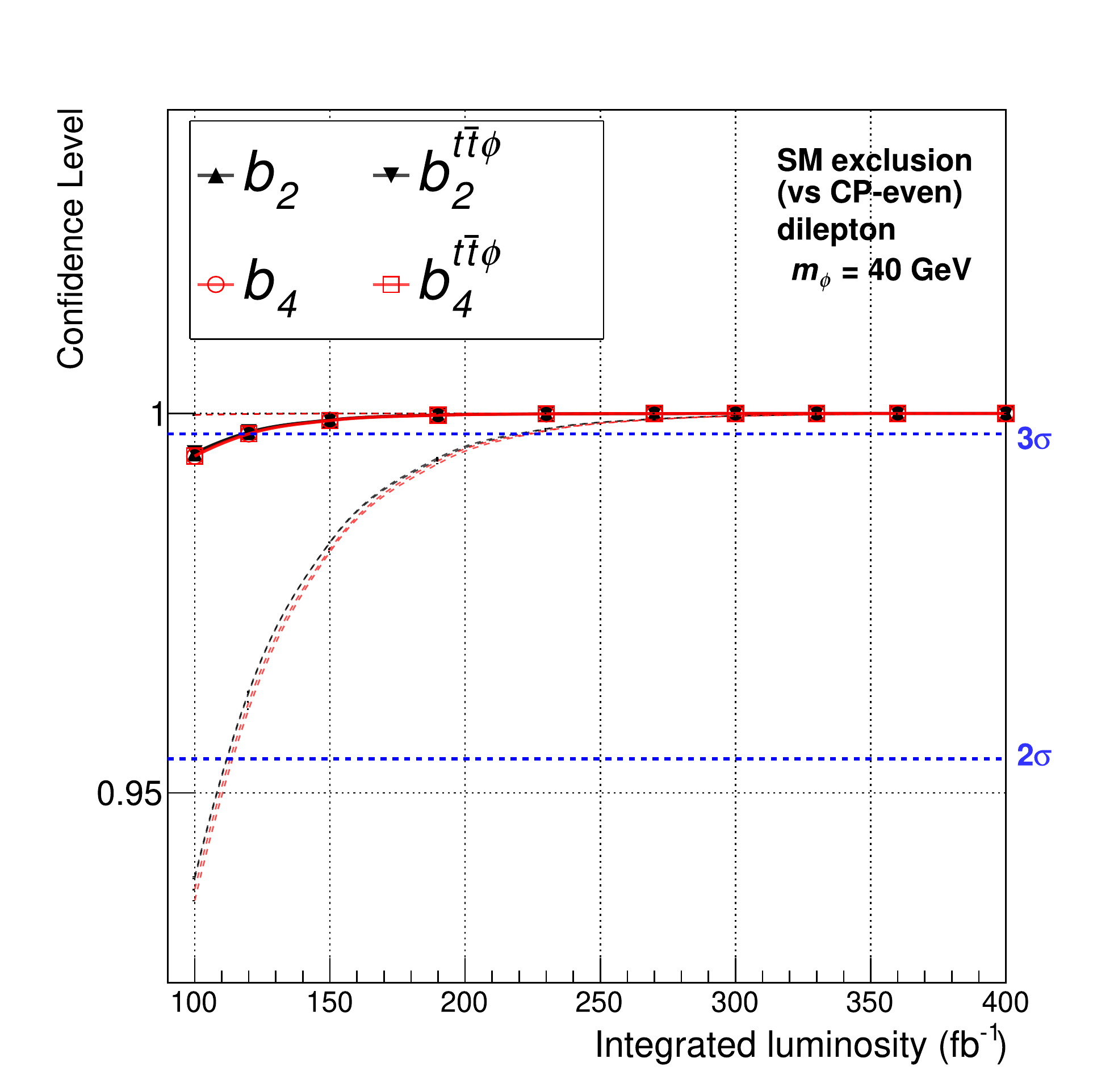}
			\includegraphics[height=7.cm]{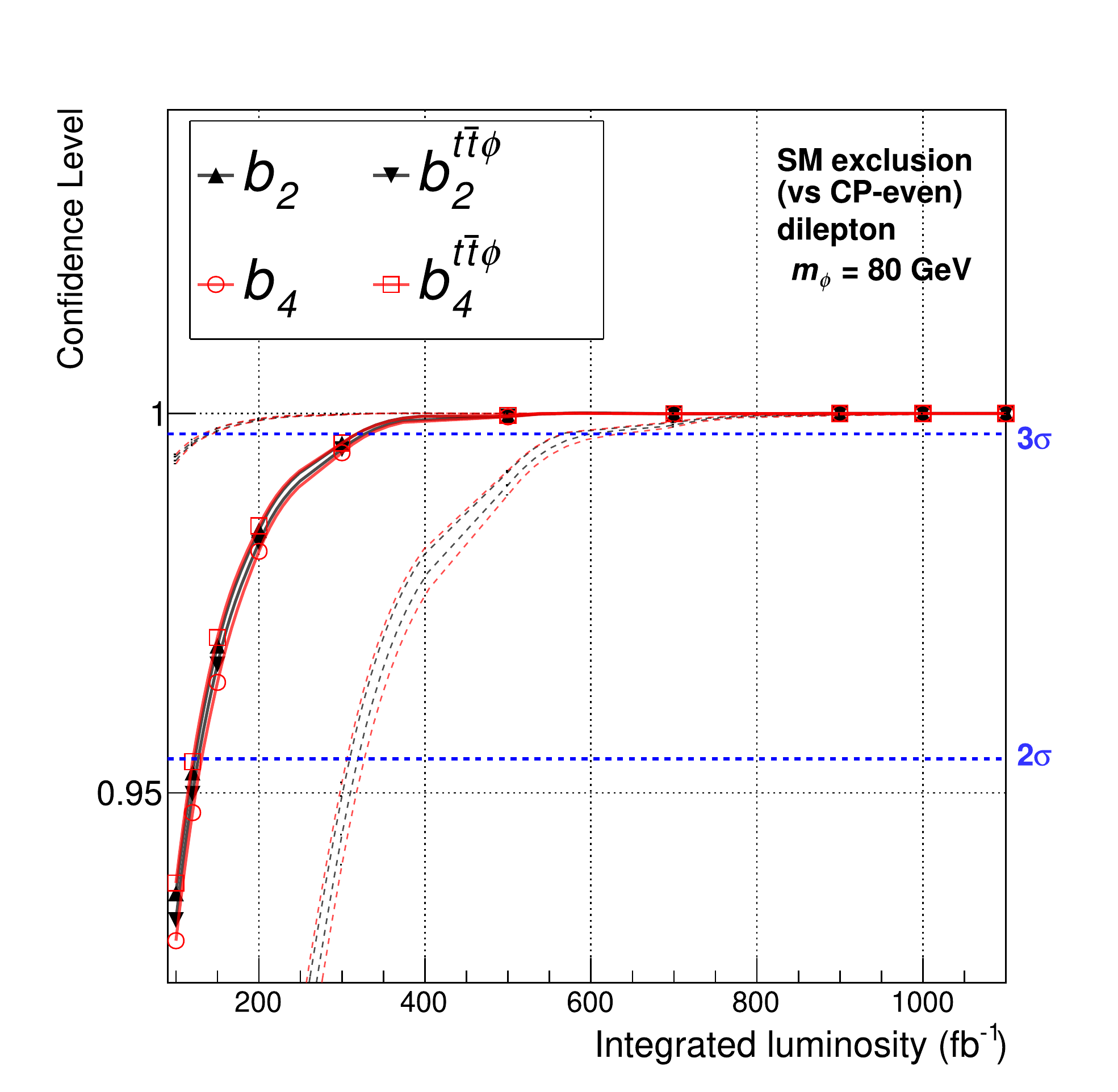}
			\\
			\includegraphics[height=7.cm]{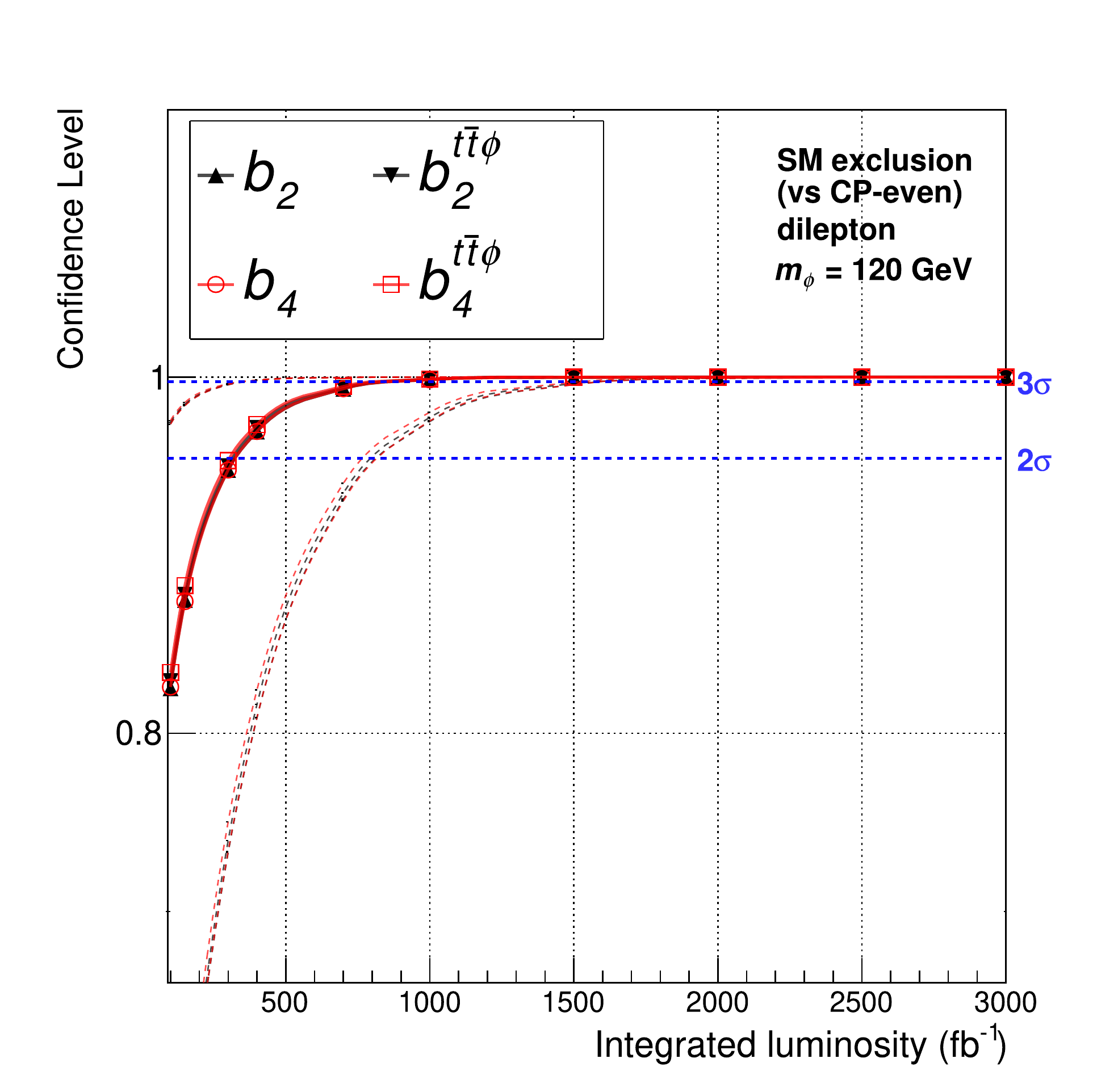}
			\includegraphics[height=7.cm]{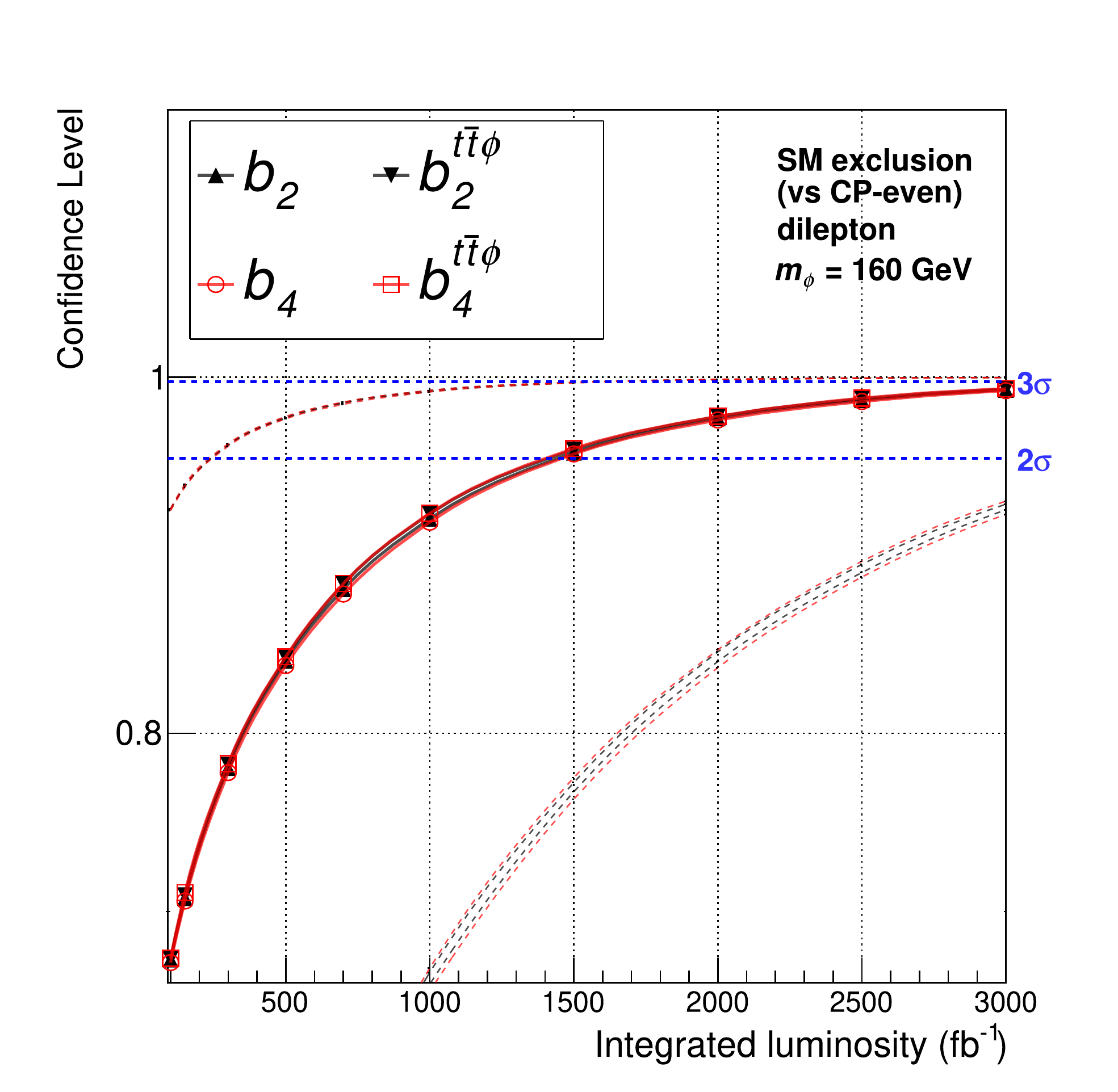}
		\end{tabular}
		\caption{Expected CLs for SM exclusion assuming the SM plus a new CP-even scalar particle (scenario 4), as a function of the integrated luminosity.}
		\label{case4}
	\end{center}
\end{figure}

Figures \ref{case1} to \ref{case4} show the expected exclusion CLs, for all four scenarios and different Higgs masses using only the dileptonic final states of $t\bar t\phi$. In each plot, the exclusion CLs using the $b_2$ and $b_4$ variables, as a function of the integrated luminosity, are shown for each given mass of the $\phi$ boson. 

Figure~\ref{case1} (Scenario 1) tells us that, with the current LHC luminosity, we can exclude a CP-even scalar boson with CLs that exceed $2\sigma$ if its mass is $m_\phi\lsim80$~GeV. 
For masses around the SM Higgs boson mass we require roughly 300~fb$^{-1}$ to achieve the $2\sigma$ CLs exclusion. This will be obtained during the incoming RUN's of the LHC. Masses of the CP-even scalar boson above 200~GeV cannot be excluded even at the end of HL-LHC, with the dileptonic channel, alone.

Figure \ref{case2} (Scenario 2) shows that the exclusion CLs are quite different for the CP-odd case, greatly due to the reduced cross section when compared with the CP-even case. To exclude pseudoscalars at $2\sigma$ with respect to the SM, and with masses in the range $m_\phi =80-200$~GeV, a luminosity of at least $\sim$1500~fb$^{-1}$ is required at the LHC. For the $m_\phi =40$ GeV case, the reconstruction efficiency is the lowest, thus the CLs are worse relative to the other CP-odd scalar boson masses. 

If a new CP-even scalar is discovered (Scenario 3), a CP-odd exclusion is possible for the mass range considered, with diminishing CLs for an increasing $	\phi$ boson mass, as shown in Figure~\ref{case3}. The exception for $m_\phi=160$ GeV, with much lower CLs, is due to both hypotheses presenting almost the same number of events distributed similarly in the variables considered, hence both hypotheses have similar results which degrades the sensitivity for exclusion. Without considering the uncertainties, the best variable is $b_4$ in the laboratory frame.

In Figure~\ref{case4} (scenario 4), we show that the SM only hypothesis will be excluded at the $5\sigma$ CL at some stage of the LHC lifetime, for masses of the discovered new CP-even scalar below $m_\phi\lsim 120$~GeV. For the case of $m_\phi=160$ GeV, it can be excluded with almost $3\sigma$ at the end of HL-LHC.

\begin{figure}[h!]
	\begin{center}
		\begin{tabular}{ccc}
			\hspace*{-3mm}
			\includegraphics[height=7.cm]{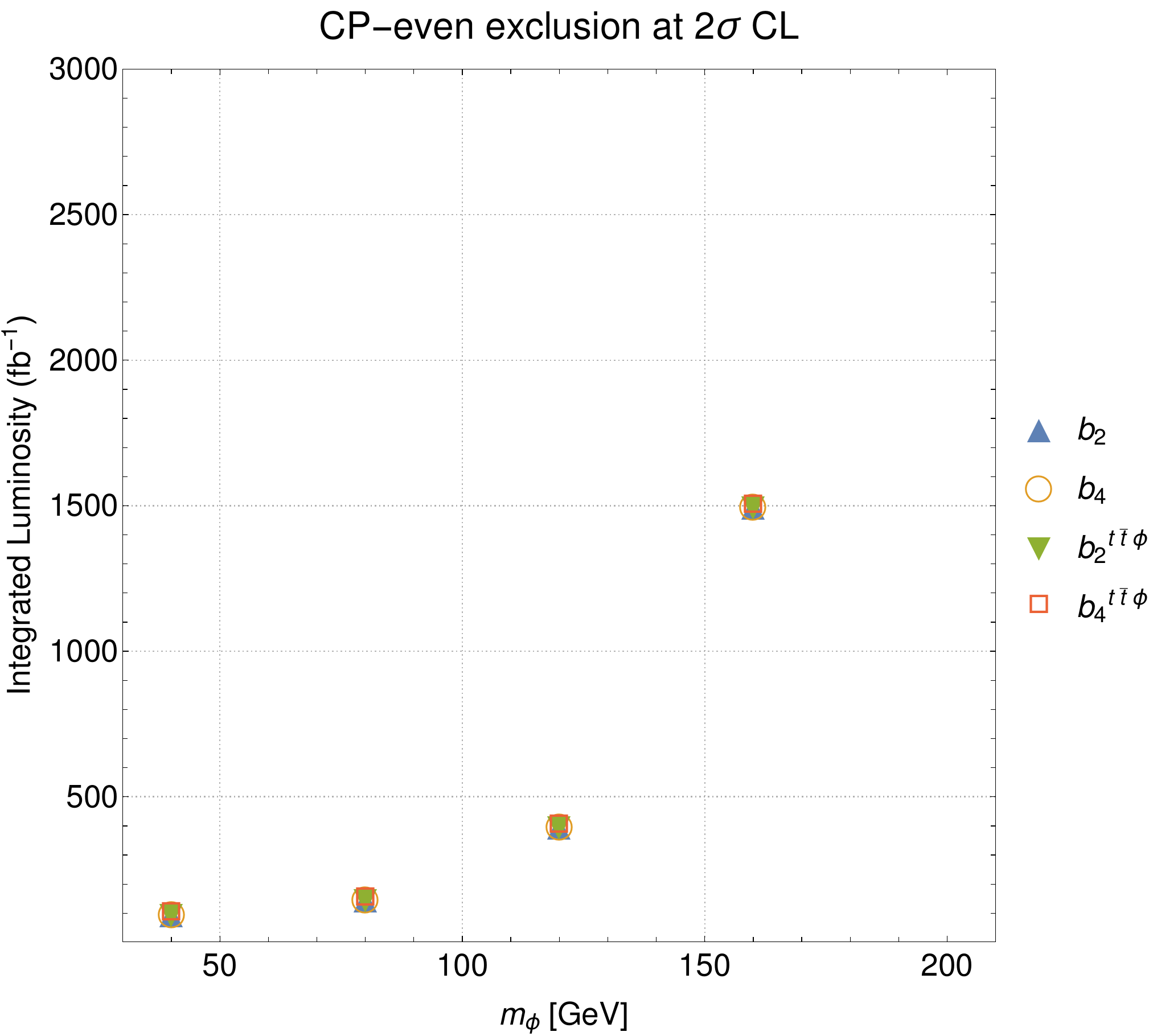}
			\includegraphics[height=7.cm]{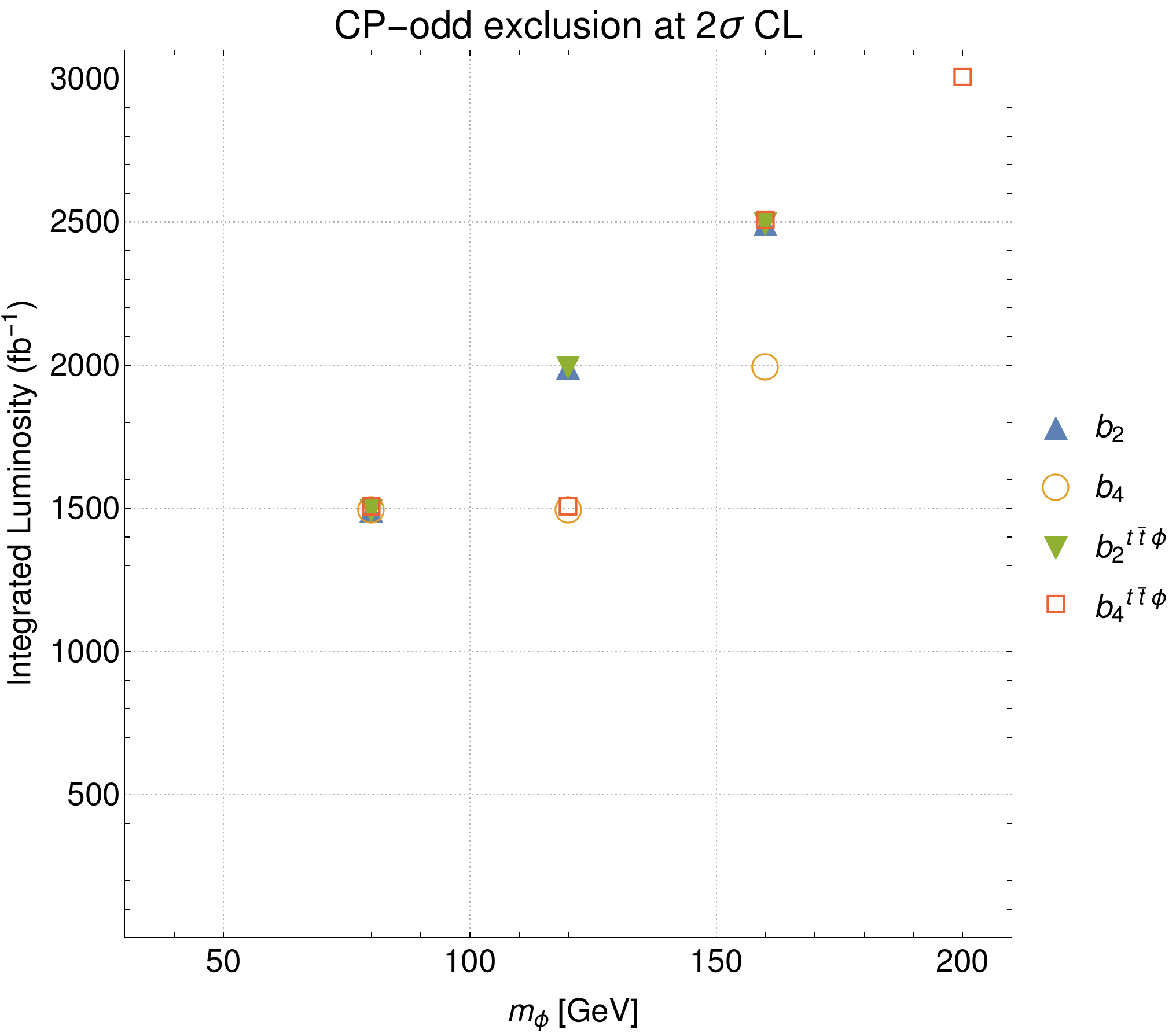}
			\\
			\hspace*{-3mm}
			\includegraphics[height=7.cm]{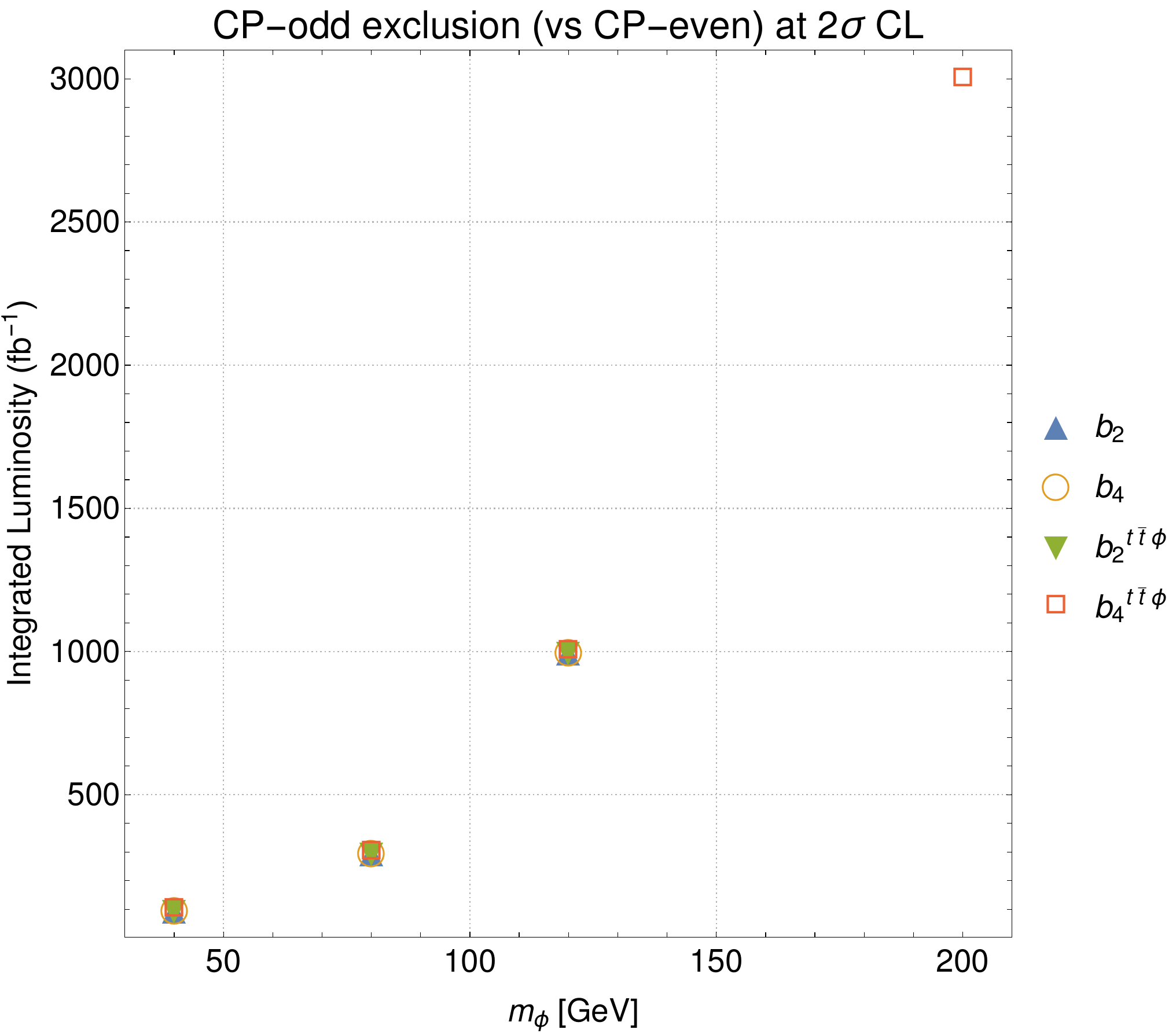}
			\includegraphics[height=7.cm]{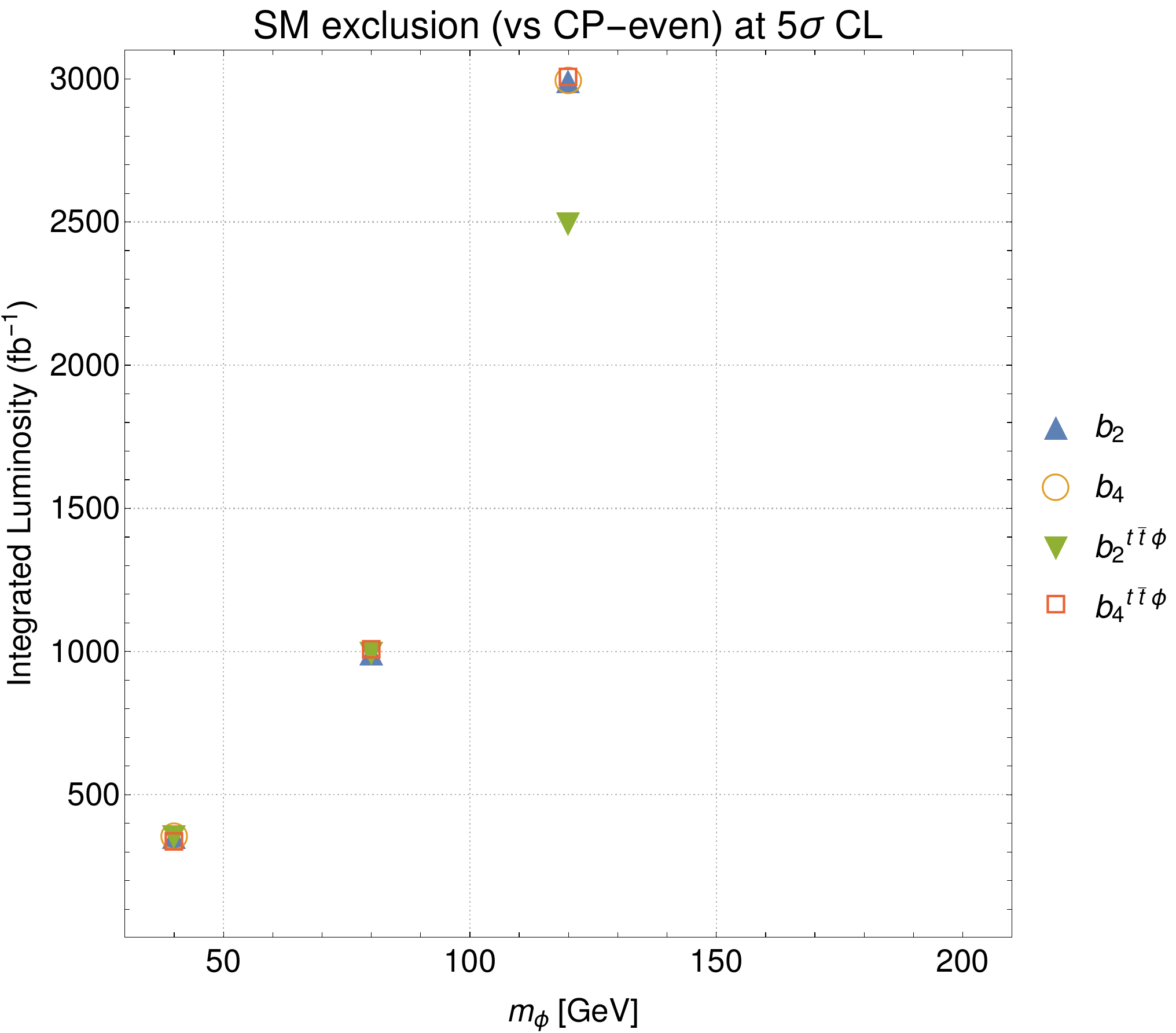}
			
		\end{tabular}
		\caption{Luminosity needed to exclude scenarios 1 (top left), 2 (top right) and 3 (bottom left) at the $2\sigma$ level, and scenario 4 (bottom right) at the $5\sigma$ level, as a function of the $\phi$ boson mass.}
		\label{summary}
	\end{center}
\end{figure}

Finally, a summary of the above conclusions is presented in Figure \ref{summary}. Each plot shows the luminosity required for exclusion, at a given CL ($2\sigma$ for the first three scenarios and $5\sigma$ for scenario 4), for each one of the scenarios considered. If no points are shown, the exclusion is not possible at the end of HL-LHC ($\mathcal{L}$ = 3000~fb$^{-1}$).

\clearpage

\clearpage

\section{Interpretation in the framework of the C2HDM}
\label{sec:C2HDM}
\hspace{\parindent} 

We will now interpret the results in the framework of the C2HDM. We will briefly review the relevant aspects of the C2HDM to be used in the discussion (for a detailed description of the model see~\cite{ Fontes:2017zfn}). 
In the C2HDM the scalar potential is explicitly CP-violating and is invariant under a 
$Z_2$ symmetry $\Phi_1 \to \Phi_1, \Phi_2 \to -\Phi_2$, softly broken by the $m_{12}^2$ term,
\beq
V &=& m_{11}^2 |\Phi_1|^2 + m_{22}^2 |\Phi_2|^2
- \left(m_{12}^2 \, \Phi_1^\dagger \Phi_2 + h.c.\right)
+ \frac{\lambda_1}{2} (\Phi_1^\dagger \Phi_1)^2 +
\frac{\lambda_2}{2} (\Phi_2^\dagger \Phi_2)^2 \nonumber \\
&& + \lambda_3
(\Phi_1^\dagger \Phi_1) (\Phi_2^\dagger \Phi_2) + \lambda_4
(\Phi_1^\dagger \Phi_2) (\Phi_2^\dagger \Phi_1) +
\left[\frac{\lambda_5}{2} (\Phi_1^\dagger \Phi_2)^2 + h.c.\right] \; ,
\eeq
where the doublets $\Phi_i$ $(i=1,2)$ develop real vacuum expectation values (VEVs) $v_{1}$ and $v_{2}$.
All parameters are real except for $m_{12}^2$ and $\lambda_5$. We define $\tan \beta \equiv \frac{v_2}{v_1} $
and the rotation matrix that takes us from the gauge to the mass eigenstates is
\beq
\left( \begin{array}{c} H_1 \\ H_2 \\ H_3 \end{array} \right) = R
\left( \begin{array}{c} \rho_1 \\ \rho_2 \\ \rho_3 \end{array} \right)
\; ,
\label{eq:c2hdmrot}
\eeq
with
\be
R =
\left(
\begin{array}{ccc}
	c_1 c_2 & s_1 c_2 & s_2\\
	-(c_1 s_2 s_3 + s_1 c_3) & c_1 c_3 - s_1 s_2 s_3  & c_2 s_3\\
	- c_1 s_2 c_3 + s_1 s_3 & -(c_1 s_3 + s_1 s_2 c_3) & c_2 c_3
\end{array}
\right)\, ,
\label{matrixR}
\ee
where $s_i = \sin{\alpha_i}$,
$c_i = \cos{\alpha_i}$ ($i = 1, 2, 3$),
and
\be
- \pi/2 < \alpha_1 \leq \pi/2,
\hspace{5ex}
- \pi/2 < \alpha_2 \leq \pi/2,
\hspace{5ex}
- \pi/2 < \alpha_3 \leq \pi/2.
\label{range_alpha}
\ee
The Higgs boson masses are ordered such that $m_{H_1} \le m_{H_2} \le m_{H_3}$. In the C2HDM, there are four types of Yukawa models. 
However, the top Yukawa couplings are the same in all four types and therefore this discussion is valid for all of them. The Yukawa Lagrangian for the
up quarks in all four types has the form
\beq
{{\cal L}_Y}_i = - \frac{{m_f}}{v} \bar{\psi}_f \left[ \frac{R_{i2}}{s_\beta} - i \frac{R_{i3}}{t_\beta}  \gamma_5 \right] \psi_f H_i \;, \label{eq:yuklag}
\eeq
where $\psi_f$ denotes the fermion fields with mass $m_f$, $i$ is the scalar index, $v^2=v_1^2 + v_2^2$ (fixed by the $W$ boson mass) and $t_\beta = \frac{v_2}{v_1}$. 

What we want to understand now is what can be concluded for the parameter space of this specific model once we have either a measurement or an exclusion
for a given $\phi$ boson mass (and luminosity). We will just analyse a simple situation where the 125 GeV Higgs is $H_2$ and the lightest Higgs is $H_1$, with a mass below 125 GeV.
We start by mapping equation~\ref{eq:yuklag} into equation~\ref{eq:higgscharacter}.,
\begin{equation}
\left\{
\begin{aligned} \kappa_t \cos \alpha & = \frac{s_1 \, c_2}{s_{\beta}}
\\ \kappa_t \sin \alpha & =  - \frac{s_2}{t_{\beta}}
\end{aligned} \right.
\qquad \qquad s_\beta ^2 \kappa_t^2  = s_1^2 c_2^2 + s_2^2 c_\beta^2 .
\end{equation}
The values of $\kappa_t$ and $\alpha$ are free to vary in their allowed range (taking into account available theoretical and experimental constraints)
because no scalar was found below 125 GeV. Let us start by noting that  $ \sin \alpha = 0$ and $ \sin \alpha_2 = 0$ are equivalent. This means that
the CP-even limit is obtained unambiguously. The $H_1 VV$ coupling, where V is a vector boson, is proportional
to $\cos \alpha_2$ which vanishes for $\alpha_2 = \pi/2$. 

What will be measured or constrained in the experiment is $\alpha$ and $\kappa_t$. Also, the limits in this work were set
for the pure scalar and pure pseudoscalar scenarios. For these scenarios we get, respectively,
\begin{equation}
\left\{
\begin{aligned} 
\sin \alpha = 0 & \implies \kappa_t = \pm \frac{s_1}{s_{\beta}} \, ,
\\ \cos \alpha= 0 & \implies \kappa_t = \pm \frac{s_2}{t_{\beta}} \,\,  (\text{if} \,\,  s_1=0)  \quad \text{or} \quad \kappa_t = \pm \frac{1}{t_{\beta}} \,\,  (\text{if} \,\,  c_2=0) \, , 
\end{aligned} \right.
\end{equation}
and a measurement or limit on $\kappa_t$ will set a limit on the parameters of the model. For the particular scenario where $c_2=0$ we obtain
a limit on $\tan \beta$. Because $\tan \beta$ is already constrained to be above one by low energy physics measurements (see~\cite{ Fontes:2017zfn})
information can only be added if we increase the limit. This is in fact the case, if the limit is for instance $\kappa_t \leq 1/10$ we get $\tan \beta \geq 10$ ($c_2=0$).
With the same limit for $\kappa_t$, in the remaining two scenarios the bound on the parameters is $s_1 \leq 1/10$ ($s_2=0$)  and $s_2 \leq t_\beta/10$ ($s_1=0$).
\begin{figure}[h!]
	\begin{center}
		\begin{tabular}{ccc}
			\hspace*{-3mm}
			\includegraphics[height=7.cm]{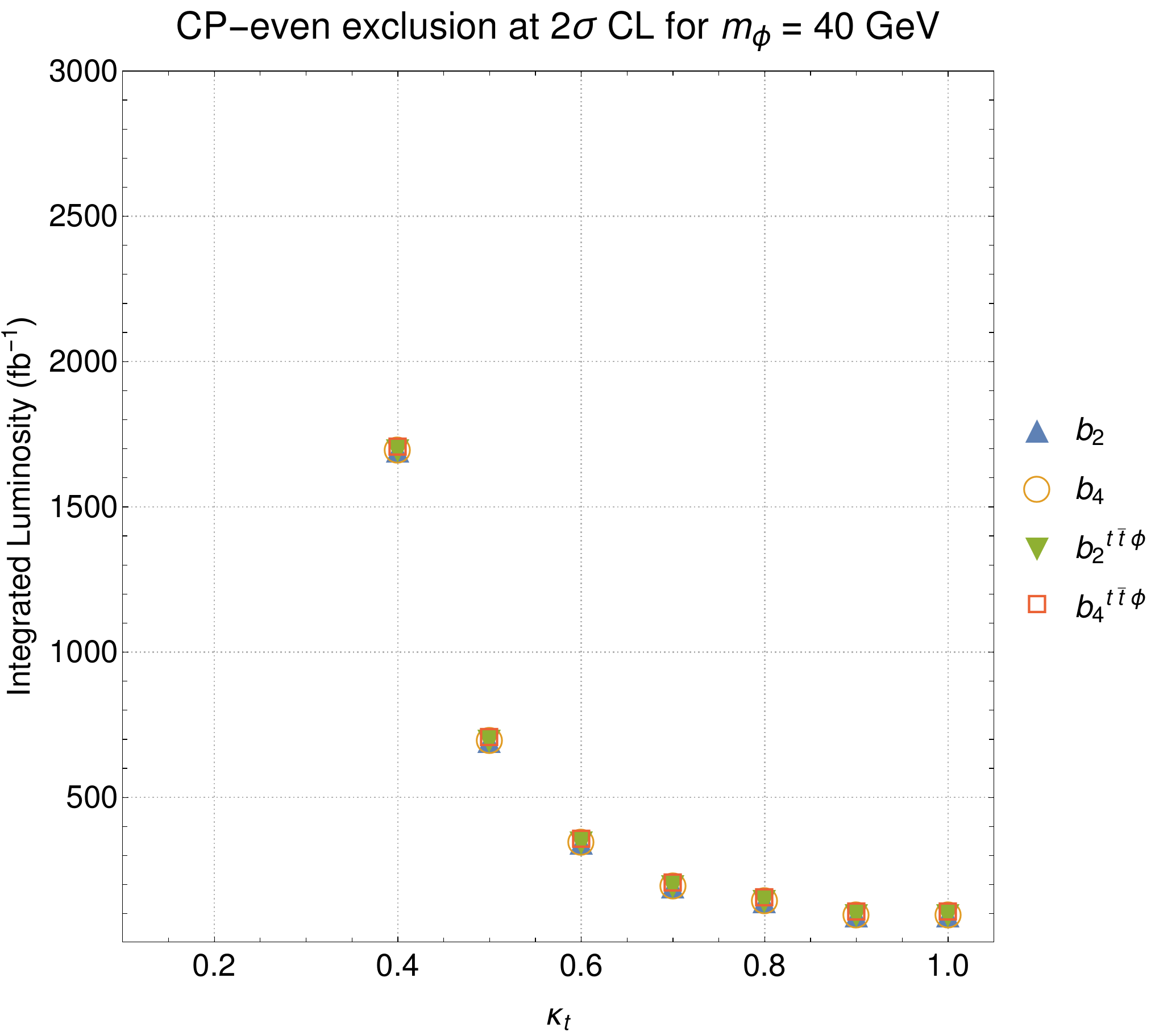}	
		\end{tabular}
		\caption{Luminosity needed to exclude $\kappa_t$ at the 2$\sigma$ level for the pure CP-even case (scenario 1), for a CP-even scalar boson mass of 40 GeV.}
		\label{last}
	\end{center}
\end{figure}
In figure~\ref{last} we present the luminosity needed to exclude $\kappa_t$ at the 2$\sigma$ level for the pure CP-even case (scenario 1), for a CP-even scalar boson mass of 40 GeV. Note
that this is the most favourable scenario for discovery (and for exclusion). As can be seen the value of $\kappa_t$ attainable is close to 0.3 by the end of the LHC run.
However, because this is a study using the dileptonic final state, we can expect to reach values of $\kappa_t$ of the order of 1/10 for an analysis which includes all other decay channels.

\begin{figure}[!h]
	\begin{center}
		\begin{tabular}{ccc}
			\includegraphics[width=7cm]{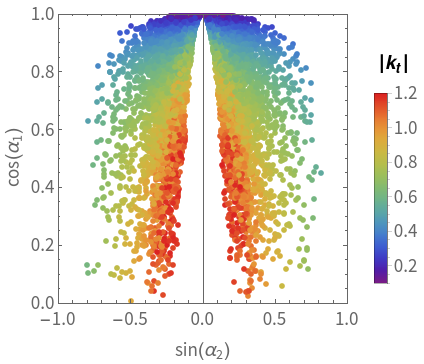}\hspace*{1cm}
			 \includegraphics[width=7cm]{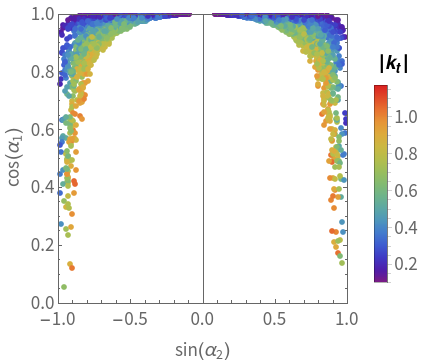}\\
			
			\includegraphics[width=7cm]{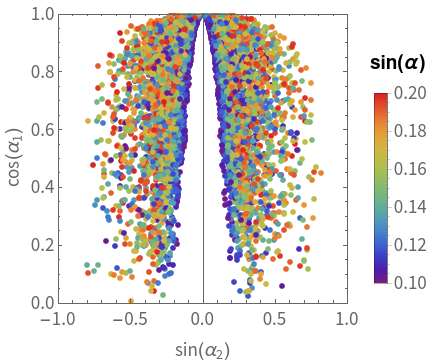}\hspace*{1cm}	 \includegraphics[width=7cm]{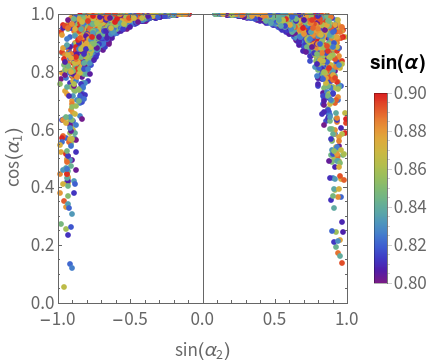}\\
			
			\includegraphics[width=7cm]{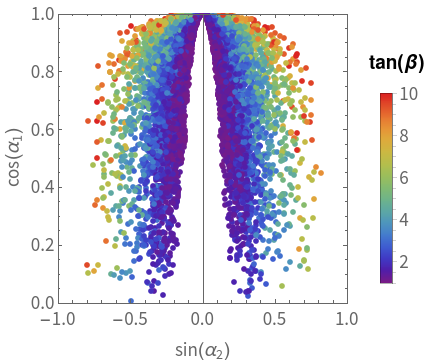}\hspace*{1cm}
			\includegraphics[width=7cm]{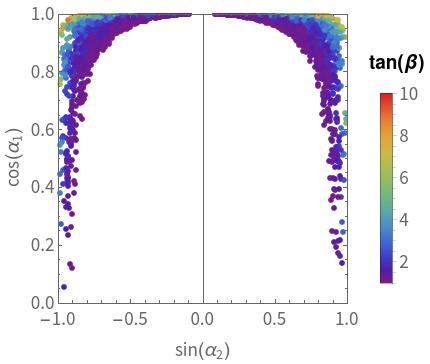}				
		
		\end{tabular}
		\caption{Points allowed in the plane $c_1$ vs. $s_2$ for $0.1 \leq \kappa_t \leq 1.2$  and  $1 \leq \tan \beta \leq 10$. In the left we impose $0.1 \leq \sin \alpha \leq 0.2$ (CP-even like) and in the right we impose $0.8 \leq \sin \alpha \leq 0.9$ (CP-odd like). 
		On the top, we color superimpose $\kappa_t$, in the middle $\sin \alpha$ and on the bottom $\tan \beta$.}
		\label{high_alt}
	\end{center}
\end{figure}

The next question to ask is what are the constraints on the parameter space in scenarios where one is either close to CP-even or to the CP-odd scenario. 
In Figure~\ref{high_alt} (left), we present the allowed points in the C2HDM parameter space ($c_1$ vs. $s_2$) if a measurement of $\kappa_t$ and $\sin \alpha$ is in the ranges
$0.1 \leq \kappa_t \leq 1.2$  and $0.1 \leq \sin \alpha \leq 0.2$. We also force $1 \leq \tan \beta \leq 10$. In the top plot we see the variation with $\kappa_t$, in the middle with $\sin \alpha$ and on the bottom with $\tan \beta$.
This is the case where we are close to the CP-even limit. 

In figure~\ref{high_alt} (right), we present the scenario when we are close to the CP-odd limit, that is $0.8 \leq \sin \alpha \leq 0.9$.
The most striking point is that although in each case we are closer to one of the limits, CP-even or CP-odd, the allowed parameter space is quite large and we clearly need some
other sources of measurement to constraint the parameter space.
%
%

\section{Conclusions \label{sec:concl}}
\hspace{\parindent} 

In this paper we examine the possibility of determining the CP nature of the heavier quarks ($b$ and $t$ quarks) Yukawa interactions with a generic scalar boson $\phi$, in $q \bar{q} \phi$ production at the LHC. We found that strategies to achieve this goal suggested in the literature for the case of the top quark do not work for the bottom quark, even at parton level. This was also confirmed for very light Higgs bosons with masses of the order of 10~GeV. The underlying reason is the interference term, responsible for the CP-asymmetries, which is proportional to $m_f^2$. Hence, this term is only meaningful when the fermion mass is of the order of $m_\phi$. 

Previous works established that several kinematic distributions for $t \bar{t} \phi$ are sensitive to the CP-components of the top quark Yukawa coupling. These studies assumed $m_\phi =$ 125~GeV and an extension of the study to other masses of scalar and pseudoscalar Higgs bosons was still missing in the literature. In this paper, we investigate the dilepton final states of $t \bar{t} \phi$ (with $\phi = H, A$) for several masses of the CP-even or CP-odd boson ($\phi$). We found that for the masses considered, there is still a good level of discrimination between scalar and pseudoscalar Yukawa interactions, at parton level. However, the differences between those cases become smaller as the Higgs mass increases, and vanish around $m_\phi = $ 450 GeV. 

A full kinematic reconstruction was applied to signal and background events, to reconstruct the four momenta of the undetected neutrinos, allowing to estimate the experimental sensitivity of the CP-search. CLs are presented for the exclusion of several scenarios as a function of the luminosity, for different Higgs boson masses. Generally, it is shown that the required luminosity for exclusion at a given CL increases with the $\phi$ boson mass. Given the current LHC luminosity, of 150 fb$^{-1}$, exclusion of the SM plus a pure CP-even Higgs with masses of 40 and 80 GeV and SM couplings, assuming the SM only, is already possible. For $m_H > 200$ GeV, CP-searches will require the inclusion of additional channels. We also found that the SM plus a CP-odd scalar exclusion, assuming the SM only, is harder than the CP-even exclusion for CP-odd Higgs masses up to 160 GeV. For higher masses, the opposite is true. If a new Higgs is found, we have enough sensitivity to exclude the possibility of the scalar being purely CP-odd in the explored mass range, again assuming SM-like couplings. In this work, only the dileptonic final states of the $t\bar t \phi$ system is considered in the CLs evaluation, at the LHC. A natural follow up would be to combine several $t \bar{t} \phi$ decay channels, to further improve the results obtained in this paper.

Finally, the impact of a new discovery was discussed for the C2HDM. If a new particle is found to be an exact CP-eigenstate, this will impose further constrains on typical 2HDM parameters such as $\tan \beta$. In case the new particle is just close to either the CP-even or the CP-odd scenarios, the allowed parameter space would still be very large and will require other measurements to further constrain it.

\clearpage

\subsubsection*{Acknowledgments}
DA, RC and RS are partially supported by the Portuguese Foundation for Science and Technology (FCT) under Contracts no. UIDB/00618/2020, UIDP/00618/2020, PTDC/FIS-PAR/31000 /2017 and CERN/FIS-PAR/0002/2017, and the HARMONIA project under contract UMO-2015/18/M/ ST2/00518. AO is partially supported by FCT, under the Contract CERN/FIS-PAR/0029/2019. DA is supported by ULisboa - BD2018.



\vspace*{1cm}
\bibliographystyle{h-physrev}
\bibliography{rodrigomaster.bib}

\end{document}